\documentclass[%
prd,
 reprint,
preprintnumbers,
 amsmath,amssymb,
 aps,
]{revtex4-2}

\usepackage{graphicx}
\usepackage{dcolumn}
\usepackage{bm}
\usepackage{svg}
\usepackage{hyperref}

\newcommand{\llangle}{{\langle\!\langle}}
\newcommand{\rrangle}{{\rangle\!\rangle}}
\newcommand{\calS}{{\cal{S}}}
\newcommand{\calO}{{\cal{O}}}
\newcommand{\tr}{{\rm{Tr}}}

\newcommand{\eff}{{\rm{eff}}}
\newcommand{\Sys}{{\rm{S}}}
\newcommand{\Env}{{\rm{E}}}
\newcommand{\rmI}{{\rm{I}}}

\newcommand{\oQ}{\overline{Q}}
\newcommand{\bfk}{{\bf{k}}}

\newcommand{\bfq}{{\bf{q}}}
\newcommand{\bfE}{{\bf{E}}}
\newcommand{\bfV}{{\bf{V}}}
\newcommand{\bfr}{{\bf{r}}}
\newcommand{\HTL}{{\rm{HTL}}}

\newcommand{\ket}[1]{\lvert #1 \rangle }
\newcommand{\norm}[1]{\lvert\rvert #1 \lvert\rvert}

\begin{document}

\preprint{TIFR/TH/25-9}

\title{Non-Markovian dynamics of bottomonia in the QGP}

\author{Vyshakh B R}
\email{vyshakh.br@theory.tifr.res.in}
\affiliation{Department of theoretical physics \\ TIFR, Mumbai, India, 400005}

\author{Rishi Sharma}
\email{rishi@theory.tifr.res.in}
\affiliation{Department of theoretical physics \\ TIFR, Mumbai, India, 400005}

\date{\today}

\begin{abstract}

The evolution of quarkonia in the QGP medium can be described through the formalism of Open Quantum Systems (OQS). In previous works with OQS, the quarkonium evolution was studied by working in either quantum Brownian or optical regime. In this paper, we set up a general non-Markovian master equation to describe the evolution of the quarkonia in the medium. Due to the non-Markovian nature, it cannot be cast in Lindblad form which makes it challenging to solve. We numerically solve the master equation for $\Upsilon(1S)$ without considering stochastic jumps for Bjorken and viscous hydrodynamic background at energies relevant to LHC and RHIC. We quantify the effect of the hierarchy between the system time scale, $\tau_{\Sys}\sim 1/E_b$, and the environment time scale, $\tau_{\Env}$, on quarkonium evolution and show that it significantly affects the nuclear modification factor. A comparison of our results with the existing experimental data from LHC and RHIC is presented.  

\end{abstract}

\maketitle



\section{\label{sec:Introduction}Introduction}
The key processes governing the dynamics of quarkonia (bound states of a heavy quark $Q$ and its antiquark $\oQ$) in the Quark-Gluon Plasma (QGP) have been recognized for a long time. They include gluo-dissociation~\cite{peskin19791,bhanot19791}, screening of the interaction between $Q$ and $\oQ$~\cite{matsui19861}, and dissociation through $t$-channel scattering with medium~\cite{Grandchamp:2001pf,laine20071}~\footnote{Additionally, regeneration~\cite{Grandchamp:2001pf,grandchamp20021,grandchamp20041,greco20031} via coalescence of $Q$ and $\oQ$ created at different hard vertices is important for charmonia. For bottomonia, this contribution is expected to be small~\cite{du20171}.}. They have been widely used to address the phenomenology of quarkonia at RHIC and LHC (see~\cite{andronic20151,Andronic:2024oxz} for reviews).

Additional insight into these processes has been obtained by studying them in an Effective Field Theory (EFT) for quarkonium states: potential Non-Relativistic QCD (pNRQCD)~\cite{brambilla20001}. pNRQCD relies on the hierarchy between the energy scales $M\gg1/r\gg E_b$, where $M$ is the mass of $Q$, $1/r$ is the inverse of the relative separation between $Q-\oQ$, and $E_b$ is the binding energy of quarkonia. It is a theory of singlet [$\calS(\bfr, t)$] and octet [$\calO(\bfr, t)$] wave functions that describe the $Q-\oQ$ state, and gluonic and light quark modes with momenta and energies $\lesssim E_b$. The Lagrangian up to order $E_b^2r$ has the form,
\begin{equation}
\begin{split}
    &L_{\rm{pNRQCD}} =
    \int d^3\bfr\; {\rm{tr}} 
    \Bigl( \calS(\bfr, t)^\dagger [i\partial_0 - h_s]\calS(\bfr, t) \\
    &
        + \calO(\bfr, t)^\dagger [iD_0 - h_o]\calO(\bfr, t)  \\ 
    & + gV_A(r)[\calO^\dagger(\bfr, t) \bfr\cdot \bfE \calS(\bfr, t)+\calS^\dagger(\bfr, t) \bfr\cdot \bfE \calO(\bfr, t)] \\ 
    & 
      + \frac{V_B(r)}{4}\{\calO^\dagger(\bfr, t),\; \lbrace\bfr\cdot g\bfE,\;\calO(\bfr, t)\rbrace\}  \Bigr)\\ 
    & 
      +L_{\rm{light}}\;,
\label{eq:LpNRQCD}
\end{split}
\end{equation}
Here $h_s=p^2/M+v_s$ ($h_o=p^2/M+v_o$) is the singlet (octet) Hamiltonian and $\bfE$ is the chromoelectric field. $v_{s, o}$  [similarly $V_{A, B}(r)$] are Wilson coefficients (which can depend on $\bfr$ and $\partial_{\bfr}$) that are determined by dynamics occurring at energy scales above $E_b$.

In the thermal medium, additional scales -- for example, the temperature $T$ and the (longitudinal) gluonic screening mass $m_D$ -- play a role. They can modify $v_{s,o}$~\cite{brambilla20081}. The EFT can be analyzed assuming different hierarchies between the intrinsic scales of the $Q-\oQ$ state, $1/r$, $E_b$ and the thermal scales, and it has been shown that the relative importance of the different processes depends on these hierarchies~\cite{brambilla20081,brambilla20101,brambilla20111,brambilla20131}. For bottomonia, $M\gg 1/r\gg T$, but the hierarchy between $T$ and $E_b$ depends on the bottomonium state and can change as the QGP cools~\cite{Sharma:2023dhj}.

The power of the EFT approach is that it allows us to separate the dynamics of the $Q-\oQ$ state from the dynamics of the modes at scales $\gtrsim E_b$. The latter can then be modeled (one can study the systematic dependence of the results on the parameters in the usual manner by varying the parameters of the models within physically motivated regions), calculated using weak-coupling techniques at high temperatures, or calculated non-perturbatively using Lattice QCD (see~\cite{petreczky20121} for reviews).

Examples already mentioned above are $v_{s,o}$. These ``potentials'' are known to 3-loop order in QCD~\cite{Smirnov:2008pn,Anzai:2009tm,Smirnov:2009fh,Pineda:2013lta}. They have also been calculated on the lattice in a thermal medium~\cite{Burnier:2016mxc,rothkopf20111,Bala:2019cqu,Bala:2020tdt,Bala:2021fkm}. $V_{A,B}(r)=1$ at leading order~\cite{Brambilla:2004jw}. 

Another illustrative example is the decay width of a state $|\psi\rangle$ in the thermal medium. This has the form~\cite{brambilla20081},
\begin{equation}
\begin{split}
    \frac{1}{\tau_{\psi}} &= \frac{g^2}{3N_c} \sum_f |\langle f|\bfr|\psi\rangle|^2 f(k^0)
    \int \frac{d^3\bfk}{(2\pi)^3} \rho_{\bfE\bfE}(k^0,\bfk)|_{k^0=\epsilon_f-\epsilon_\psi}
\label{eq:Gamma}
\end{split}
\end{equation}
where $f(k^0)$ is the Bose distribution and $\rho_{\bfE\bfE}$ is a Chromoeletric spectral function in the medium. Matrix elements like $\langle f|\bfr|\psi\rangle$ can be computed within the EFT (Eq.~\ref{eq:LpNRQCD}). On the other hand, the spectral function is a property of the thermal medium. It is proportional to the real part of the Fourier transform of a gauge-invariant Chromoelectric field correlator (Eq.~\ref{eq:Gammadef}). One can estimate the correlator using weak coupling techniques~\cite{brambilla20081,Eller:2019spw,Binder:2021otw,Scheihing-Hitschfeld:2022xqx}, constrain it using the measurement of quarkonium spectral functions on the lattice~\cite{Brambilla:2020siz,Brambilla:2023hkw,Brambilla:2024tqg}, or calculate it~\cite{Nijs:2023dks,Nijs:2023dbc} for strongly coupled supersymmetric theories sharing features with QCD using AdS/CFT techniques (see~\cite{Casalderrey-Solana:2011dxg} for a review). Ideally, one would like to compute it non-perturbatively using Lattice QCD techniques, though it is a challenging calculation because it is a real-time correlation function~\cite{Leino:2024pen,Mayer-Steudte:2025gvz}. 

The Wilson coefficients of the EFT and the correlators of the Chromoelectric field complete the microscopic input required to describe the dynamics of $Q-\oQ$ in a static medium. Going from the microscopic theory to the phenomenology of quarkonia at LHC and RHIC requires two additional technical pieces.
\begin{enumerate}
\item{The background medium evolves hydrodynamically after quickly reaching local thermal equilibrium (see Refs.~\cite{Romatschke:2009im,Teaney:2009qa,Hirano:2012qz,Song:2013gia,Gale:2013da,Heinz:2013th,Jaiswal:2016hex} for reviews). Assuming local thermal equilibrium, the temperature in the neighbourhood of the $Q-\oQ$ state determines the dynamics of the state. In this paper, we use the results from the hydrodynamic model described in~\cite{PhysRevC.97.034915}, which describes several aspects of the phenomenology of low-energy hadrons observed in heavy-ion collisions at RHIC~\cite{Bhalerao:2015iya} and at LHC~\cite{Bhalerao:2015iya,PhysRevC.97.034915} very well.}
\item{For temperatures encountered by the $Q-\oQ$ in the QGP (few $100$ MeV) the width of the states is not much smaller~\cite{Sharma:2023dhj} than the binding energy. Furthermore, the time scale on which the medium evolves is $\sim$ fm, again comparable to the inverse of the binding energy. Hence, one needs to follow the quantum evolution of the $Q-\oQ$ state~\cite{young20131,akamatsu20121,akamatsu20131}. Quantum systems ($\Sys$) in contact with an external environment ($\Env$) are described using the OQS framework~\cite{breuer20021}, in which master equations for the evolution equations for the $Q-\oQ$ density matrix ($\rho_{\Sys}$) are derived. We use this framework in this paper. (See Refs.~\cite{Akamatsu:2020ypb,Yao:2021lus,Sharma:2021vvu} for reviews of applications OQS approaches to quarkonia.) 
}
\end{enumerate}

Previously, several works have derived and solved master equations for the evolution of $\rho_{\Sys}$ under various assumptions and approximations, and we will briefly review some of their key features to set the context for our calculation. The most convenient choice of the basis for describing the quantum system depends on the hierarchy between the intrinsic time scale ($\tau_{\Sys}\sim 1/E_b$) and the system relaxation time [$\tau_R \sim \tau_{\psi}$ (Eq.~\ref{eq:Gamma})]. 

For $\tau_{R}\gg\tau_{\Sys}$, the widths of the bound energy eigenstates are small compared to the splitting between them, and the energy eigenfunctions of the system Hamiltonian form a convenient basis. This is known as the quantum optical regime~\cite{breuer20021,Akamatsu:2020ypb} and was studied in~\cite{borghini20111,borghini20121}. Furthermore, if we assume that the time scale for relaxation of the excitations in the environment $\tau_{\Env}$ is shorter than $\tau_{R}$, by expanding the master equations second order in $\tau_{\Env}/\tau_{R}$ (NLO) one can write the master equations in a Lindblad~\cite{lindblad19761,Gorini:1975nb} form. This Lindblad equation in the optical regime was derived and solved by converting it to coupled Boltzmann equations in Refs.~\cite{yao20171,yao20184,yao20191,Yao:2021lus}.

If $\tau_{R}$ is not much less than $\tau_{\Sys}$ a more convenient basis is a basis of a continuum of states, for example wavefunctions in position space. Furthermore, if we assume $\tau_{\Env}$ is shorter than $\tau_{\Sys}$, then by expanding the master equations at leading (LO) or at the next order (NLO) order in $\tau_{\Env}/\tau_{\Sys}$ one can write the master equations in a Lindblad form (Quantum-Brownian regime). The Lindblad equations were derived in the weak-coupling approximation in Ref.~\cite{akamatsu20151} (LO) and Ref.~\cite{akamatsu20181} (NLO) and solved in Refs.~\cite{akamatsu20191,Sharma:2019xum}. Lindblad equations in the Quantum-Brownian regime were also derived in the weak coupling approximation by a different group in Refs.~\cite{blaizot20151,blaizot20171,DeBoni:2017ocl,blaizot20181} and solved by making semi-classical approximations~\cite{Katz:2013rpa,blaizot20151,blaizot20171,Gossiaux:2016htk,DeBoni:2017ocl,blaizot20181} and numerically~\cite{Delorme:2024rdo}. Using 
pNRQCD, Lindblad equations for $Q-\oQ$ were derived in the Quantum-Brownian regime and solved numerically in Refs. ~\cite{Brambilla:2016wgg,brambilla20171,Brambilla:2020qwo,Brambilla:2021wkt} (LO) and Refs.
~\cite{Brambilla:2022ynh,Brambilla:2023hkw,Brambilla:2024tqg} (NLO). 

The starting point for all these analyses is an operator equation for the time derivative of $\rho_{\Sys}$ of the form~\cite{Redfield:1957,Blum1981DensityMT} described below in Eq.~\ref{eq:Redfieldv2}, which we show here for convenience,
\begin{equation}
\begin{split}
\frac{\partial \rho_{\Sys} (t)}{\partial t}= &-\sum_{m,n}\int_0^t ds\;\Gamma_{mn}(t-s)[\bfV^m_{\Sys}(t),\bfV^n_{\Sys}(s)\rho_{\Sys}(t)]\; \\
   & + \;\textrm{H.C},
\end{split}
~\label{eq:Redfieldv2Intro}
\end{equation}
where $\Gamma_{mn}(t-s)$ are environmental correlation functions (in our case chromoelectric correlators) which decay on time scales $t-s\sim \tau_{\Env}$.

If this time scale is short compared to the time scale of the variation of $\bfV^n_{\Sys}(s)$ ($\tau_{\Sys}$) and the observation time ($t$), then the time evolution of $\rho_{\Sys}$ can be written as a Lindblad equation~\cite{lindblad19761,Gorini:1975nb}. (See Appendix~\ref{sec:Markov} for a brief review of the derivation.) Physically, if $\tau_{\Env}\ll \tau_{\Sys}$, any memory of the $\Sys-\Env$ exchanges is quickly lost, and the $\Sys$ evolution is Markovian. Evolution in this regime is sensitive only to the zero-frequency limit of the environment correlators, $\lim_{\omega\rightarrow 0}\tilde{\Gamma}_{mn}(\omega)$ [we will refer to the Fourier transform of $\Gamma_{mn}(t)$ as $\tilde{\Gamma}_{mn}(\omega)$]. 

However, for bottomonia in the QGP, $1/\tau_{\Sys}\sim $ few $100$ MeV is comparable to $1/\tau_{\Env}\sim T$ which also has the magnitude of a few $100$ MeV~\cite{Sharma:2023dhj}. Thus it is necessary to solve the $Q-\oQ$ master equations for $\tau_{\Env}\sim\tau_{\Sys}$, using the master equation with memory (Eq.~\ref{eq:Redfieldv2Intro}). In particular, this can be important for
$\Upsilon(1S)$, whose binding energy is the largest. The need for solving non-Markovian master equations in strongly coupled theories has also been recently emphasized in Ref.~\cite{Nijs:2023dbc}.

This is the main goal of this paper. The main technical advance in the paper is the numerical solution of the non-Markovian master equation using the method of stochastic unraveling \cite{PhysRevA.59.1633} (see Sec.~\ref{sec:Numerical} for details) and calculating the survival probability for the $\Upsilon(1S)$ state. To our knowledge, this is the first calculation of the suppression of $\Upsilon$ states using a master equation with memory.

Thinking about Eq.~\ref{eq:Redfieldv2Intro} in Fourier space, the evolution of $\rho_{\Sys}$ is sensitive to $\tilde{\Gamma}_{mn}(\omega)$ up to frequencies $\omega \sim 1/\tau_{\Env}$. It has previously been shown that keeping the dependence of the chromoelectric correlator on $\omega$ in the calculation of the decay rate (Eq.~\ref{eq:Gamma}) leads to a substantial reduction of the decay rate~\cite{Blaizot:2021xqa,Sharma:2023dhj} when compared to a calculation that approximates the correlator with its zero-frequency limit. In this paper, we show that in the OQS framework, using the non-Markovian master equation also leads to a much smaller suppression than a LO calculation with otherwise identical system and medium parameters. We also show that the NLO Markovian equation~\cite{Brambilla:2022ynh,Brambilla:2023hkw} captures this important effect and gives results comparable to those obtained from the full master equation for most of the evolution (unless $\tau_{\Env}$ is larger than $1/T$), except for $T\lesssim 250$ MeV. The master equation is better controlled than the NLO Markovian equation at these lower temperatures. 

Using medium and system parameters that describe the phenomenology of bottomonia at LHC~\cite{chatrchyan20111,CMS:2012gvv,CMS:2016rpc,ATLAS:2019can,ATLAS:2022exb,CMS:2022rna,CMS:2023lfu} using NLO Markovian equations~\cite{Brambilla:2023hkw}, we present results for the $R_\textrm{AA}$ of $\Upsilon(1S)$ in $200$ GeV Au-Au collisions at RHIC, and $2.76$ and $5.02$ TeV Pb-Pb collisions at LHC. For $\tau_{\Env}\sim 1/T$ our calculation gives results for $R_\textrm{AA}$ comparable in magnitude to that of NLO Markovian equations and gives results broadly in agreement with the suppression of $\Upsilon(1S)$ states obtained at LHC, and underpredict~\cite{Strickland:2023nfm} the suppression at RHIC~\cite{STAR:2022rpk}. We analyze the systematic dependence of the survival probability on $\tau_{\Env}$.   

However, addressing the phenomenology of quarkonia in heavy-ion collisions is not the central goal of this paper. There are some effects that have not been included in our calculations that might be important for that purpose. We list a few below.
\begin{enumerate}
\item{Gluo-dissociation is an effect that has not been included beyond the static limit in master equations so far (see Ref.~\cite{yao20171,yao20184,yao20191,Yao:2021lus} for a semi-classical implementation, see Ref.~\cite{Sharma:2019xum} for incorporation of gluo-dissociation in the stochastic Schrödinger equation), and may play an important role at lower temperatures.}
\item{Cold nuclear matter effects on the parton density functions (PDFs) might affect~\cite{LHCb:2018psc,ATLAS:2017prf,ALICE:2019qie,CMS:2022wfi,Strickland:2024oat} the initial hard production of bottomonia in heavy ion collisions.}
\item{As we have not (yet) calculated the survival probability of the excited states, we can not calculate the contribution to $R_\textrm{AA}$ from their feed-down here.}
\end{enumerate}
A systematic analysis of the phenomenology will require including these effects and then refitting the medium transport coefficients (in particular the zero-frequency limit of the chromoelectric correlator, as it is not yet known independently). We leave this detailed program for the future and restrict ourselves to understanding the systematic effect of considering different hierarchies between $\tau_{\Env}$ and $\tau_{\Sys}$ in this paper.

The plan of the paper is as follows. In Sec.~\ref{sec:Formalism}, we review the formalism and derive the master equations. In Sec.~\ref{sec:Numerical} the algorithm and the numerical setup used for solving the master equations. 

In Sec.~\ref{sec:results}, we describe our results. To get a clearer understanding, we first analyze the effect of memory in a time-independent medium in Sec.~\ref{subsec:constantT}. Effects in a simple Bjorken flow are analyzed in Sec.~\ref{subsec:bjorken}. Results for realistic hydrodynamic backgrounds, with the closest connection to phenomenology at RHIC and LHC, are shown in Sec.~\ref{subsec:hydro}.

Some technical details have been moved to the Appendix. In Appendix~\ref{sec:Markov} we briefly review the derivation of the LO Markovian equation. In Appendix~\ref{sec:constantTadd} we explore how memory effects change with $T$ in a time-independent medium. In Appendix~\ref {sec:HTLcorrelator} we motivate the choice of the form of the time-dependent correlator. Results for PbPb collisions at $2.76$ TeV are given in Appendix~\ref{sec:lhc276}. 

\section{\label{sec:Formalism}Formalism}
In this section, we review the derivation of the general master equation for quarkonium dynamics. 

Consider a system (\Sys) interacting with the environment (\Env). The combined system and environment are isolated, so the density matrix of the total system evolves as,
\begin{align}
	\frac{\partial \rho_{\rm{tot}}(t)}{\partial t} = -i \left[ H_{\text{tot}},\rho_\text{tot} \right] 	\;,\;\; H_{\rm{tot}}= H_{\Sys} + H_{\Env} + H_{\rm{I}} \;,
\end{align} 
where $H_{\rm{\Sys}}$ and $H_{\rm{\Env}}$ are the system and environment Hamiltonians, respectively. $\Sys$ and $\Env$ interact through an interaction term $H_{\rm{I}}$. The reduced density matrix of the system is calculated by tracing out the environment variables,
\begin{align}
	\rho_{\Sys}= \rm{Tr}_{\Sys} \left[ \rho_{\rm{tot}} \right] \;.
\end{align}

In the interaction picture (denoted with a superscript $I$ on the density matrix), taking $H_{\rm{I}}$ as the interaction, the evolution equation of $\rho_\Sys$ can be written as,
\begin{align}\label{eq:main}
	\frac{\partial \rho^{I}_{\Sys}(t)}{\partial t} = 
    - \int\limits_{0}^{t} ds\; \textrm{Tr}_{\Env}\left( \left[ H_{\rmI}(t),\left[H_{\rmI}(s),\rho^I_{\rm{tot}}(s)\right] \right] \right)\;.
\end{align}

At the initial time, $\Sys$ and $\Env$ are typically not entangled (often the system starts in a pure state, as we assume here). Therefore, $\rho_{\rm{tot}}(0)= \rho_\Sys(0) \otimes \rho_\Env$. Since $\Env$ is much larger than $\Sys$, its density matrix does not change substantially during evolution~\footnote{The master equations are derived for a medium in global equilibrium. For phenomenology in an expanding medium, local thermodynamic variables are used to describe $\Env$ in the neighbourhood of $\Sys$.}, although $\Sys$ becomes entangled with $\Env$. If $H_{\rm{I}}$ is weak, however, $\rho_{\rm{tot}}(s)$ can be approximated by $ \rho_\Sys(s)\otimes\rho_{\Env}$. This is known as the Born approximation~\cite{PhysRevA.59.1633}. Furthermore, if $H_{\rm{I}}^2\tau_{\Env}$ is much smaller than the inverse of the time scale on which $\rho_{\Sys}^I$ changes, then $\rho_{\Sys}^I(s)$ in the $s$ integrand can be replaced by $\rho_{\Sys}^I(t)$~\cite{PhysRevA.59.1633}. Corrections to these approximations go as $H_{\rmI}^4$ and are small if $\tau_R\gg \tau_{\Env}$ since $\tau_R$ scales as $H_{\rmI}^2$. For quarkonia, this hierarchy follows from the multipole expansion as long as $1/r\gg T$.

With these approximations, Eq.~\ref{eq:main} gives a master equation of the form,
\begin{equation}
\begin{split}
\frac{\partial \rho^{I}_{\Sys} (t)}{\partial t}&= -\int\limits_0^t ds\;\tr_{\Env} \left([H_{\rmI}(t), [H_{\rmI}(s),\rho^{I}_{\rm{\Sys}}(t)\otimes \rho_{\Env}]]\right)\;.
~\label{eq:Redfield}
\end{split}
\end{equation}
It is known in the literature as the Redfield equation~\cite{Redfield:1957,Blum1981DensityMT}. Although the right hand side (RHS) of Eq.~\ref{eq:Redfield} depends only on $\rho_{\rm{\Sys}}$ at the instant $t$, Eq.~\ref{eq:Redfield} is not a Markovian equation since the operator $\tr_{\Env} \left(H_{\rmI}(t)H_{\rmI}(s)\rho_{\Env}\right)$ depends on times $s$ between $0$ and $t$.  

$H_{\rmI}$ can be written as the sum of tensor products of $\Env$ and $\Sys$ operators, $H_{\rmI}=\sum_n {\bfE}^n_{\Env}\otimes {\bfV}^n_{\Sys}$. After taking the trace over $\Env$, Eq.~\ref{eq:Redfield} can be rewritten as,
\begin{equation}
\begin{split}
\frac{\partial \rho^{I}_{\Sys} (t)}{\partial t}= -\sum_{m,n}&\int\limits_0^t ds\;\Gamma_{mn}(t,s)\Bigl(\bfV^{m}_{\Sys}(t)\bfV^{n}_{\Sys}(s)\rho_{\Sys}^I(t) \\
&  -\bfV^{n}_{\Sys}(s)\rho_{\Sys}^I(t)\bfV^{m}_{\Sys}(t) \Bigr)  
+\;\;\textrm{H.C.} 
\end{split}
~\label{eq:Redfieldv2}
\end{equation}
where $\Gamma_{mn}(t,s)=\langle {\bfE}^{m}_{\Env}(t){\bfE}^{n}_{\Env}(s)\rangle_{\Env}$ and only depends on $t-s$ for a static medium.  

To find the specific form of $\Gamma_{mn}$ and $\bfV_{\Sys}^n$ for quarkonia, we use the pNRQCD Lagrangian (Eq.~\ref{eq:LpNRQCD}). In the centre of mass (CM) of the quarkonium state, the system Hilbert space features the singlet states $|s\rangle$ (the separation $\bfr$ is suppressed for clarity) and octet states $|a\rangle$ ($a=1,.8$).
\begin{equation}
\begin{split}
  H_{\Sys} = &\;
  h_s |s\rangle\langle{s}| +
  h_o |a\rangle\langle{a}|\\
 H_{\rmI} = &- g{\bf{E}}^{a}\cdot\;{\bf{r}}\Bigl(\frac{1}{\sqrt{2N_{c}}} |s\rangle \langle{a}| +\frac{1}{\sqrt{2N_{c}}} |a\rangle \langle{s}| +\frac{d_{abc}}{2} |b\rangle \langle{c}| \Bigr)\;.
 \end{split}
\end{equation}

The colour structure of $\rho_\Sys$ can be further simplified by reducing it to blocks of singlet and octet components. The singlet and octet reduced density matrices are defined as $\rho_s=\langle s | \rho_\Sys |s \rangle$ and $\rho_o=\sum_{a} \langle a | \rho_\Sys |a \rangle$. Going back to the Schr\"{o}dinger picture and changing the integration variable from $s$ to $t-s$, the evolution equation for $\rho_\Sys$ (Eq.~\ref{eq:Redfieldv2}) can be written as,

\begin{equation}
\begin{split}
\frac{\partial{\rho_{\Sys}(t)}}{\partial t} &= -i\Bigl(H_{\rm{eff}}\rho_{\Sys}(t)-\rho_{\Sys}(t)H_{\rm{eff}}^\dagger \Bigr)\\
						& +  \int\limits_0^t ds\; \Bigl( \sum\limits_{n=1}^{3} 
\Gamma_n(t,t-s)\bfV_{n}(-s) \rho_{\Sys}(t)\bfV_{n}^\dagger (0) + \rm{H.C.}\Bigr)\;,\\
& \hspace{-1cm}{\rm{where}},\;
    H_{\rm{eff}}= H_{\Sys}-i\int\limits_0^t ds\;  \sum\limits_{n=1}^{3} \Gamma_n(t,t-s)\bfV_{n}^\dagger(0) \bfV_{n} (-s)\;.
    \label{eq:master}
\end{split}
\end{equation}
$\{\bfV_{n}(t)\}$ (Eq.~\ref{eq:master}) are time-dependent operators and the index $n$ labels their structure in colour space: $n=1$ corresponds to $s\rightarrow o$, $n=2$ to $o\rightarrow s$, and $n=3$ to $o\rightarrow o$ transitions. The explicit form of $\bfV_n$ in $s$, $o$ space is as follows:
\begin{align}
	\bfV_{n}(t) = \begin{cases}
    e^{ih_ot}\bfr e^{-ih_st}  \left(\begin{array}{cc} 0 & 0\\1 & 0\end{array}\right) \;\;\;\;\; n=1 \\
        e^{ih_st}\bfr e^{-ih_ot}  
    \sqrt{\frac{1}{(N_c^2-1)}}\left(\begin{array}{cc} 0 & 1\\0 & 0\end{array}\right)   \;\; n=2\\
        e^{ih_ot}\bfr e^{-ih_ot} 
    \sqrt{\frac{N_c^2-4}{2(N_c^2-1)}}
    \left(\begin{array}{cc} 0 & 0\\0 & 1\end{array}\right)   \;\; n=3\;.
	\end{cases}
\end{align}
The spatial indices of $\bfV_n$ are summed in Eq.~\ref{eq:master}. 

The medium response functions $\Gamma_n$ with comes from tracing out the $\bfE$ fields. For $n=1$ we have,
\begin{equation}
\Gamma_1(t,t')=
\frac{g^2}{6N_c}\tr_{\Env}
\Bigl({\bfE}^a(t,{\vec{0}}) {\cal{W}}_{ab}(t,t')
{\bfE}^b(t',{\vec{0}})\rho_{\Env}\Bigr)
\label{eq:Gammadef}
\end{equation}
where ${\cal{W}}_{ab}(t,t')$ is the adjoint Wilson line~\cite{Brambilla:2024tqg} joining $(t',\vec{0})$ and $(t,\vec{0})$. The correlators $\Gamma_{2}$ and $\Gamma_3$ have ${\cal{W}}$ at the ends (see \cite{Yao:2020eqy} for a derivation of the correlators). The insertion of ${\cal{W}}_{ab}$ makes $\Gamma_n$ Gauge-invariant. Since our results does not require $\Gamma_{2,3}$, for a simpler notation we drop the subscript $n$ and identify $\Gamma$ with $\Gamma_1$.

The operators ${\bf{V}}_n(t)$ evolve on the system timescale $\tau_{\rm{S}}\sim {1}/{E_b}$. The correlator $\Gamma(t)$ is significant when $t\lesssim\tau_{\Env}\sim {1}/{T}$. In the quantum Brownian approximation, one assumes the hierarchy $\tau_{\rm{E}}\ll\tau_{\rm{S}}$. The operators ${\bf{V}}_n(t)$ can therefore be expanded as 

\begin{align}
    {\bf{V}}_{n}(t) \sim e^{ih_\alpha t} {\bf{r}} e^{-ih_\beta t}
    \approx {\bf{r}} + it (h_{\alpha}{\bf{r}} - {\bf{r}} h_{\beta})
    + {\cal{O}}\Bigl[\Bigl(\frac{\tau_{\Env}}{\tau_{\Sys}}\Bigr)^2\Bigr]\;,~\label{eq:Vnexpansion}
\end{align}
where $\{\alpha=o,\;\beta=s\}$ for $n=1$,  $\{\alpha=s,\;\beta=o\}$ for $n=2$,  $\{\alpha=o,\;\beta=o\}$ for $n=3$.

As reviewed in the introduction, truncating the expansion in Eq.~\ref{eq:Vnexpansion} at the first term gives the leading order (LO)~\cite{Brambilla:2020qwo} Lindblad equation and keeping the term proportional to $t$ gives the next-to-leading order (NLO)~\cite{Brambilla:2022ynh} Lindblad equation. These Lindblad equations feature two real transport coefficients $\kappa$ and $\gamma$ defined as,

\begin{align}
	\tilde{\Gamma}(\omega)= \int\limits_{0}^{\infty} dt \; e^{i\omega t}\;\Gamma(t,0) \;,\;\; \lim_{\omega\rightarrow 0}\tilde{\Gamma}(\omega)=\frac{1}{2}(\kappa+i\gamma)\;.
 \label{eq:tildegamma}
\end{align}
These are often written in terms of dimensionless variables $\hat{\kappa}=\kappa/T^3$, $\hat{\gamma}=\gamma/T^3$. For chromoelectric correlators with fundamental connections that appear in the context of heavy quark diffusion, $\hat{\kappa}$ has been calculated over a wide range of $T$'s and is known to have a weak~\cite{Banerjee:2022gen,Brambilla:2020siz} $T$ dependence, showing an increase near the critical temperature (in the pure glue theory) as it is approached from above. Here we will assume that $\hat{\kappa}$ and $\hat{\gamma}$ are $T$ independent.

Our goal is to go beyond these approximations and directly solve Eq.~\ref{eq:master}. Although linear, it is still formidable to solve due to the large number of degrees of freedom associated with the spatial coordinates in three dimensions. Since we will ultimately look at only states with a definite angular-momentum quantum number $l$, it is useful to work in the angular-momentum basis $\ket{l,m}$. An additional benefit of using this basis is that the transition rules can be written in a simpler form~\cite{brambilla20171,brambilla20081,Brambilla:2022ynh}. Furthermore, it can be shown that when the initial density matrix $\rho_\Sys(0)$ is block diagonal in $|l,m\rangle$ basis, then for the evolution equation Eq.~\ref{eq:master} it will be block diagonal for subsequent times. Defining the blocks by 
\begin{equation}
\rho^{l}(t)= \sum\limits_{m=-l}^{l} \langle l,m \rvert \rho_S(t)  \lvert l,m \rangle\;, 
\end{equation}
where we have summed over the azimuthal quantum number $m$ since we are not considering any observables that are sensitive to transitions between different $m$ values for a given $l$. The evolution equation for $\rho^{l}(t)$ can be written as,

\begin{widetext}
\begin{align}\label{eq:ang}
    \frac{\partial \rho^{l}(t)}{\partial t} = -i \left[ h^l,\rho^{l}(t) \right] 
	- \int\limits_{0}^{t} ds\sum\limits_{n,l'}^{} 
    &\Bigl(
     \Gamma(t,t-s){T}^{\dagger}_{n}(l \to l',0)T_{n}(l \to l',-s)\rho^{l}(t)\\
     \nonumber
    &- \Gamma(t,t-s){T}_{n}(l' \to l,-s)\rho^{l'}(t)T^{\dagger}_{n}(l' \to l,0) 
    +\textrm{H.C.}\Bigr)\;,	
\end{align}
\end{widetext}
where,
\begin{equation}
\begin{split}
h^l&=\left(\begin{array}{cc} h_s^l & 0\\0 & h_o^l\end{array}\right)\;,\;\;{\rm{with}}\\
h_{s,o}^l&=\frac{1}{M}\Bigl(-\frac{1}{r}\frac{\partial^2}{\partial r^2}r + \frac{l(l+1)}{r^2}\Bigr)+v_{s,o}(r)\;.
\end{split}
\end{equation}

The summations over $n$ and $l'$ in Eq.~\ref{eq:ang} indicates a sum over transitions between states with different colour quantum numbers ($s,\;o$) and orbital angular momentum ($l,\;l'$), and the corresponding time-dependent transition operators at time $t$ are $T_{n}(l\to l',t)$. To the order in pNRQCD that we are working in (Eq.~\ref{eq:LpNRQCD}) only transitions between angular momentum states $l'=l \pm 1$ are allowed due to the dipole interaction. The explicit form of the transition operators is given by:

\begin{align}
	T_{n}(l\to l',t)= C_{l,l'}\times \begin{cases}
		e^{i h_{o}^{l'} t} r e^{-i h_{s}^{l} t} \left( \begin{array}{cc} 0 & 0 \\ 1 & 0\end{array} \right) \;\; \\
        e^{i h_{s}^{l'} t} r e^{-i h_{o}^{l} t} \left( \begin{array}{cc} 0 & \sqrt{\frac{1}{(N_c^2-1)}} \\ 0 & 0\end{array} \right)  \;\;\\
        e^{i h_{o}^{l'} t} r e^{-i h_{o}^{l} t} \left( \begin{array}{cc} 0 & 0 \\ 0 & \sqrt{\frac{N_c^2-4}{2(N_c^2-1)}}\end{array} \right)  \;\;
	\end{cases}\;,
\end{align}
for $n=1,2,3$ respectively and $C_{l,l'}$ is given by:

\begin{align}
    C_{l,l'}= \sqrt{\frac{l+1}{2l+1}}\delta_{l',l+1} + \sqrt{\frac{l}{2l+1}}\delta_{l',l-1}
\end{align}

The master equation \ref{eq:ang} can be given a simple interpretation similar to that of a rate equation. In Eq.~\ref{eq:ang}, the first pair of $T_n$'s correspond to the process $l\to l'=l\pm 1\to l$ and the second term corresponds to the contribution from the transitions $l'=l\pm1 \to l$. There will be higher-order transitions if one goes beyond the leading-order multipole expansion.
\section{\label{sec:Numerical}Numerical Simulation}
Master equations of the form Eq.~\ref{eq:ang} can be solved using stochastic methods such as stochastic unraveling \cite{PhysRevA.59.1633}. We review this technique here.

Consider a general master equation,
 \begin{equation}\label{eq:res1}
 \begin{split}
	 \frac{\partial \rho_\Sys(t)}{\partial t} &= 
     A(t)\rho_\Sys(t) 
     + \rho_\Sys B^\dagger(t) \\ 
      &+ \sum_i \Bigl(C_i(t) \rho_\Sys(t) D_i^\dagger(t) +  \textrm{H.C.}\Bigr)\;.
 \end{split}    
 \end{equation}
 
 For Eq.~\ref{eq:ang}, 
 \begin{equation}
 \begin{split}
 A(t)&=B(t)=-ih^l_{\rm{eff}}(t)\\
 &=-ih^l-\int\limits_{0}^{t} ds\; \Gamma(t,t-s) \times \\
 & \qquad\sum\limits_{n,l'}^{} {T}^{\dagger}_{n}(l \to l',0)T_{n}(l \to l',-s)\;,\\
 C_{n,l'}(t)&=\int\limits_{0}^{t} ds\; \Gamma(t,t-s)  T_{n}(l \to l',-s)\;,\\
 D_{n,l'}(t)&={T}_{n}(l \to l',0)\;.
 ~\label{eq:ABCD}
 \end{split}
 \end{equation}

Before describing the formalism of the general case, it is useful to review it for the Lindblad equation (Eq.~\ref{eq:Lindblad}) for which $A(t)=B(t)=-iH_{\rm{eff}}=-i(H_{\rm{h}}-iL_i^\dagger L_i/2)$, and $C_i(t)=D_i(t)=L_i/\sqrt{2}$. The master equation can be solved by considering an ensemble of wavefunctions $\{|\psi(t)\rangle\}$ each evolving according to the stochastic equation,
\begin{equation}
\begin{split}
	d\ket{\psi(t)}=& -i \,dt\; H_{\eff}|\psi(t)\rangle \\
    &+ \sum\limits_i \left(
    \frac{L_i(t)|\psi(t)\rangle}{\norm{L_i(t)|\psi(t)\rangle}} -|\psi\rangle\right) dN_i(t)\;,
    ~\label{eq:LindblabUnfolding}
\end{split}
\end{equation}
where $\{dN_i(t)\}$ are a set of infinitesimal independent Poisson processes (called ``jumps'') with an ensemble average $ dN_i(t)dN_j(t)=\delta_{ij}\;dN_i(t)$, satisfying $\llangle dN_i(t) \rrangle=\norm{L_i(t)|\psi(t)\rangle}^2/\norm{|\psi(t)\rangle}^2$. The density matrix $\rho_\Sys$ is given by the ensemble average $\llangle |\psi(t)\rangle \langle \psi(t)| \rrangle$ and satisfies the required Lindblad equation. See Ref.~\cite{Brambilla:2020qwo} for an implementation of this technique to quarkonia.

To unravel the general master equation (Eq.~\ref{eq:res1}) a two- component stochastic wavefunction $\ket{\psi(t)}= \bigl(\ket{\psi_1(t)}, \ket{\psi_2(t)}\bigr)$ needs to be defined. The evolution equation for $\ket{\psi(t)}$ is given by, 
\begin{equation}
\begin{split}
	d|\psi(t)\rangle=& -i \,dt\;{\overline{H}}|\psi(t)\rangle \\
    &+ \sum\limits_i \left(
    \frac{J_i(t)|\psi(t)\rangle}{\norm{J_i(t)\ket{\psi(t)}}} -|\psi\rangle\right) dN_i(t)\; \\
    &+ \sum\limits_i \left(
    \frac{\overline{J}_i(t)|\psi(t)\rangle}{\norm{\overline{J}_i(t)\ket{\psi(t)}}} -|\psi\rangle\right) d\overline{N}_i(t)\;,
\end{split}\label{eq:dpsi}
\end{equation}
where, 
\begin{align}
    \nonumber
&\overline{H}=\left(\begin{array}{cc} A(t) & 0 \\ 0 & B(t)\end{array}\right), 
\;\;\; J_i(t)=\left(\begin{array}{cc} C_i(t) & 0 \\ 0 & D_i(t)\end{array}\right)\\
&\qquad\qquad \overline{J}_i(t)=\left(\begin{array}{cc} D_i(t) & 0 \\ 0 & C_i(t)\end{array}\right)
\end{align}
and $\{dN_i(t),d\overline{N}_i(t)\}$ are infinitesimal independent Poisson processes as defined above, satisfying
\begin{equation}
\begin{split}
\llangle d N_i(t)\rrangle &= \norm{J_i(t)|\psi(t)\rangle}^2/\norm{|\psi(t)\rangle}^2,
\end{split}
\end{equation}
with a similar equation for $\llangle d \overline{N}_i(t)\rrangle$.
The density matrix $\rho_\Sys(t)$ is given by the ensemble average $\llangle |\psi_1(t)\rangle \langle \psi_2(t)| \rrangle$ and satisfies Eq.~\ref{eq:res1}.

For LO~\cite{Brambilla:2021wkt} and NLO~\cite{Brambilla:2022ynh,Brambilla:2023hkw} Lindblad equations, it has been found previously that evolution with $H_{\rm{eff}}$ (Eq.~\ref{eq:LindblabUnfolding}) is sufficient to describe the survival probability of $\Upsilon(1S)$ states, and results obtained by solving the Lindblad equation with and without ``jumps'' are very close. However, for $\Upsilon(2S)$ and $\Upsilon(3S)$ which are not as strongly bound, ``jumps'' are necessary to obtain the accurate survival probability and is essential for capturing the phenomenon of regeneration. Regeneration is a key process for obtaining a realistic phenomenological description of the suppression of the excited states~\cite{Brambilla:2022ynh,Brambilla:2023hkw}. The physical reason behind this observation is that ``jumps'' from the $1S$ states contribute significantly to the depleted number of excited states, while the reverse process is less likely. 

Motivated by this observation, in this paper, we focus on $\Upsilon(1S)$ states and simplify our calculation by neglecting the ``jump'' terms in Eq.~\ref{eq:dpsi} (formally taking $C, D$ to $0$ in Eq.~\ref{eq:ABCD}). In this approximation $|\psi_1(t)\rangle=|\psi_2(t)\rangle$ so evolving a single component is enough. The calculation is still numerically challenging because for each value of $s$, operators like $T_n(l\rightarrow l',-s)$ in Eq.~\ref{eq:ABCD} are proportional to $e^{-ih^{l'}_\alpha s}re^{ih_\beta^{l} s}$. The operators of the form $e^{ih_\alpha^{l} t}$ can be evaluated once in the energy basis for a range of $t$ values of interest, and saved for efficient implementation. At each time $t$, $T_n(l\rightarrow l',-s)$ is convolved with $\Gamma(t,s)$ and this is the most computationally challenging step when the background medium evolves with time since $\Gamma(t,s)$ depends on the evolution history of the temperature $T$ along the trajectory of the CM of the quarkonium state.

We end this section by giving some information about the numerical implementation of Eq.~\ref{eq:dpsi} [with $A(t), B(t)$ from Eqn.~\ref{eq:ABCD} and $C=D=0$ (without ``jumps'')], and some parameter values for the system. For a given $l$, the radial wavefunction is discretised in the range $[0,L_r]$ with $L_r=6$ fm and spacing between points is $\Delta r=0.02$ fm. We work with the radial part of the stochastic wavefunction (Eq.~\ref{eq:dpsi}) multiplied by $r$, as usual~\cite{Messiah:1979eg}, $y_l(r)=rR_l(r)$, where $R_l(r)=\langle r|\psi_l\rangle$. We use zero boundary conditions for $y_l(r)$.

To make a direct comparison with the NLO results of~\cite{Brambilla:2022ynh,Brambilla:2023hkw}, in this study we take the potentials be Coulombic,
\begin{equation}
v_s(r)=-\frac{C_{F}\alpha_{s}}{r},\;\;
v_o(r)=\frac{\alpha_{s}}{2N_{c}r}\;.
~\label{eq:potential}
\end{equation}
$\alpha_s$ and $M$ (see Table.~\ref{table:parameters} for a summary) are taken to be the same as those used in Ref.~\cite{Brambilla:2022ynh}. With these parameters, $E_b=460.4$ MeV, giving $\tau_\Sys\sim 0.4$fm. The initial state is taken to be the $1S$ eigenstate with the singlet Coulomb potential $v_s(r)$. In \cite{Brambilla:2020qwo}, it is found that for LO Lindblad evolution, initial states of a Gaussian form with widths $\sim1/M$, and Coulombic eigenstates give the same evolution for $1S$ state. Hence, we do not investigate these systematics here. 

The final ingredient we need in order to evaluate $A(t), B(t)$ (Eq.~\ref{eq:ABCD}) is $\Gamma(t,s)$. Motivated by the arguments of Appendix~\ref{sec:HTLcorrelator}, we take its form in a static medium to be 
\begin{align}
    \Gamma(t,s)= \frac{\kappa}{2\tau_E}e^{-(t-s)/\tau_E},\;\; t>s\;.
    \label{eq:Gammat}
\end{align}
Values for $t<s$ are not needed in our calculation, but the real part of $\Gamma$ is symmetric under $t\rightarrow s$.

For all the results presented in this Section, we use the transport coefficient values to be $\hat{\kappa}=4.0$ and $\hat{\gamma}=0$ \cite{Brambilla:2023hkw}. $\tau_\Env\sim 1/T$ and we vary it in a reasonable range to explore the sensitivity of our results to this parameter. The wavefunction is evolved using the second-order Runge-Kutta method with step size $\delta t=0.001$ fm which is much smaller than $\tau_R$ as well as $\tau_\Env$.

\section{\label{sec:results} Results}

\begin{center}
\begin{table}[htb]
\begin{tabular}{ |c|c|c|c|c| } 
 \hline
 $M$ & $\alpha_s$ & $\;\hat{\kappa}\;$ & $\;\hat{\gamma}\;$ & $T_{\rm{min}}$ \\ 
 \hline
 $4.73$ GeV& $0.468$ & $4.0$ & $0$ & $160$ MeV\\
 \hline
\end{tabular}
\caption{Parameters of the model. $\alpha_s$ determines the potentials (Eq.~\ref{eq:potential}) and the choice of $M$ and $\alpha_s$ was made to facilitate comparisons with Ref.~\cite{Brambilla:2022ynh}. We evolve the states till the medium in the neighbourhood of the state reaches a temperature $T_{\rm{min}}$ (see discussion below).}
\label{table:parameters}
\end{table}
\end{center}

In this section, we describe results for $\Upsilon(1S)$ suppression calculated using the method described in the previous section. Our main goal is to illustrate the dependence of suppression on the hierarchy between $\tau_\Sys$ and $\tau_\Env$. Since the binding energy and therefore $\tau_\Sys$ is fixed, we vary $\tau_\Env$. Intuitively, $\tau_\Env$ should scale as the inverse of the medium temperature. Therefore, we vary $\tau_\Env$ by changing $\xi_{\Env}=\tau_\Env\cdot T$. We calculate results for three different values of $\xi_{\Env}$:  $\{1/(1.5),\;1,\;1.5\}$.

We consider three different models for the hydrodynamic background. First, to understand the effect of varying $\tau_\Env$ in a simple setting, we will take the medium to be at a fixed temperature and see how the survival probability evolves with time. We then consider the evolution of the survival probability in a simple Bjorken expanding medium. Finally, we show results obtained by taking the medium evolution to be described by a phenomenologically validated model of viscous $2+1$ hydrodynamic evolution at LHC and RHIC~\cite{Bhalerao:2015iya,PhysRevC.97.034915}, and compare our results for $R_\textrm{AA}$ for $\Upsilon(1S)$ with those obtained by these experiments. 

For Bjorken and hydrodynamic backgrounds, we evolve the quantum state until the medium temperature drops below $T_{\rm{min}}=160$ MeV, close to the freeze-out temperature. We will see that the survival probability stops changing well before this, and hence increasing or decreasing $T_{\rm{min}}$ by a few MeV will not change our results for $R_\textrm{AA}$.

\subsection{\label{subsec:constantT} Static medium}

\begin{figure}[h]
\includegraphics{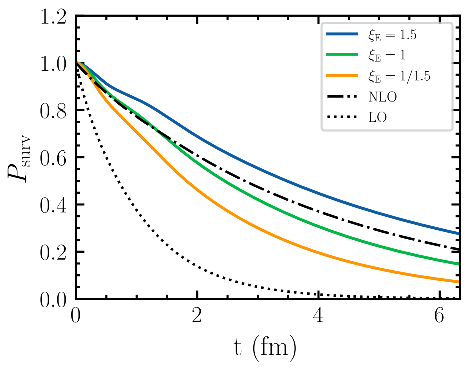}
\caption{(Colour online) The time dependence of the survival probabilities of $\Upsilon(1S)$ in a static medium at a temperature of $350$ MeV. The three solid curves correspond to the calculations using the master equation Eq.~\ref{eq:ang}. The dotted curve (black online) corresponds to LO~\cite{Brambilla:2020qwo} and the dot-dashed curve (black online) to NLO~\cite{Brambilla:2022ynh}.\label{fig:constantT}}
\end{figure}

In Fig.~\ref{fig:constantT}, we present $\Upsilon(1S)$ survival probabilities for a static background at a constant temperature of $T=350$ MeV. Along with the three values of $\xi_{\Env}$, we plot LO and NLO results calculated from the effective Hamiltonian given in \cite{Brambilla:2022ynh}. We observe a strong dependence of the survival probabilities on $\xi_{\Env}$ and therefore on the value of the $\tau_\Sys/\tau_\Env$, keeping $T$ and hence $\kappa$ fixed. 

Let us first consider the difference between the results obtained by solving the master equation (solid curves in Fig.~\ref{fig:constantT}) and LO (dotted black curve in Fig.~\ref{fig:constantT}). The suppression using LO is much larger. This can be understood from Eq.~\ref{eq:Gamma}. The spectral function corresponding to Eq.~\ref{eq:Gammat} can be approximated by the form~\cite{Akamatsu:2020ypb}
\begin{equation}
\frac{\rho(\omega)}{\omega}\sim 
\frac{\kappa\Omega^2}{(\omega^2+\Omega^2)T},~\label{eq:rhomodel}
\end{equation} 
with $\Omega=1/\tau_{\Env}$. The LO calculation corresponds to assuming $\Omega\gg E_b$, in which case one can set $\omega\sim E_b=0$ in Eq.~\ref{eq:rhomodel}. However, for $\omega\sim E_b=460$ MeV and $\Omega\sim T=350$ MeV the RHS of Eq.~\ref{eq:rhomodel} is smaller than its zero frequency value, leading to higher survival probability~\cite{Blaizot:2021xqa,Sharma:2023dhj}.

The relative suppression for the three $\xi_{\Env}$ values in Fig.~\ref{fig:constantT} is also clear from the above argument. Increasing $\Omega$ keeping $T$, $\kappa$ and $E_b$ unchanged will lead to a larger suppression, with $\Omega\rightarrow \infty$ corresponding to the LO result. 

We can also anticipate the effect of increasing $T$ keeping $\xi_{\Env}$ and $E_b$ unchanged. The overall suppression will increase for all the calculations, because of a larger $\kappa$. However, we expect the LO and the master equation to be relatively closer together, corresponding to a smaller $E_b/\Omega$. The opposite effect is expected if we decrease $T$. Detailed calculations bear out this intuition. We have relegated these results to Fig.~\ref{fig:constantThighlow} in Appendix.~\ref{sec:constantTadd}. The trend is as expected.

Now we consider the results of the NLO calculation (dash-dotted black curve in Fig.~\ref{fig:constantT}). As expected, the NLO corrections capture the effect of a finite $E_b/T$. It is interesting that for $T=350$ MeV, and $\xi_{\Env}=1$, the NLO results lie very close to the result from the master equation, especially after about $2$ fm when the transient oscillations in the master equation have died down. However, this is a numerical coincidence. Based on the argument above, we expect the NLO results to lie closer to results from the master equation with a larger $\xi_{\Env}$ if we increase $T$. This is because we expect the NLO calculation to capture the dynamics for longer environmental relaxation times ($E_b/T$ is smaller). Similarly, for a smaller $T$ we expect NLO results to be closer to results from the master equation with a smaller $\xi_{\Env}$. This trend is seen in  Fig.~\ref{fig:constantThighlow} in Appendix.~\ref{sec:constantTadd}.

It is worth remarking on the transient oscillations in the survival probability seen at early times in the solution of the master equation. These arise from the convolution of $e^{\pm i (h_s-h_o)t}\sim e^{\pm iE_bt}$ with $e^{-t/\tau_{\Env}}$ and decay after a few $\tau_{\Env}$.

\subsection{\label{subsec:bjorken}Bjorken background}

\begin{figure*}
\includegraphics{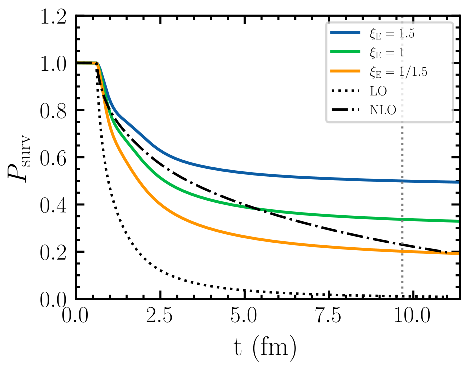}
\includegraphics{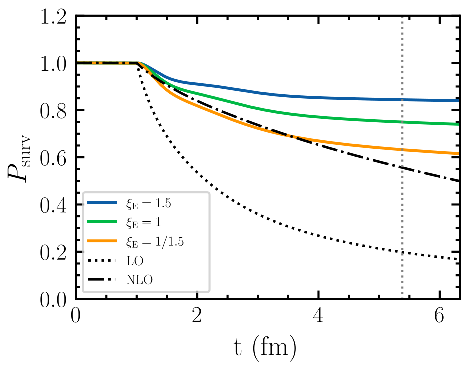}%
\caption{$\Upsilon(1S)$ survival probabilities as a function of time in Bjorken expanding media with $(t_0,~T(t_0))$ taken to be $(0.6~\textrm{fm},~480~\textrm{MeV})$ (left panel) and $(1.0~\textrm{fm},~333~\textrm{MeV})$ (right panel). The vertical line corresponds to the time at which the temperature 190 MeV, which corresponds to the $T_\textrm{min}$ used for NLO simulations in \cite{Brambilla:2022ynh}. The choice of the style of curves for the curves is the same as Fig.~\ref{fig:constantT}. \label{fig:bjorken} }
\end{figure*}

In this Section, we analyse the evolution of the survival probability as a function of time in a Bjorken expanding medium with a temperature evolution given by
\begin{align}
	T(\tau)=T(\tau_0)\left( \frac{\tau_0}{\tau} \right) ^{\frac{1}{3}}
    \label{eq:bjorken}
\end{align}
where $\tau_0$ is the initial time of the start of hydrodynamic evolution. In Fig.~\ref{fig:bjorken} we show two plots for the survival probability of $\Upsilon(1S)$ with initial temperatures $T(\tau_0)$ $480$~MeV and $333$~MeV. The values of $\tau_0$ are $0.6$~fm and $1.0$ fm, respectively. These temperatures and times correspond to the typical values in the most central collisions at $5.02$~TeV and $200$~GeV energies~\cite{Sharma:2023dhj,alberico20131}. As in the static case, we show the survival probability for three values of $\xi_{\Env}\in \{1.5,\;1,\;1/1.5\}$. Along with these curves, we plot the LO and NLO probabilities.

We consider $\Upsilon(1S)$ states at rest in the medium or moving slowly at central rapidity in the medium, and hence $t$ in Eq.~\ref{eq:ang} is the same as $\tau$. The $T$ at time $t$ needed for estimating $\Gamma$ is obtained from Eq.~\ref{eq:bjorken}. However, there is a subtlety in determining $\Gamma(t,s)$ in the evolving medium since $\Gamma(t,s)$ is sensitive to temperatures over a time interval $\tau_{\Env}$ around $t$ and modeling $\Gamma(t,s)$ requires further assumptions. In fact, in principle $\Gamma(t,s)$ is not necessarily a function only of $t-s$. However, here we assume that the evolution is slow enough that $\Gamma(t,s)$ still has the form given by Eq.~\ref{eq:Gammat} with the change that $\kappa$ is given by $\hat{\kappa}\left(\frac{T(t)+T(s)}{2}\right)^3$ and $\tau_E$ given by 
\begin{align}
    {\tau_\Env(t,s)}=\xi_{\Env} \Bigl(\frac{2}{T(t)+T(s)}\Bigr)\;.
    \label{eq:Tprescription}
\end{align}

We have seen in a simple case that using $T(t)$ instead of the average of $T(t)$ and $T(s)$ changes the result by less than the range obtained by using $\xi_{\Env}\in \{1/1.5,\;1.5\}$. We leave a more detailed analysis of this systematic for future work.

The relative order (Fig.~\ref{fig:bjorken}) between the results obtained from the master equation, and the results from the LO and the NLO calculations are similar to those observed and described in Section~\ref{subsec:constantT}. However, there are a few additional features to highlight. 

First, in contrast to the static medium, the survival probabilities in the master equation change very gradually after about $t\sim 8$ fm at LHC and $t\sim 5$ fm (corresponding to $T\sim 200$ MeV in both cases) . (This is not an artifact of the choice of the prescription Eq.~\ref{eq:Tprescription}, since, for example, using $T(t)$ to determine the environment correlator will lead to an even smaller decay probability at late times.) 

It simply stems from the fact that for a given $\xi_{\Env}$ at small $T$, $\omega/\Omega\sim E_b/\Omega$ is large, and the factor $\Omega^2/(\omega^2+\Omega^2)$ in Eq.~\ref{eq:rhomodel} is smaller. This further reduces the probability of decay, on top of the overall reduction in the strength of $\Gamma$ due to a smaller $\kappa$.

This effect is absent in LO. In fact, for RHIC with the parameters above, the decay probability remains substantial for LO even at $6$ fm. As anticipated in Section~\ref{subsec:constantT}, while the NLO calculation captures this effect at high $T$, at low $T$ (late times), the NLO calculation is expected to break down and receive large corrections from higher order correction terms in the $E_b/T$ expansion. The effect is more pronounced at RHIC because of the smaller $T$. This late time behaviour is tamed in Refs.~\cite{Brambilla:2022ynh,Brambilla:2023hkw} by taking a larger $T_{\rm{min}}\sim 190$ MeV. 

\subsection{\label{subsec:hydro} Realistic hydrodynamic background}

\begin{figure*}[t]
\includegraphics[width=0.45\textwidth]{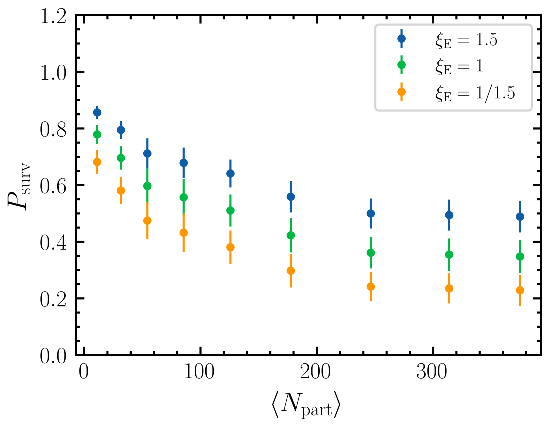}
\includegraphics[width=0.45\textwidth]{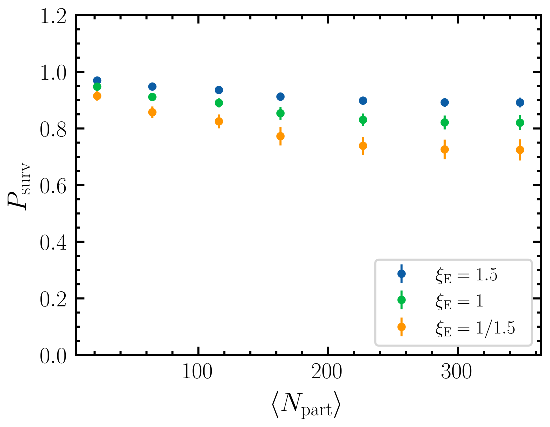}
\caption{\label{fig:lhc502200} Final survival probability as a function of $\langle N_\textrm{part}\rangle$ for $\Upsilon(1S)$ with a viscous hydrodynamic background for Pb-Pb collisions at $5.02$ TeV (left) and Au-Au collisions at $200$ GeV (right). In each plot, three sets of results corresponding to $\xi_{\Env}=\{1.5,\;1,\;1/1.5\}$ are shown which were obtained from  the master equation Eq.~\ref{eq:ang}. The error bars correspond to the statistical errors from the sampling of quarkonium trajectories.}
\end{figure*}

\begin{figure*}[t]
\includegraphics[width=0.45\textwidth]{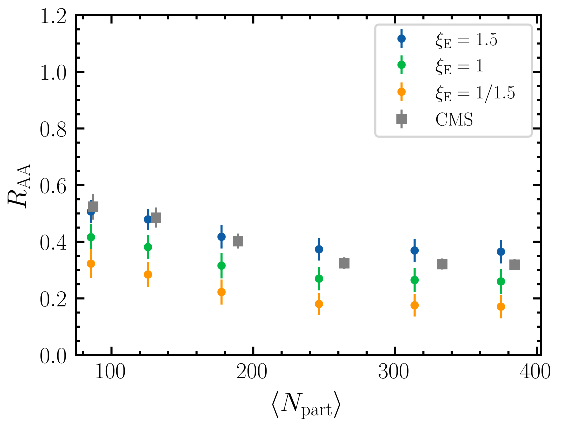}
\includegraphics[width=0.45\textwidth]{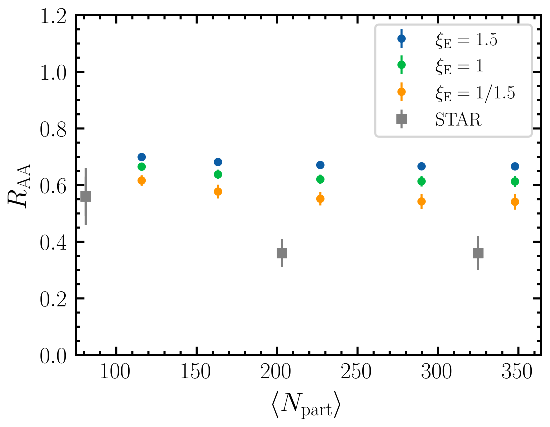}
\caption{\label{fig:lhc502200_feeddown} $R_\textrm{AA}$ as a function of $\langle N_\textrm{part}\rangle$ for $\Upsilon(1S)$ in Pb-Pb collisions at 5.02 TeV (left) and Au-Au collisions at 200 GeV energy (right) after estimating the feed-down contributions as discussed in the text. Experimental data from CMS~\cite{CMS:2019270} and STAR~\cite{STAR:2022rpk} is plotted for comparison. Data from ATLAS~\cite{ATLAS:2022exb} at $5.02$ TeV is consistent with the CMS data and are not shown here for cleaner plots.}
\end{figure*}

In this Section, we present the results for $\Upsilon(1S)$ suppression for $200$ GeV Au-Au and $5.02$ TeV Pb-Pb collisions. Results for $2.76$ TeV Pb-Pb collisions are similar to $5.02$ TeV Pb-Pb collisions, and are shown in Appendix~\ref{sec:lhc276}. We take the evolution of the background thermal medium to be given by the model used for the phenomenology of light hadrons at RHIC and LHC~\cite{Bhalerao:2015iya,PhysRevC.97.034915}. In this model, for a given impact parameter $b$, the initial distribution of nucleons is given by the Glauber model~\cite{Miller:2007ri}, and the collision and the initial dynamics are described by AMPT~\cite{PhysRevC.72.064901}. This is coarse-grained into a hydrodynamic state, which is evolved using (2+1) dimensional viscous hydrodynamics. The transport coefficients are tuned to fit the spectra of soft particles.

For each decile of centrality bins, we sample $10$ $b$'s from the distribution given by the Glauber model. For a given $b$, the hydrodynamic simulation provides the temperature profile of the medium as a function of position and time.  The model incorporates initial-state fluctuations since it uses fluctuating Glauber initial conditions, and since the transport model AMPT is probabilistic, and hence for each $b$ value we use $5$ different hydrodynamic realisations and average over them to obtain our results. 

For a given hydrodynamic realization, the $\Upsilon(1S)$ can move along various trajectories. In pNRQCD, the CM dynamics decouples from that of the relative motion of the $Q-\oQ$ and since bottomonia do not get strong enough kicks from the medium for it to alter the CM trajectory, each trajectory is taken to be a straight line. The trajectory is fixed by the initial conditions: position $(x_0,y_0)$, transverse momentum $p_T$ and direction of velocity given by the azimuthal angle $\phi$ with respect to the reaction plane in the Glauber model. We assume that the generated quarkonia have zero momentum rapidity.  Therefore,
\begin{equation}
    \bigl(x(t), y(t)\bigr)=\bigl(x_0,y_0\bigr)+t\frac{p_T}{E}\bigl(\cos(\phi), \sin(\phi)\bigr)\;,
\end{equation}
where $p_T$ is the transverse momentum and $E$ the energy.

The production points $(x_0,y_0)$ of the quarkonium are sampled from the distribution of binary collisions, 
\begin{align}
    N_{\textrm{bin}}(x,y,b)= T_{AB}(x,y,b)\sigma^{\textrm{NN}}_{in}
\end{align}
where $\sigma^{\textrm{NN}}_{in}$ is the nucleon-nucleon inelastic cross-section and $T_{AB}$ is the overlap distribution. The velocity angle $\phi$ is sampled uniformly from the interval $[0,2\pi)$. 

The transverse momentum is sampled from the distribution
\begin{align}
    P(p_{T})=\frac{2M_{Q\overline{Q}}^2 \;p_T}{\left(p_T^2+ M_{Q\overline{Q}}^2\right)^2}
\end{align}
where $M_{Q\overline{Q}}$ is the mass of the quarkonium under consideration. For a given impact parameter, we sample $2000$ trajectories.

The quarkonium state experiences a time-dependent temperature as it moves through the medium along a trajectory obtained as described above. For a given trajectory, the temperature is given as a function of time by the hydrodynamic realisation. This determines the correlator $\Gamma(t,s)$ according to Eq.~\ref{eq:Tprescription}.  

In Fig.~\ref{fig:lhc502200}) we show the final survival probability $P_\textrm{surv}$ for $\Upsilon(1S)$ in $5.02$ TeV Pb-Pb and $200$ GeV Au-Au collisions. The impact parameters are binned into centrality classes, and $P_\textrm{surv}$ is plotted for each centrality bin as a function of the averaged $N_\textrm{part}$. As in the case of static and Bjorken backgrounds, there is a strong dependence of the suppression on $\xi_\Env$.  

For a comparison with the experimental data, late-time feed-down must be included. For more central collisions, we can make a rough estimate of the contributions from feed-down. Since the direct contribution to $\Upsilon(1S)$ state is about $75\%$~\cite{Brambilla:2020qwo}, if we assume that for $N_{\textrm{part}}\gtrsim 80$ all the excited states [in particular $\chi_{b0}(1P)$ and $\chi_{b1}(1P)$] are much more suppressed (a plausible hypothesis based on the observed $R_\textrm{AA}$ of $\Upsilon(2,3S)$~\cite{CMS:2019270,CMS:2016rpc,ATLAS:2022exb}), then the $R_\textrm{AA}$ for $\Upsilon(1S)$ can be roughly estimated from our results in Fig.~\ref{fig:lhc502200}, by scaling them by $0.75$ and this is shown in Fig.~\ref{fig:lhc502200_feeddown} for LHC (left) and RHIC (right).

The model agrees with the experimental data at LHC energies (also see Fig.~\ref{fig:lhc276_feeddown} in Appendix.~\ref{sec:lhc276}) for reasonable values of $\xi_{\Env}$. RHIC measurements of the $\Upsilon(1S)$ $R_\textrm{AA}$ indicate the same degree of suppression as obtained for LHC energies. Intuitively, one would expect a lower suppression due to the lower temperatures of the QGP produced at RHIC. Our values underestimate the $\Upsilon(1S)$ suppression compared to the experimental values.

We note that one can reconcile the two measurements by taking $\xi_{\Env}$ to be around $1.5$ at LHC and even smaller than $1/1.5$ at RHIC. However, a reduction of $\xi_{\Env}$ at RHIC by a factor of over $2.3$ is not very easy to understand, and hence we surmise that additional effects may need to be taken into account to get to a realistic phenomenology. Some of these have been listed in the Introduction (Sec.~\ref{sec:Introduction}).

For example, we have not considered the initial and final state cold nuclear matter effects such as nuclear absorption, which might give additional contributions to the observed suppression. The results from Ref.~\cite{PhysRevC.111.014902} indicate that nuclear adsorption may play a role at $200$ GeV energies. 

Another effect that might play a role in understanding the larger-than- expected suppression observed at RHIC could be an increase in $\hat{\kappa}$ as $T$ approaches the crossover temperature from above (see discussion below Eq.~\ref{eq:tildegamma}). This may play a more important role at RHIC than at LHC because of the overall lower temperature of the medium.

\section{\label{sec:Conclusion}Conclusions}
We solve non-Markovian master equations for the evolution of the $\Upsilon(1S)$ state in the QGP. For $\tau_{\Env}/\tau_{\Sys}\rightarrow 0$, these equations reduce to LO Lindblad equations. To study the dependence on $\tau_{\Env}$ ($\tau_{\Sys}\sim 1/E_b$ is fixed) we analyze the survival probability of $\Upsilon(1S)$ for $\tau_{\Env}=\xi_{\Env}/T$ for three values of $\xi_{\Env}$: $\xi_{\Env}=\{1/1.5,1,1.5\}$.  We find that as $\xi_{\Env}$ decreases, the survival probability decreases, but for all $\xi_{\Env}\sim 1$, LO~\cite{Brambilla:2020qwo,Brambilla:2021wkt} Lindblad equations significantly overestimate the suppression of $\Upsilon(1S)$ (see. Figs.~\ref{fig:constantT},~\ref{fig:bjorken}). 

We find that the NLO ~\cite{Brambilla:2022ynh,Brambilla:2023hkw} equations capture much of this reduction of the suppression for realistic $\xi_{\Env}$ at $T$'s well above the crossover temperature, but our calculations in Bjorken backgrounds (Fig.~\ref{fig:bjorken}) suggest that at temperatures $\lesssim250$ MeV the NLO equations do not agree with non-Markovian master equations suggesting higher-order corrections might be necessary. $R_\textrm{AA}$ values found using the non-Markovian master equation are not very sensitive to the choice of the freezeout time of the $\Upsilon(1S)$, as long as it is roughly around $200$ MeV or below. (Interestingly, temperatures of about $200$ MeV are not inconsistent with the extraction of the freezeout time for bottomonium states from statistical models~\cite{gupta20141,Kumar:2023acr}.) 

The key ingredient in our calculation is the chromoelectric correlator whose overall strength is governed by the transport coefficient $\kappa$~(Eq.~\ref{eq:tildegamma}) (we take $\gamma=0$). From our results, one can see that a larger $\tau_{\Env}$ can be compensated for by a larger $\kappa$, and hence an independent determination of $\kappa$ is needed to narrow down the parameter space in the absence of other independent observables in this sector.

Using a realistic hydrodynamic medium for RHIC (right) and LHC (left) in Fig.~\ref{fig:lhc502200_feeddown}, we find the final probability of survival of the $\Upsilon(1S)$ states in these experiments. While a smaller ``effective'' $\xi_{\Env}$ might partly explain the suppression observed at RHIC, there is room for additional effects. We leave a more detailed attempt at phenomenology for future work.

Finally, an obvious next step can be including ``jumps'' and addressing the phenomenology of the excited bottomonium states, and their feed-down.

\begin{acknowledgments}
We would like to thank Subrata Pal for providing us with results for the temperature evolution of the thermal medium obtained using their transport+viscous hydro code. We are also grateful for valuable discussions with Saumen Datta and Subarta Pal throughout the project. We also acknowledge insightful discussions with Nora Brambilla, Sourendu Gupta, Felix Ringer, Michael Strickland, Balbeer Singh, Anurag Tiwari, and Antonio Vairo. We thank Subikash Choudhury for discussions on the observation of Cold Nuclear Matter effects for $\Upsilon$'s in $pA$ collisions. We acknowledge the use of computing resources of the
Department of Theoretical Physics, TIFR. We acknowledge the support of the Department of Atomic Energy, Government of India, under Project Identification No. RTI 4002.
\end{acknowledgments}

\appendix

\section{\label{sec:Markov} LO Markovian approximation}
In this section we review how the master equation, Eq.~\ref{eq:Redfieldv2}, can be rewritten as a LO Markovian equation if $\tau_{\Env}\ll \tau_{\Sys}$.

The environmental correlation functions $\Gamma_{mn}(t,s)$ in Eq.~\ref{eq:Redfieldv2} decay on time scales $t-s\sim \tau_{\Env}$. If this time scale is short compared to the time scale of the variation of $\bfV^n_{\Sys}(s)$ ($\tau_{\Sys}$) and the observation time ($t$), then in the RHS of Eq.~\ref{eq:Redfieldv2}, the lower limit of the $s$ integration can be set to $-\infty$ and the variation of $\bfV^n_{\Sys}(s)$ can be ignored (LO in $\tau_{\Env}/\tau_{\Sys}$). This is known as the Markovian approximation. In this approximation, the integral in $s$ simply gives the $0$ frequency limit of $\Gamma_{mn}(t)$ and the RHS of Eq.~\ref{eq:Redfieldv2} has a structure,
\begin{equation}
\begin{split}
&-\sum_{m,n} \tilde{\Gamma}_{m,n}(0) \Bigl(\bfV^m_{\Sys}(0)\bfV^n_{\Sys}(0)\rho_{\Sys}(t)-\bfV^m_{\Sys}(0)\rho_{\Sys}(t)\bfV^n_{\Sys}(0)
\Bigr)\\
&+{\rm{H.C.}} \;,~\label{eq:RHSMarkovian}
\end{split}
\end{equation}
where $\textrm{H.C.}$ is the hermitian conjugate of the operator in the previous line.

The form in Eq.~\ref{eq:RHSMarkovian} is not an accident. Going to the Schr\"{o}dinger picture and defining a new set of operators $L_k$ by taking appropriate linear combinations of $\bfV_{\Sys}^n$, Eq.~\ref{eq:Redfieldv2Intro} in the Markovian approximation can be rewritten in a more familiar form,
\begin{equation}
\frac{\partial\rho_{\Sys}(t)}{\partial t}
    =-i[H_{\rm{h}},\rho_{\Sys}]+\sum_{k}\Bigl(L_k\rho_{\Sys}L_k^\dagger-\frac{1}{2}\{L_k^\dagger L_k,\rho_{\Sys}\}\Bigr)\;,
    ~\label{eq:Lindblad}
\end{equation}
where $H_{\rm{h}}$ gives the Hermitian evolution of $\rho_{\Sys}$ and $L_k$ is a set of operators that encode its non-Hermitian evolution in a trace-preserving manner. The evolution of the state depends only on the state at that instant, and the evolution operator does not depend on the history. Eq.~\ref{eq:Lindblad} is the well-known Lindblad form. For the specific forms of $H_{\rm{h}}$ and $L_k$ calculated for quarkonia under various assumptions, we refer the reader to Refs.~\cite{Akamatsu:2020ypb,Yao:2021lus,blaizot20181,brambilla20181,Brambilla:2022ynh}.

The physical picture behind the mathematical form above is that if $\tau_{\Env}\ll \tau_{\Sys}$, any memory of the $\Sys-\Env$ exchanges is quickly lost, and the $\Sys$ evolution is Markovian, and it is known that any positive definite, trace-preserving, Markovian equation for the time derivative of $\rho_{\Sys}$ can be written as a Lindblad equation~\cite{lindblad19761,Gorini:1975nb}.

\begin{figure*}[t]
\includegraphics[width=0.45\textwidth]{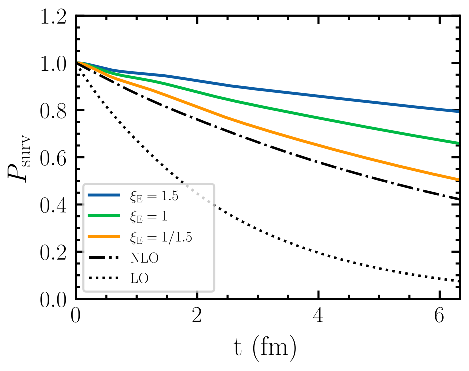}
\includegraphics[width=0.45\textwidth]{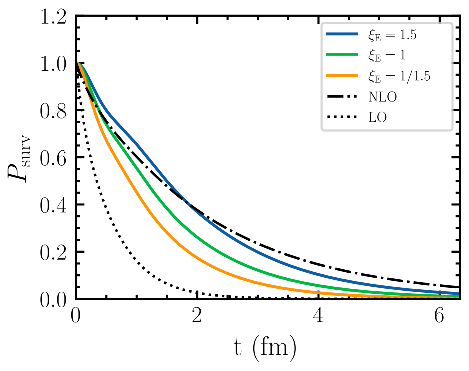}
\caption{Static medium at $T=250$ MeV (left) and $T=450$ MeV (right).~\label{fig:constantThighlow}}
\end{figure*}
\section{\label{sec:constantTadd} Effect of $T$ on the evolution for a static medium}
In this Section, we show results for the survival probability as a function of time for the $\Upsilon(1S)$ state in a static medium at temperatures $T=250$ MeV (left) and $450$ MeV (right) (Fig.~\ref{fig:constantThighlow}). 

Comparing these with Fig.~\ref{fig:constantT} in Sec.~\ref{subsec:constantT}, we notice the following pattern. As $T$ is increased, the overall suppression increases for all the master equations since $\kappa$ increases. Comparing the NLO calculation with the master equation, we note that the NLO calculation comes closer to the solution of the master equation with a larger $\xi_{\Env}$, as $T$ is increased.

\section{\label{sec:HTLcorrelator}Chromoelectric correlator in the Hard Thermal Loop (HTL) approximation}
The chromoelectric-correlator, $\Gamma(t)$ (Eq.~\ref{eq:Gammadef}), in a deconfined medium at temperature $T$ governs the evolution of the quarkonium density matrix in the QGP. Since it is a real-time quantity, it is challenging to compute non-perturbatively using lattice QCD. Nonetheless, significant progress has been made in understanding this object. The real and imaginary parts of the zero frequency limit of the (causal) correlator define two important transport coefficients~\cite{brambilla20181,Brambilla:2022ynh}, $\kappa$ and $\gamma$, defined in Eq.~\ref{eq:tildegamma}.

One can estimate~\cite{Brambilla:2019tpt} $\gamma$ using the shift in the peak of the quarkonium spectral function computed on the lattice~\cite{Kim:2018yhk}. More recent lattice results~\cite{Larsen:2019bwy} for the shift suggest that $\hat{\gamma}$~\cite{Brambilla:2024tqg} for temperatures of a few $100$MeV, is small. This motivates us in this study to ignore the imaginary part of $\Gamma(t,0)$, an assumption worth closer investigation in the future. The direct computation of $\hat{\gamma}$ is an open problem.  

Similarly, a non-perturbative calculation of $\hat{\kappa}$ is an open problem. A related quantity, the heavy quark momentum diffusion constant is defined as,
\begin{widetext}
\begin{equation}
    \kappa_{\rm{F}}=\frac{g^2}{3N_c}\int dt\; \tr_{\Env}
    \left({\cal{U}}(-\infty,t;{\vec{0}})\bfE(t,{\vec{0}}){\cal{U}}(t,0;{\vec{0}})\bfE(0,{\vec{0}}){\cal{U}}(0,\infty;{\vec{0}})\right)\;,
\label{eq:kappaF}
\end{equation}
\end{widetext}
where $\bfE=\bfE^a t^a$ and ${\cal{U}}(t_1,t_2;{\vec{r}})$ is the fundamental Wilson line from time $t_2$ to $t_1$ at the spatial location ${\vec{r}}$. This has been computed on the lattice~\cite{datta20121,Ding:2011hr}. (See Refs.~\cite{Banerjee:2022uge,Banerjee:2022gen,Brambilla:2020siz,Brambilla:2021wqs,Brambilla:2021egm} for more recent results.)

It is well known that the leading order (in the strong coupling $g$) perturbative~\cite{moore20051,mustafa20051} estimate of ${\kappa_{\rm{F}}}$ in the QGP phase is roughly an order of magnitude smaller than the non-perturbative result~\cite{datta20121,Ding:2011hr}. The next-to-leading order result~\cite{Caron-Huot:2007rwy,huot20081} is closer, but the large correction suggests that the series in $g$ converges slowly for temperatures of interest for heavy ion phenomenology (few $100$ MeV). 

Little is known about $\Gamma$ at finite frequency non-perturbatively. Therefore, it is useful to gain insight from weak-coupling approaches, even though we do not expect these approaches to be quantitatively reliable. Formally,
\begin{equation}
\Re e[\Gamma(t,0)]= \frac{g^2 C_F}{3}\int \frac{d^4q}{(2\pi)^4}
e^{-iq^0 t} 
\frac{e^{q^0/T}}{e^{q^0/T}-1} \rho(q^\mu)\;,~\label{eq:Gammavsrho}
\end{equation}
where, $C_F=(N_c^2-1)/(2N_c)$ is the colour factor and $\rho(q^\mu)$ is the chromoelectric spectral density. 

In the Hard Thermal Loop (HTL) formalism at leading order~\cite{kapusta_gale_2006}, the contribution to the spectral density from longitudinal gluons is
\begin{equation}
\rho(q^\mu)=\frac{\pi m_D^2 q q^0}{(\bfq^2+m_D^2)^2+(\pi m_D^2 q^0/2q)^2}\;,~\label{eq:rhoLHTL}
\end{equation}
where $m_D\sim gT$ is the screening mass. 

Performing the $q^0$ integration in Eq.~\ref{eq:Gammavsrho} via contour integration for $t>0$, the integrand in the lower complex plane picks residues from the pole of $\rho$ [$q^0=-i2q(q^2+m_D^2)/m_D^2$ in HTL] and the residues from the poles of the Bose-Einstein distribution [$q^0=-i2n\pi/T$, for integral $n\geq 1$]. The real part of the result, therefore, has the form
\begin{widetext}
\begin{equation}
\begin{split}
    \Re e [\Gamma(t,0)]_{\HTL} &=  \int \frac{d^3q}{(2\pi)^3} q^3 \Bigl\{\exp[-2t q(q^2+m_D^2)/(\pi m_D^2)]
    \frac{q^2 \cot[q(q^2+m_D^2)/(\pi m_D^2T)]}{\pi m_D^2}
    \\
    +&\sum_{n\geq 1}
    \exp[-2t n\pi T] \frac{2 \pi^2 q^3 {m_D}^2 n T^2 }
    {-q^6-2k^4 {m_D}^2-k^2{m_D}^4+\pi^4{m_D}^4 n^2T^2}\Bigr\} \;.
\end{split}
\label{eq:GammatHTL}
\end{equation}
\end{widetext}
The terms in the sum (second term in Eq.~\ref{eq:GammatHTL}) fall off with $t$ as $\exp[-2t n\pi T]$, while the contribution from the pole of $\rho$ (first term in Eq.~\ref{eq:GammatHTL}) falls off as $\sim \exp[-m_Dt]$ for $q\sim m_D$ and is the dominant contribution in weak coupling when $m_D\ll \pi T$. For $g\gtrsim 1$, both the first and the second term decay on a time scale $\sim 1/T$. Finally, it is easy to verify that~\footnote{The analysis is a little formal since the integral over $d^3q$ is UV-divergent and needs to be regularized by an ultraviolet completion of the theory~\cite{brambilla20131}. This detail does not change the arguments.}
\begin{equation}
    \int_0^{\infty} dt\; \Re e [\Gamma(t,0)]_{\HTL}=  
    \frac{1}{2}\int \frac{d^3q}{(2\pi)^3} \frac{qm_D^2\pi T}{(q^2+m_D^2)^2}=\frac{1}{2}\kappa_{\HTL}\;.~\label{eq:GammatIntegral}
\end{equation}

This motivates our choice of the time-dependent form of the correlator,
\begin{equation}
\Gamma(t,0)=\frac{\kappa}{2} \frac{e^{-t/\tau_{\Env}}}{\tau_{\Env}},\;\; t>0 ~\label{eq:GammatChoice}
\end{equation}
with $\tau_{\Env}\sim 1/T$, a simple form which satisfies the  integral constraint Eq.~\ref{eq:tildegamma} and captures the expected exponential decay of the correlator. We note that $\Gamma(t,0)$ will have an imaginary part~\cite{Bellac:2011kqa}, even if one takes $\gamma=0$ as we do in our paper. However, we simplify our calculation by neglecting this.

\section{\label{sec:lhc276}$\Upsilon(1S)$ suppression at $2.76$ TeV}
\begin{figure}[t]
\includegraphics[width=0.45\textwidth]{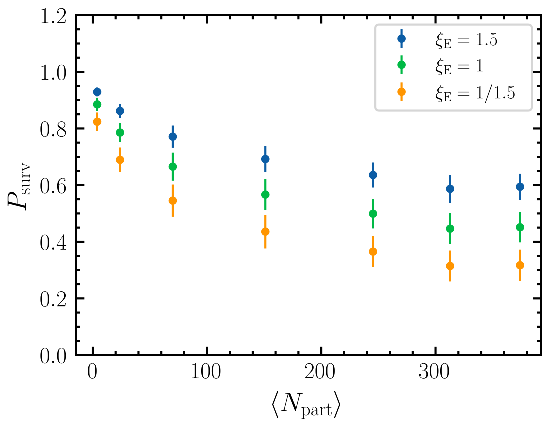}
\caption{\label{fig:lhc276} Survival probability as function of $\langle N_\textrm{part}\rangle$ for $\Upsilon(1S)$ with a viscous hydrodynamic background for Pb-Pb collisions at $2.76$ TeV.}
\end{figure}

In this Section, we show results for the $\Upsilon(1S)$ suppression in the presence of a viscous hydrodynamic background for Pb-Pb collisions at $2.76$ TeV energies. In Fig.~\ref{fig:lhc276}, we show the final survival probabilities, $P_\textrm{surv}$, as a function of $\langle N_\textrm{part}\rangle$. As in Fig.~\ref{fig:lhc502200}, feed-down contributions at late time are not included in this plot.

\begin{figure}[t]
\includegraphics[width=0.45\textwidth]{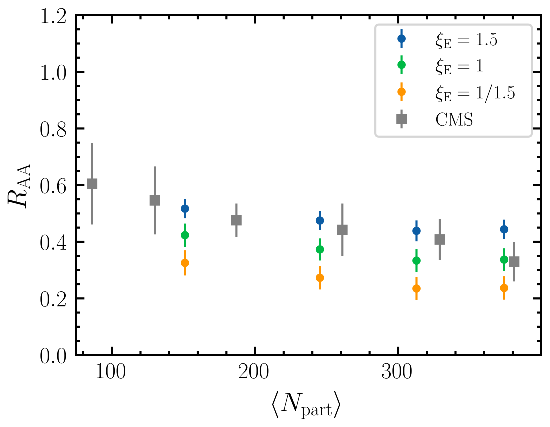}
\caption{\label{fig:lhc276_feeddown} $R_\textrm{AA}$ as a function of $\langle N_\textrm{part}\rangle$ for $\Upsilon(1S)$ after estimating the feed-down contributions in Pb-Pb collisions at 2.76 TeV energy as discussed in Sec.~\ref{subsec:hydro}. Experimental data from CMS~\cite{CMS:2016rpc} is plotted for comparison.}
\end{figure}

Following the discussion in Sec.~\ref{subsec:hydro} about the estimation of feed-down contributions, we show in Fig.~\ref{fig:lhc276_feeddown} the $R_\textrm{AA}$ obtained by scaling $P_\textrm{surv}$ by $0.75$.

\clearpage


\bibliography{refs}

\providecommand{\noopsort}[1]{}\providecommand{\singleletter}[1]{#1}%
\begin{thebibliography}{124}%
\makeatletter
\providecommand \@ifxundefined [1]{%
 \@ifx{#1\undefined}
}%
\providecommand \@ifnum [1]{%
 \ifnum #1\expandafter \@firstoftwo
 \else \expandafter \@secondoftwo
 \fi
}%
\providecommand \@ifx [1]{%
 \ifx #1\expandafter \@firstoftwo
 \else \expandafter \@secondoftwo
 \fi
}%
\providecommand \natexlab [1]{#1}%
\providecommand \enquote  [1]{``#1''}%
\providecommand \bibnamefont  [1]{#1}%
\providecommand \bibfnamefont [1]{#1}%
\providecommand \citenamefont [1]{#1}%
\providecommand \href@noop [0]{\@secondoftwo}%
\providecommand \href [0]{\begingroup \@sanitize@url \@href}%
\providecommand \@href[1]{\@@startlink{#1}\@@href}%
\providecommand \@@href[1]{\endgroup#1\@@endlink}%
\providecommand \@sanitize@url [0]{\catcode `\\12\catcode `\$12\catcode `\&12\catcode `\#12\catcode `\^12\catcode `\_12\catcode `\%12\relax}%
\providecommand \@@startlink[1]{}%
\providecommand \@@endlink[0]{}%
\providecommand \url  [0]{\begingroup\@sanitize@url \@url }%
\providecommand \@url [1]{\endgroup\@href {#1}{\urlprefix }}%
\providecommand \urlprefix  [0]{URL }%
\providecommand \Eprint [0]{\href }%
\providecommand \doibase [0]{https://doi.org/}%
\providecommand \selectlanguage [0]{\@gobble}%
\providecommand \bibinfo  [0]{\@secondoftwo}%
\providecommand \bibfield  [0]{\@secondoftwo}%
\providecommand \translation [1]{[#1]}%
\providecommand \BibitemOpen [0]{}%
\providecommand \bibitemStop [0]{}%
\providecommand \bibitemNoStop [0]{.\EOS\space}%
\providecommand \EOS [0]{\spacefactor3000\relax}%
\providecommand \BibitemShut  [1]{\csname bibitem#1\endcsname}%
\let\auto@bib@innerbib\@empty
\bibitem [{\citenamefont {Peskin}(1979)}]{peskin19791}%
  \BibitemOpen
  \bibfield  {author} {\bibinfo {author} {\bibfnamefont {M.~E.}\ \bibnamefont {Peskin}},\ }\bibfield  {title} {\bibinfo {title} {Short-distance analysis for heavy-quark systems: (i). diagrammatics},\ }\href {https://doi.org/https://doi.org/10.1016/0550-3213(79)90199-8} {\bibfield  {journal} {\bibinfo  {journal} {Nuclear Physics B}\ }\textbf {\bibinfo {volume} {156}},\ \bibinfo {pages} {365 } (\bibinfo {year} {1979})}\BibitemShut {NoStop}%
\bibitem [{\citenamefont {{G. Bhanot and M. E. Peskin }}(1979)}]{bhanot19791}%
  \BibitemOpen
  \bibfield  {author} {\bibinfo {author} {\bibnamefont {{G. Bhanot and M. E. Peskin }}},\ }\bibfield  {title} {\bibinfo {title} {{Short Distance Analysis for Heavy Quark Systems. 2. Applications}},\ }\href {https://doi.org/10.1016/0550-3213(79)90200-1} {\bibfield  {journal} {\bibinfo  {journal} {Nucl. Phys.}\ }\textbf {\bibinfo {volume} {B156}},\ \bibinfo {pages} {391} (\bibinfo {year} {1979})}\BibitemShut {NoStop}%
\bibitem [{\citenamefont {{T. Matsui and H. Satz}}(1986)}]{matsui19861}%
  \BibitemOpen
  \bibfield  {author} {\bibinfo {author} {\bibnamefont {{T. Matsui and H. Satz}}},\ }\bibfield  {title} {\bibinfo {title} {{$J/\psi$ Suppression by Quark-Gluon Plasma Formation}},\ }\href {https://doi.org/10.1016/0370-2693(86)91404-8} {\bibfield  {journal} {\bibinfo  {journal} {Phys. Lett.}\ }\textbf {\bibinfo {volume} {B178}},\ \bibinfo {pages} {416} (\bibinfo {year} {1986})}\BibitemShut {NoStop}%
\bibitem [{\citenamefont {{L. Grandchamp and R. Rapp}}(2001)}]{Grandchamp:2001pf}%
  \BibitemOpen
  \bibfield  {author} {\bibinfo {author} {\bibnamefont {{L. Grandchamp and R. Rapp}}},\ }\bibfield  {title} {\bibinfo {title} {{Thermal versus direct J / Psi production in ultrarelativistic heavy ion collisions}},\ }\href {https://doi.org/10.1016/S0370-2693(01)01311-9} {\bibfield  {journal} {\bibinfo  {journal} {Phys. Lett.}\ }\textbf {\bibinfo {volume} {B523}},\ \bibinfo {pages} {60} (\bibinfo {year} {2001})},\ \Eprint {https://arxiv.org/abs/hep-ph/0103124} {arXiv:hep-ph/0103124 [hep-ph]} \BibitemShut {NoStop}%
\bibitem [{\citenamefont {Laine}\ \emph {et~al.}(2007)\citenamefont {Laine}, \citenamefont {Philipsen}, \citenamefont {Tassler},\ and\ \citenamefont {Romatschke}}]{laine20071}%
  \BibitemOpen
  \bibfield  {author} {\bibinfo {author} {\bibfnamefont {M.}~\bibnamefont {Laine}}, \bibinfo {author} {\bibfnamefont {O.}~\bibnamefont {Philipsen}}, \bibinfo {author} {\bibfnamefont {M.}~\bibnamefont {Tassler}},\ and\ \bibinfo {author} {\bibfnamefont {P.}~\bibnamefont {Romatschke}},\ }\bibfield  {title} {\bibinfo {title} {Real-time static potential in hot {QCD}},\ }\href {https://doi.org/10.1088/1126-6708/2007/03/054} {\bibfield  {journal} {\bibinfo  {journal} {Journal of High Energy Physics}\ }\textbf {\bibinfo {volume} {2007}},\ \bibinfo {pages} {054} (\bibinfo {year} {2007})}\BibitemShut {NoStop}%
\bibitem [{Note1()}]{Note1}%
  \BibitemOpen
  \bibinfo {note} {Additionally, regeneration~\cite {Grandchamp:2001pf,grandchamp20021,grandchamp20041,greco20031} via coalescence of $Q$ and $\protect \overline {Q}$ created at different hard vertices is important for charmonia. For bottomonia, this contribution is expected to be small~\cite {du20171}.}\BibitemShut {Stop}%
\bibitem [{\citenamefont {Andronic}\ \emph {et~al.}(2016)\citenamefont {Andronic} \emph {et~al.}}]{andronic20151}%
  \BibitemOpen
  \bibfield  {author} {\bibinfo {author} {\bibfnamefont {A.}~\bibnamefont {Andronic}} \emph {et~al.},\ }\bibfield  {title} {\bibinfo {title} {{Heavy-flavour and quarkonium production in the LHC era: from proton–proton to heavy-ion collisions}},\ }\href {https://doi.org/10.1140/epjc/s10052-015-3819-5} {\bibfield  {journal} {\bibinfo  {journal} {Eur. Phys. J.}\ }\textbf {\bibinfo {volume} {C76}},\ \bibinfo {pages} {107} (\bibinfo {year} {2016})},\ \Eprint {https://arxiv.org/abs/1506.03981} {arXiv:1506.03981 [nucl-ex]} \BibitemShut {NoStop}%
\bibitem [{\citenamefont {Andronic}\ \emph {et~al.}(2024)\citenamefont {Andronic} \emph {et~al.}}]{Andronic:2024oxz}%
  \BibitemOpen
  \bibfield  {author} {\bibinfo {author} {\bibfnamefont {A.}~\bibnamefont {Andronic}} \emph {et~al.},\ }\bibfield  {title} {\bibinfo {title} {{Comparative study of quarkonium transport in hot QCD matter}},\ }\href {https://doi.org/10.1140/epja/s10050-024-01306-6} {\bibfield  {journal} {\bibinfo  {journal} {Eur. Phys. J. A}\ }\textbf {\bibinfo {volume} {60}},\ \bibinfo {pages} {88} (\bibinfo {year} {2024})},\ \Eprint {https://arxiv.org/abs/2402.04366} {arXiv:2402.04366 [nucl-th]} \BibitemShut {NoStop}%
\bibitem [{\citenamefont {{N. Brambilla, A. Pineda, J. Soto and A. Vairo}}(2000)}]{brambilla20001}%
  \BibitemOpen
  \bibfield  {author} {\bibinfo {author} {\bibnamefont {{N. Brambilla, A. Pineda, J. Soto and A. Vairo}}},\ }\bibfield  {title} {\bibinfo {title} {Potential nrqcd: an effective theory for heavy quarkonium},\ }\href {https://doi.org/https://doi.org/10.1016/S0550-3213(99)00693-8} {\bibfield  {journal} {\bibinfo  {journal} {Nuclear Physics B}\ }\textbf {\bibinfo {volume} {566}},\ \bibinfo {pages} {275 } (\bibinfo {year} {2000})}\BibitemShut {NoStop}%
\bibitem [{\citenamefont {{N. Brambilla, J. Ghiglieri, A. Vairo and P. Petreczky}}(2008)}]{brambilla20081}%
  \BibitemOpen
  \bibfield  {author} {\bibinfo {author} {\bibnamefont {{N. Brambilla, J. Ghiglieri, A. Vairo and P. Petreczky}}},\ }\bibfield  {title} {\bibinfo {title} {Static quark-antiquark pairs at finite temperature},\ }\href {https://doi.org/10.1103/PhysRevD.78.014017} {\bibfield  {journal} {\bibinfo  {journal} {Phys. Rev. D}\ }\textbf {\bibinfo {volume} {78}},\ \bibinfo {pages} {014017} (\bibinfo {year} {2008})}\BibitemShut {NoStop}%
\bibitem [{\citenamefont {Brambilla}\ \emph {et~al.}(2010)\citenamefont {Brambilla}, \citenamefont {Escobedo}, \citenamefont {Ghiglieri}, \citenamefont {Soto},\ and\ \citenamefont {Vairo}}]{brambilla20101}%
  \BibitemOpen
  \bibfield  {author} {\bibinfo {author} {\bibfnamefont {N.}~\bibnamefont {Brambilla}}, \bibinfo {author} {\bibfnamefont {M.~{\'A}.}\ \bibnamefont {Escobedo}}, \bibinfo {author} {\bibfnamefont {J.}~\bibnamefont {Ghiglieri}}, \bibinfo {author} {\bibfnamefont {J.}~\bibnamefont {Soto}},\ and\ \bibinfo {author} {\bibfnamefont {A.}~\bibnamefont {Vairo}},\ }\bibfield  {title} {\bibinfo {title} {Heavy quarkonium in a weakly-coupled quark-gluon plasma below the melting temperature},\ }\href {https://doi.org/10.1007/JHEP09(2010)038} {\bibfield  {journal} {\bibinfo  {journal} {Journal of High Energy Physics}\ }\textbf {\bibinfo {volume} {2010}},\ \bibinfo {pages} {38} (\bibinfo {year} {2010})}\BibitemShut {NoStop}%
\bibitem [{\citenamefont {Brambilla}\ \emph {et~al.}(2011)\citenamefont {Brambilla}, \citenamefont {Escobedo}, \citenamefont {Ghiglieri},\ and\ \citenamefont {Vairo}}]{brambilla20111}%
  \BibitemOpen
  \bibfield  {author} {\bibinfo {author} {\bibfnamefont {N.}~\bibnamefont {Brambilla}}, \bibinfo {author} {\bibfnamefont {M.~{\'A}.}\ \bibnamefont {Escobedo}}, \bibinfo {author} {\bibfnamefont {J.}~\bibnamefont {Ghiglieri}},\ and\ \bibinfo {author} {\bibfnamefont {A.}~\bibnamefont {Vairo}},\ }\bibfield  {title} {\bibinfo {title} {Thermal width and gluo-dissociation of quarkonium in pnrqcd},\ }\href {https://doi.org/10.1007/JHEP12(2011)116} {\bibfield  {journal} {\bibinfo  {journal} {Journal of High Energy Physics}\ }\textbf {\bibinfo {volume} {2011}},\ \bibinfo {pages} {116} (\bibinfo {year} {2011})}\BibitemShut {NoStop}%
\bibitem [{\citenamefont {Brambilla}\ \emph {et~al.}(2013)\citenamefont {Brambilla}, \citenamefont {Escobedo}, \citenamefont {Ghiglieri},\ and\ \citenamefont {Vairo}}]{brambilla20131}%
  \BibitemOpen
  \bibfield  {author} {\bibinfo {author} {\bibfnamefont {N.}~\bibnamefont {Brambilla}}, \bibinfo {author} {\bibfnamefont {M.~A.}\ \bibnamefont {Escobedo}}, \bibinfo {author} {\bibfnamefont {J.}~\bibnamefont {Ghiglieri}},\ and\ \bibinfo {author} {\bibfnamefont {A.}~\bibnamefont {Vairo}},\ }\bibfield  {title} {\bibinfo {title} {{Thermal width and quarkonium dissociation by inelastic parton scattering}},\ }\href {https://doi.org/10.1007/JHEP05(2013)130} {\bibfield  {journal} {\bibinfo  {journal} {JHEP}\ }\textbf {\bibinfo {volume} {05}},\ \bibinfo {pages} {130}},\ \Eprint {https://arxiv.org/abs/1303.6097} {arXiv:1303.6097 [hep-ph]} \BibitemShut {NoStop}%
\bibitem [{\citenamefont {Sharma}\ and\ \citenamefont {Singh}(2024)}]{Sharma:2023dhj}%
  \BibitemOpen
  \bibfield  {author} {\bibinfo {author} {\bibfnamefont {R.}~\bibnamefont {Sharma}}\ and\ \bibinfo {author} {\bibfnamefont {B.}~\bibnamefont {Singh}},\ }\bibfield  {title} {\bibinfo {title} {{Energy hierarchies governing quarkonium dynamics in heavy ion collisions}},\ }\href {https://doi.org/10.1103/PhysRevC.109.054905} {\bibfield  {journal} {\bibinfo  {journal} {Phys. Rev. C}\ }\textbf {\bibinfo {volume} {109}},\ \bibinfo {pages} {054905} (\bibinfo {year} {2024})},\ \Eprint {https://arxiv.org/abs/2302.00508} {arXiv:2302.00508 [hep-ph]} \BibitemShut {NoStop}%
\bibitem [{\citenamefont {Petreczky}(2012)}]{petreczky20121}%
  \BibitemOpen
  \bibfield  {author} {\bibinfo {author} {\bibfnamefont {P.}~\bibnamefont {Petreczky}},\ }\bibfield  {title} {\bibinfo {title} {Lattice {QCD} at non-zero temperature},\ }\href {https://doi.org/10.1088/0954-3899/39/9/093002} {\bibfield  {journal} {\bibinfo  {journal} {Journal of Physics G: Nuclear and Particle Physics}\ }\textbf {\bibinfo {volume} {39}},\ \bibinfo {pages} {093002} (\bibinfo {year} {2012})}\BibitemShut {NoStop}%
\bibitem [{\citenamefont {Smirnov}\ \emph {et~al.}(2008)\citenamefont {Smirnov}, \citenamefont {Smirnov},\ and\ \citenamefont {Steinhauser}}]{Smirnov:2008pn}%
  \BibitemOpen
  \bibfield  {author} {\bibinfo {author} {\bibfnamefont {A.~V.}\ \bibnamefont {Smirnov}}, \bibinfo {author} {\bibfnamefont {V.~A.}\ \bibnamefont {Smirnov}},\ and\ \bibinfo {author} {\bibfnamefont {M.}~\bibnamefont {Steinhauser}},\ }\bibfield  {title} {\bibinfo {title} {{Fermionic contributions to the three-loop static potential}},\ }\href {https://doi.org/10.1016/j.physletb.2008.08.070} {\bibfield  {journal} {\bibinfo  {journal} {Phys. Lett. B}\ }\textbf {\bibinfo {volume} {668}},\ \bibinfo {pages} {293} (\bibinfo {year} {2008})},\ \Eprint {https://arxiv.org/abs/0809.1927} {arXiv:0809.1927 [hep-ph]} \BibitemShut {NoStop}%
\bibitem [{\citenamefont {Anzai}\ \emph {et~al.}(2010)\citenamefont {Anzai}, \citenamefont {Kiyo},\ and\ \citenamefont {Sumino}}]{Anzai:2009tm}%
  \BibitemOpen
  \bibfield  {author} {\bibinfo {author} {\bibfnamefont {C.}~\bibnamefont {Anzai}}, \bibinfo {author} {\bibfnamefont {Y.}~\bibnamefont {Kiyo}},\ and\ \bibinfo {author} {\bibfnamefont {Y.}~\bibnamefont {Sumino}},\ }\bibfield  {title} {\bibinfo {title} {{Static QCD potential at three-loop order}},\ }\href {https://doi.org/10.1103/PhysRevLett.104.112003} {\bibfield  {journal} {\bibinfo  {journal} {Phys. Rev. Lett.}\ }\textbf {\bibinfo {volume} {104}},\ \bibinfo {pages} {112003} (\bibinfo {year} {2010})},\ \Eprint {https://arxiv.org/abs/0911.4335} {arXiv:0911.4335 [hep-ph]} \BibitemShut {NoStop}%
\bibitem [{\citenamefont {Smirnov}\ \emph {et~al.}(2010)\citenamefont {Smirnov}, \citenamefont {Smirnov},\ and\ \citenamefont {Steinhauser}}]{Smirnov:2009fh}%
  \BibitemOpen
  \bibfield  {author} {\bibinfo {author} {\bibfnamefont {A.~V.}\ \bibnamefont {Smirnov}}, \bibinfo {author} {\bibfnamefont {V.~A.}\ \bibnamefont {Smirnov}},\ and\ \bibinfo {author} {\bibfnamefont {M.}~\bibnamefont {Steinhauser}},\ }\bibfield  {title} {\bibinfo {title} {{Three-loop static potential}},\ }\href {https://doi.org/10.1103/PhysRevLett.104.112002} {\bibfield  {journal} {\bibinfo  {journal} {Phys. Rev. Lett.}\ }\textbf {\bibinfo {volume} {104}},\ \bibinfo {pages} {112002} (\bibinfo {year} {2010})},\ \Eprint {https://arxiv.org/abs/0911.4742} {arXiv:0911.4742 [hep-ph]} \BibitemShut {NoStop}%
\bibitem [{\citenamefont {Pineda}\ and\ \citenamefont {Segovia}(2013)}]{Pineda:2013lta}%
  \BibitemOpen
  \bibfield  {author} {\bibinfo {author} {\bibfnamefont {A.}~\bibnamefont {Pineda}}\ and\ \bibinfo {author} {\bibfnamefont {J.}~\bibnamefont {Segovia}},\ }\bibfield  {title} {\bibinfo {title} {{Improved determination of heavy quarkonium magnetic dipole transitions in potential nonrelativistic QCD}},\ }\href {https://doi.org/10.1103/PhysRevD.87.074024} {\bibfield  {journal} {\bibinfo  {journal} {Phys. Rev. D}\ }\textbf {\bibinfo {volume} {87}},\ \bibinfo {pages} {074024} (\bibinfo {year} {2013})},\ \Eprint {https://arxiv.org/abs/1302.3528} {arXiv:1302.3528 [hep-ph]} \BibitemShut {NoStop}%
\bibitem [{\citenamefont {Burnier}\ and\ \citenamefont {Rothkopf}(2017)}]{Burnier:2016mxc}%
  \BibitemOpen
  \bibfield  {author} {\bibinfo {author} {\bibfnamefont {Y.}~\bibnamefont {Burnier}}\ and\ \bibinfo {author} {\bibfnamefont {A.}~\bibnamefont {Rothkopf}},\ }\bibfield  {title} {\bibinfo {title} {{Complex heavy-quark potential and Debye mass in a gluonic medium from lattice QCD}},\ }\href {https://doi.org/10.1103/PhysRevD.95.054511} {\bibfield  {journal} {\bibinfo  {journal} {Phys. Rev. D}\ }\textbf {\bibinfo {volume} {95}},\ \bibinfo {pages} {054511} (\bibinfo {year} {2017})},\ \Eprint {https://arxiv.org/abs/1607.04049} {arXiv:1607.04049 [hep-lat]} \BibitemShut {NoStop}%
\bibitem [{\citenamefont {Rothkopf}\ \emph {et~al.}(2012)\citenamefont {Rothkopf}, \citenamefont {Hatsuda},\ and\ \citenamefont {Sasaki}}]{rothkopf20111}%
  \BibitemOpen
  \bibfield  {author} {\bibinfo {author} {\bibfnamefont {A.}~\bibnamefont {Rothkopf}}, \bibinfo {author} {\bibfnamefont {T.}~\bibnamefont {Hatsuda}},\ and\ \bibinfo {author} {\bibfnamefont {S.}~\bibnamefont {Sasaki}},\ }\bibfield  {title} {\bibinfo {title} {{Complex Heavy-Quark Potential at Finite Temperature from Lattice QCD}},\ }\href {https://doi.org/10.1103/PhysRevLett.108.162001} {\bibfield  {journal} {\bibinfo  {journal} {Phys. Rev. Lett.}\ }\textbf {\bibinfo {volume} {108}},\ \bibinfo {pages} {162001} (\bibinfo {year} {2012})},\ \Eprint {https://arxiv.org/abs/1108.1579} {arXiv:1108.1579 [hep-lat]} \BibitemShut {NoStop}%
\bibitem [{\citenamefont {Bala}\ and\ \citenamefont {Datta}(2020)}]{Bala:2019cqu}%
  \BibitemOpen
  \bibfield  {author} {\bibinfo {author} {\bibfnamefont {D.}~\bibnamefont {Bala}}\ and\ \bibinfo {author} {\bibfnamefont {S.}~\bibnamefont {Datta}},\ }\bibfield  {title} {\bibinfo {title} {{Nonperturbative potential for the study of quarkonia in QGP}},\ }\href {https://doi.org/10.1103/PhysRevD.101.034507} {\bibfield  {journal} {\bibinfo  {journal} {Phys. Rev. D}\ }\textbf {\bibinfo {volume} {101}},\ \bibinfo {pages} {034507} (\bibinfo {year} {2020})},\ \Eprint {https://arxiv.org/abs/1909.10548} {arXiv:1909.10548 [hep-lat]} \BibitemShut {NoStop}%
\bibitem [{\citenamefont {Bala}\ and\ \citenamefont {Datta}(2021)}]{Bala:2020tdt}%
  \BibitemOpen
  \bibfield  {author} {\bibinfo {author} {\bibfnamefont {D.}~\bibnamefont {Bala}}\ and\ \bibinfo {author} {\bibfnamefont {S.}~\bibnamefont {Datta}},\ }\bibfield  {title} {\bibinfo {title} {{Interaction potential between heavy $Q\overline{Q}$ in a color octet configuration in the QGP from a study of hybrid Wilson loops}},\ }\href {https://doi.org/10.1103/PhysRevD.103.014512} {\bibfield  {journal} {\bibinfo  {journal} {Phys. Rev. D}\ }\textbf {\bibinfo {volume} {103}},\ \bibinfo {pages} {014512} (\bibinfo {year} {2021})},\ \Eprint {https://arxiv.org/abs/2009.00773} {arXiv:2009.00773 [hep-lat]} \BibitemShut {NoStop}%
\bibitem [{\citenamefont {Bala}\ \emph {et~al.}(2022)\citenamefont {Bala}, \citenamefont {Kaczmarek}, \citenamefont {Larsen}, \citenamefont {Mukherjee}, \citenamefont {Parkar}, \citenamefont {Petreczky}, \citenamefont {Rothkopf},\ and\ \citenamefont {Weber}}]{Bala:2021fkm}%
  \BibitemOpen
  \bibfield  {author} {\bibinfo {author} {\bibfnamefont {D.}~\bibnamefont {Bala}}, \bibinfo {author} {\bibfnamefont {O.}~\bibnamefont {Kaczmarek}}, \bibinfo {author} {\bibfnamefont {R.}~\bibnamefont {Larsen}}, \bibinfo {author} {\bibfnamefont {S.}~\bibnamefont {Mukherjee}}, \bibinfo {author} {\bibfnamefont {G.}~\bibnamefont {Parkar}}, \bibinfo {author} {\bibfnamefont {P.}~\bibnamefont {Petreczky}}, \bibinfo {author} {\bibfnamefont {A.}~\bibnamefont {Rothkopf}},\ and\ \bibinfo {author} {\bibfnamefont {J.~H.}\ \bibnamefont {Weber}} (\bibinfo {collaboration} {HotQCD}),\ }\bibfield  {title} {\bibinfo {title} {{Static quark-antiquark interactions at nonzero temperature from lattice QCD}},\ }\href {https://doi.org/10.1103/PhysRevD.105.054513} {\bibfield  {journal} {\bibinfo  {journal} {Phys. Rev. D}\ }\textbf {\bibinfo {volume} {105}},\ \bibinfo {pages} {054513} (\bibinfo {year} {2022})},\ \Eprint {https://arxiv.org/abs/2110.11659} {arXiv:2110.11659 [hep-lat]} \BibitemShut {NoStop}%
\bibitem [{\citenamefont {Brambilla}\ \emph {et~al.}(2005)\citenamefont {Brambilla}, \citenamefont {Pineda}, \citenamefont {Soto},\ and\ \citenamefont {Vairo}}]{Brambilla:2004jw}%
  \BibitemOpen
  \bibfield  {author} {\bibinfo {author} {\bibfnamefont {N.}~\bibnamefont {Brambilla}}, \bibinfo {author} {\bibfnamefont {A.}~\bibnamefont {Pineda}}, \bibinfo {author} {\bibfnamefont {J.}~\bibnamefont {Soto}},\ and\ \bibinfo {author} {\bibfnamefont {A.}~\bibnamefont {Vairo}},\ }\bibfield  {title} {\bibinfo {title} {{Effective Field Theories for Heavy Quarkonium}},\ }\href {https://doi.org/10.1103/RevModPhys.77.1423} {\bibfield  {journal} {\bibinfo  {journal} {Rev. Mod. Phys.}\ }\textbf {\bibinfo {volume} {77}},\ \bibinfo {pages} {1423} (\bibinfo {year} {2005})},\ \Eprint {https://arxiv.org/abs/hep-ph/0410047} {arXiv:hep-ph/0410047} \BibitemShut {NoStop}%
\bibitem [{\citenamefont {Eller}\ \emph {et~al.}(2019)\citenamefont {Eller}, \citenamefont {Ghiglieri},\ and\ \citenamefont {Moore}}]{Eller:2019spw}%
  \BibitemOpen
  \bibfield  {author} {\bibinfo {author} {\bibfnamefont {A.~M.}\ \bibnamefont {Eller}}, \bibinfo {author} {\bibfnamefont {J.}~\bibnamefont {Ghiglieri}},\ and\ \bibinfo {author} {\bibfnamefont {G.~D.}\ \bibnamefont {Moore}},\ }\bibfield  {title} {\bibinfo {title} {{Thermal Heavy Quark Self-Energy from Euclidean Correlators}},\ }\href {https://doi.org/10.1103/PhysRevD.99.094042} {\bibfield  {journal} {\bibinfo  {journal} {Phys. Rev. D}\ }\textbf {\bibinfo {volume} {99}},\ \bibinfo {pages} {094042} (\bibinfo {year} {2019})},\ \bibinfo {note} {[Erratum: Phys.Rev.D 102, 039901 (2020)]},\ \Eprint {https://arxiv.org/abs/1903.08064} {arXiv:1903.08064 [hep-ph]} \BibitemShut {NoStop}%
\bibitem [{\citenamefont {Binder}\ \emph {et~al.}(2022)\citenamefont {Binder}, \citenamefont {Mukaida}, \citenamefont {Scheihing-Hitschfeld},\ and\ \citenamefont {Yao}}]{Binder:2021otw}%
  \BibitemOpen
  \bibfield  {author} {\bibinfo {author} {\bibfnamefont {T.}~\bibnamefont {Binder}}, \bibinfo {author} {\bibfnamefont {K.}~\bibnamefont {Mukaida}}, \bibinfo {author} {\bibfnamefont {B.}~\bibnamefont {Scheihing-Hitschfeld}},\ and\ \bibinfo {author} {\bibfnamefont {X.}~\bibnamefont {Yao}},\ }\bibfield  {title} {\bibinfo {title} {{Non-Abelian electric field correlator at NLO for dark matter relic abundance and quarkonium transport}},\ }\href {https://doi.org/10.1007/JHEP01(2022)137} {\bibfield  {journal} {\bibinfo  {journal} {JHEP}\ }\textbf {\bibinfo {volume} {01}},\ \bibinfo {pages} {137}},\ \Eprint {https://arxiv.org/abs/2107.03945} {arXiv:2107.03945 [hep-ph]} \BibitemShut {NoStop}%
\bibitem [{\citenamefont {Scheihing-Hitschfeld}\ and\ \citenamefont {Yao}(2023)}]{Scheihing-Hitschfeld:2022xqx}%
  \BibitemOpen
  \bibfield  {author} {\bibinfo {author} {\bibfnamefont {B.}~\bibnamefont {Scheihing-Hitschfeld}}\ and\ \bibinfo {author} {\bibfnamefont {X.}~\bibnamefont {Yao}},\ }\bibfield  {title} {\bibinfo {title} {{Gauge Invariance of Non-Abelian Field Strength Correlators: The Axial Gauge Puzzle}},\ }\href {https://doi.org/10.1103/PhysRevLett.130.052302} {\bibfield  {journal} {\bibinfo  {journal} {Phys. Rev. Lett.}\ }\textbf {\bibinfo {volume} {130}},\ \bibinfo {pages} {052302} (\bibinfo {year} {2023})},\ \Eprint {https://arxiv.org/abs/2205.04477} {arXiv:2205.04477 [hep-ph]} \BibitemShut {NoStop}%
\bibitem [{\citenamefont {Brambilla}\ \emph {et~al.}(2020)\citenamefont {Brambilla}, \citenamefont {Leino}, \citenamefont {Petreczky},\ and\ \citenamefont {Vairo}}]{Brambilla:2020siz}%
  \BibitemOpen
  \bibfield  {author} {\bibinfo {author} {\bibfnamefont {N.}~\bibnamefont {Brambilla}}, \bibinfo {author} {\bibfnamefont {V.}~\bibnamefont {Leino}}, \bibinfo {author} {\bibfnamefont {P.}~\bibnamefont {Petreczky}},\ and\ \bibinfo {author} {\bibfnamefont {A.}~\bibnamefont {Vairo}},\ }\bibfield  {title} {\bibinfo {title} {{Lattice QCD constraints on the heavy quark diffusion coefficient}},\ }\href {https://doi.org/10.1103/PhysRevD.102.074503} {\bibfield  {journal} {\bibinfo  {journal} {Phys. Rev. D}\ }\textbf {\bibinfo {volume} {102}},\ \bibinfo {pages} {074503} (\bibinfo {year} {2020})},\ \Eprint {https://arxiv.org/abs/2007.10078} {arXiv:2007.10078 [hep-lat]} \BibitemShut {NoStop}%
\bibitem [{\citenamefont {Brambilla}\ \emph {et~al.}(2023)\citenamefont {Brambilla}, \citenamefont {Escobedo}, \citenamefont {Islam}, \citenamefont {Strickland}, \citenamefont {Tiwari}, \citenamefont {Vairo},\ and\ \citenamefont {Vander~Griend}}]{Brambilla:2023hkw}%
  \BibitemOpen
  \bibfield  {author} {\bibinfo {author} {\bibfnamefont {N.}~\bibnamefont {Brambilla}}, \bibinfo {author} {\bibfnamefont {M.~A.}\ \bibnamefont {Escobedo}}, \bibinfo {author} {\bibfnamefont {A.}~\bibnamefont {Islam}}, \bibinfo {author} {\bibfnamefont {M.}~\bibnamefont {Strickland}}, \bibinfo {author} {\bibfnamefont {A.}~\bibnamefont {Tiwari}}, \bibinfo {author} {\bibfnamefont {A.}~\bibnamefont {Vairo}},\ and\ \bibinfo {author} {\bibfnamefont {P.}~\bibnamefont {Vander~Griend}},\ }\bibfield  {title} {\bibinfo {title} {{Regeneration of bottomonia in an open quantum systems approach}},\ }\href {https://doi.org/10.1103/PhysRevD.108.L011502} {\bibfield  {journal} {\bibinfo  {journal} {Phys. Rev. D}\ }\textbf {\bibinfo {volume} {108}},\ \bibinfo {pages} {L011502} (\bibinfo {year} {2023})},\ \Eprint {https://arxiv.org/abs/2302.11826} {arXiv:2302.11826 [hep-ph]} \BibitemShut {NoStop}%
\bibitem [{\citenamefont {Brambilla}\ \emph {et~al.}(2024)\citenamefont {Brambilla}, \citenamefont {Magorsch}, \citenamefont {Strickland}, \citenamefont {Vairo},\ and\ \citenamefont {Vander~Griend}}]{Brambilla:2024tqg}%
  \BibitemOpen
  \bibfield  {author} {\bibinfo {author} {\bibfnamefont {N.}~\bibnamefont {Brambilla}}, \bibinfo {author} {\bibfnamefont {T.}~\bibnamefont {Magorsch}}, \bibinfo {author} {\bibfnamefont {M.}~\bibnamefont {Strickland}}, \bibinfo {author} {\bibfnamefont {A.}~\bibnamefont {Vairo}},\ and\ \bibinfo {author} {\bibfnamefont {P.}~\bibnamefont {Vander~Griend}},\ }\bibfield  {title} {\bibinfo {title} {{Bottomonium suppression from the three-loop QCD potential}},\ }\href {https://doi.org/10.1103/PhysRevD.109.114016} {\bibfield  {journal} {\bibinfo  {journal} {Phys. Rev. D}\ }\textbf {\bibinfo {volume} {109}},\ \bibinfo {pages} {114016} (\bibinfo {year} {2024})},\ \Eprint {https://arxiv.org/abs/2403.15545} {arXiv:2403.15545 [hep-ph]} \BibitemShut {NoStop}%
\bibitem [{\citenamefont {Nijs}\ \emph {et~al.}(2023)\citenamefont {Nijs}, \citenamefont {Scheihing-Hitschfeld},\ and\ \citenamefont {Yao}}]{Nijs:2023dks}%
  \BibitemOpen
  \bibfield  {author} {\bibinfo {author} {\bibfnamefont {G.}~\bibnamefont {Nijs}}, \bibinfo {author} {\bibfnamefont {B.}~\bibnamefont {Scheihing-Hitschfeld}},\ and\ \bibinfo {author} {\bibfnamefont {X.}~\bibnamefont {Yao}},\ }\bibfield  {title} {\bibinfo {title} {{Chromoelectric field correlator for quarkonium transport in the strongly coupled $ \mathcal{N} $ = 4 Yang-Mills plasma from AdS/CFT}},\ }\href {https://doi.org/10.1007/JHEP06(2023)007} {\bibfield  {journal} {\bibinfo  {journal} {JHEP}\ }\textbf {\bibinfo {volume} {06}},\ \bibinfo {pages} {007}},\ \Eprint {https://arxiv.org/abs/2304.03298} {arXiv:2304.03298 [hep-ph]} \BibitemShut {NoStop}%
\bibitem [{\citenamefont {Nijs}\ \emph {et~al.}(2024)\citenamefont {Nijs}, \citenamefont {Scheihing-Hitschfeld},\ and\ \citenamefont {Yao}}]{Nijs:2023dbc}%
  \BibitemOpen
  \bibfield  {author} {\bibinfo {author} {\bibfnamefont {G.}~\bibnamefont {Nijs}}, \bibinfo {author} {\bibfnamefont {B.}~\bibnamefont {Scheihing-Hitschfeld}},\ and\ \bibinfo {author} {\bibfnamefont {X.}~\bibnamefont {Yao}},\ }\bibfield  {title} {\bibinfo {title} {{Generalized gluon distribution for quarkonium dynamics in strongly coupled N=4 Yang-Mills theory}},\ }\href {https://doi.org/10.1103/PhysRevD.109.094043} {\bibfield  {journal} {\bibinfo  {journal} {Phys. Rev. D}\ }\textbf {\bibinfo {volume} {109}},\ \bibinfo {pages} {094043} (\bibinfo {year} {2024})},\ \Eprint {https://arxiv.org/abs/2310.09325} {arXiv:2310.09325 [hep-ph]} \BibitemShut {NoStop}%
\bibitem [{\citenamefont {Casalderrey-Solana}\ \emph {et~al.}(2014)\citenamefont {Casalderrey-Solana}, \citenamefont {Liu}, \citenamefont {Mateos}, \citenamefont {Rajagopal},\ and\ \citenamefont {Achim~Wiedemann}}]{Casalderrey-Solana:2011dxg}%
  \BibitemOpen
  \bibfield  {author} {\bibinfo {author} {\bibfnamefont {J.}~\bibnamefont {Casalderrey-Solana}}, \bibinfo {author} {\bibfnamefont {H.}~\bibnamefont {Liu}}, \bibinfo {author} {\bibfnamefont {D.}~\bibnamefont {Mateos}}, \bibinfo {author} {\bibfnamefont {K.}~\bibnamefont {Rajagopal}},\ and\ \bibinfo {author} {\bibfnamefont {U.}~\bibnamefont {Achim~Wiedemann}},\ }\href {https://doi.org/10.1017/9781009403504} {\emph {\bibinfo {title} {{Gauge/String Duality, Hot QCD and Heavy Ion Collisions}}}}\ (\bibinfo  {publisher} {Cambridge University Press},\ \bibinfo {year} {2014})\ \Eprint {https://arxiv.org/abs/1101.0618} {arXiv:1101.0618 [hep-th]} \BibitemShut {NoStop}%
\bibitem [{\citenamefont {Leino}(2024)}]{Leino:2024pen}%
  \BibitemOpen
  \bibfield  {author} {\bibinfo {author} {\bibfnamefont {V.}~\bibnamefont {Leino}},\ }\bibfield  {title} {\bibinfo {title} {{Adjoint chromoelectric and -magnetic correlators with gradient flow}},\ }\href {https://doi.org/10.22323/1.453.0385} {\bibfield  {journal} {\bibinfo  {journal} {PoS}\ }\textbf {\bibinfo {volume} {LATTICE2023}},\ \bibinfo {pages} {385} (\bibinfo {year} {2024})},\ \Eprint {https://arxiv.org/abs/2401.06733} {arXiv:2401.06733 [hep-lat]} \BibitemShut {NoStop}%
\bibitem [{\citenamefont {Mayer-Steudte}(2025)}]{Mayer-Steudte:2025gvz}%
  \BibitemOpen
  \bibfield  {author} {\bibinfo {author} {\bibfnamefont {J.}~\bibnamefont {Mayer-Steudte}},\ }\bibfield  {title} {\bibinfo {title} {{Adjoint chromoelectric correlators for heavy quarkonium diffusion}},\ }\href {https://doi.org/10.22323/1.466.0205} {\bibfield  {journal} {\bibinfo  {journal} {PoS}\ }\textbf {\bibinfo {volume} {LATTICE2024}},\ \bibinfo {pages} {205} (\bibinfo {year} {2025})}\BibitemShut {NoStop}%
\bibitem [{\citenamefont {Romatschke}(2010)}]{Romatschke:2009im}%
  \BibitemOpen
  \bibfield  {author} {\bibinfo {author} {\bibfnamefont {P.}~\bibnamefont {Romatschke}},\ }\bibfield  {title} {\bibinfo {title} {{New Developments in Relativistic Viscous Hydrodynamics}},\ }\href {https://doi.org/10.1142/S0218301310014613} {\bibfield  {journal} {\bibinfo  {journal} {Int. J. Mod. Phys. E}\ }\textbf {\bibinfo {volume} {19}},\ \bibinfo {pages} {1} (\bibinfo {year} {2010})},\ \Eprint {https://arxiv.org/abs/0902.3663} {arXiv:0902.3663 [hep-ph]} \BibitemShut {NoStop}%
\bibitem [{\citenamefont {Teaney}(2010)}]{Teaney:2009qa}%
  \BibitemOpen
  \bibfield  {author} {\bibinfo {author} {\bibfnamefont {D.~A.}\ \bibnamefont {Teaney}},\ }\bibinfo {title} {{Viscous Hydrodynamics and the Quark Gluon Plasma}},\ in\ \href {https://doi.org/10.1142/9789814293297_0004} {\emph {\bibinfo {booktitle} {{Quark-gluon plasma 4}}}},\ \bibinfo {editor} {edited by\ \bibinfo {editor} {\bibfnamefont {R.~C.}\ \bibnamefont {Hwa}}\ and\ \bibinfo {editor} {\bibfnamefont {X.-N.}\ \bibnamefont {Wang}}}\ (\bibinfo {year} {2010})\ pp.\ \bibinfo {pages} {207--266},\ \Eprint {https://arxiv.org/abs/0905.2433} {arXiv:0905.2433 [nucl-th]} \BibitemShut {NoStop}%
\bibitem [{\citenamefont {Hirano}(2012)}]{Hirano:2012qz}%
  \BibitemOpen
  \bibfield  {author} {\bibinfo {author} {\bibfnamefont {T.}~\bibnamefont {Hirano}},\ }\bibfield  {title} {\bibinfo {title} {{Dynamics of relativistic heavy ion collisions and the quark gluon plasma}},\ }\href {https://doi.org/10.1143/PTPS.195.1} {\bibfield  {journal} {\bibinfo  {journal} {Prog. Theor. Phys. Suppl.}\ }\textbf {\bibinfo {volume} {195}},\ \bibinfo {pages} {1} (\bibinfo {year} {2012})}\BibitemShut {NoStop}%
\bibitem [{\citenamefont {Song}(2015)}]{Song:2013gia}%
  \BibitemOpen
  \bibfield  {author} {\bibinfo {author} {\bibfnamefont {H.}~\bibnamefont {Song}},\ }\bibfield  {title} {\bibinfo {title} {{Hydrodynamic modelling for relativistic heavy-ion collisions at RHIC and LHC}},\ }\href {https://doi.org/10.1007/s12043-015-0971-2} {\bibfield  {journal} {\bibinfo  {journal} {Pramana}\ }\textbf {\bibinfo {volume} {84}},\ \bibinfo {pages} {703} (\bibinfo {year} {2015})},\ \Eprint {https://arxiv.org/abs/1401.0079} {arXiv:1401.0079 [nucl-th]} \BibitemShut {NoStop}%
\bibitem [{\citenamefont {Gale}\ \emph {et~al.}(2013)\citenamefont {Gale}, \citenamefont {Jeon},\ and\ \citenamefont {Schenke}}]{Gale:2013da}%
  \BibitemOpen
  \bibfield  {author} {\bibinfo {author} {\bibfnamefont {C.}~\bibnamefont {Gale}}, \bibinfo {author} {\bibfnamefont {S.}~\bibnamefont {Jeon}},\ and\ \bibinfo {author} {\bibfnamefont {B.}~\bibnamefont {Schenke}},\ }\bibfield  {title} {\bibinfo {title} {{Hydrodynamic Modeling of Heavy-Ion Collisions}},\ }\href {https://doi.org/10.1142/S0217751X13400113} {\bibfield  {journal} {\bibinfo  {journal} {Int. J. Mod. Phys. A}\ }\textbf {\bibinfo {volume} {28}},\ \bibinfo {pages} {1340011} (\bibinfo {year} {2013})},\ \Eprint {https://arxiv.org/abs/1301.5893} {arXiv:1301.5893 [nucl-th]} \BibitemShut {NoStop}%
\bibitem [{\citenamefont {Heinz}\ and\ \citenamefont {Snellings}(2013)}]{Heinz:2013th}%
  \BibitemOpen
  \bibfield  {author} {\bibinfo {author} {\bibfnamefont {U.}~\bibnamefont {Heinz}}\ and\ \bibinfo {author} {\bibfnamefont {R.}~\bibnamefont {Snellings}},\ }\bibfield  {title} {\bibinfo {title} {{Collective flow and viscosity in relativistic heavy-ion collisions}},\ }\href {https://doi.org/10.1146/annurev-nucl-102212-170540} {\bibfield  {journal} {\bibinfo  {journal} {Ann. Rev. Nucl. Part. Sci.}\ }\textbf {\bibinfo {volume} {63}},\ \bibinfo {pages} {123} (\bibinfo {year} {2013})},\ \Eprint {https://arxiv.org/abs/1301.2826} {arXiv:1301.2826 [nucl-th]} \BibitemShut {NoStop}%
\bibitem [{\citenamefont {Jaiswal}\ and\ \citenamefont {Roy}(2016)}]{Jaiswal:2016hex}%
  \BibitemOpen
  \bibfield  {author} {\bibinfo {author} {\bibfnamefont {A.}~\bibnamefont {Jaiswal}}\ and\ \bibinfo {author} {\bibfnamefont {V.}~\bibnamefont {Roy}},\ }\bibfield  {title} {\bibinfo {title} {{Relativistic hydrodynamics in heavy-ion collisions: general aspects and recent developments}},\ }\href {https://doi.org/10.1155/2016/9623034} {\bibfield  {journal} {\bibinfo  {journal} {Adv. High Energy Phys.}\ }\textbf {\bibinfo {volume} {2016}},\ \bibinfo {pages} {9623034} (\bibinfo {year} {2016})},\ \Eprint {https://arxiv.org/abs/1605.08694} {arXiv:1605.08694 [nucl-th]} \BibitemShut {NoStop}%
\bibitem [{\citenamefont {Chattopadhyay}\ \emph {et~al.}(2018)\citenamefont {Chattopadhyay}, \citenamefont {Bhalerao}, \citenamefont {Ollitrault},\ and\ \citenamefont {Pal}}]{PhysRevC.97.034915}%
  \BibitemOpen
  \bibfield  {author} {\bibinfo {author} {\bibfnamefont {C.}~\bibnamefont {Chattopadhyay}}, \bibinfo {author} {\bibfnamefont {R.~S.}\ \bibnamefont {Bhalerao}}, \bibinfo {author} {\bibfnamefont {J.-Y.}\ \bibnamefont {Ollitrault}},\ and\ \bibinfo {author} {\bibfnamefont {S.}~\bibnamefont {Pal}},\ }\bibfield  {title} {\bibinfo {title} {Effects of initial-state dynamics on collective flow within a coupled transport and viscous hydrodynamic approach},\ }\href {https://doi.org/10.1103/PhysRevC.97.034915} {\bibfield  {journal} {\bibinfo  {journal} {Phys. Rev. C}\ }\textbf {\bibinfo {volume} {97}},\ \bibinfo {pages} {034915} (\bibinfo {year} {2018})}\BibitemShut {NoStop}%
\bibitem [{\citenamefont {Bhalerao}\ \emph {et~al.}(2015)\citenamefont {Bhalerao}, \citenamefont {Jaiswal},\ and\ \citenamefont {Pal}}]{Bhalerao:2015iya}%
  \BibitemOpen
  \bibfield  {author} {\bibinfo {author} {\bibfnamefont {R.~S.}\ \bibnamefont {Bhalerao}}, \bibinfo {author} {\bibfnamefont {A.}~\bibnamefont {Jaiswal}},\ and\ \bibinfo {author} {\bibfnamefont {S.}~\bibnamefont {Pal}},\ }\bibfield  {title} {\bibinfo {title} {{Collective flow in event-by-event partonic transport plus hydrodynamics hybrid approach}},\ }\href {https://doi.org/10.1103/PhysRevC.92.014903} {\bibfield  {journal} {\bibinfo  {journal} {Phys. Rev. C}\ }\textbf {\bibinfo {volume} {92}},\ \bibinfo {pages} {014903} (\bibinfo {year} {2015})},\ \Eprint {https://arxiv.org/abs/1503.03862} {arXiv:1503.03862 [nucl-th]} \BibitemShut {NoStop}%
\bibitem [{\citenamefont {{C. Young and K. Dusling}}(2013)}]{young20131}%
  \BibitemOpen
  \bibfield  {author} {\bibinfo {author} {\bibnamefont {{C. Young and K. Dusling}}},\ }\bibfield  {title} {\bibinfo {title} {Quarkonium above deconfinement as an open quantum system},\ }\href {https://doi.org/10.1103/PhysRevC.87.065206} {\bibfield  {journal} {\bibinfo  {journal} {Phys. Rev. C}\ }\textbf {\bibinfo {volume} {87}},\ \bibinfo {pages} {065206} (\bibinfo {year} {2013})}\BibitemShut {NoStop}%
\bibitem [{\citenamefont {{Y. Akamatsu and A. Rothkopf}}(2012)}]{akamatsu20121}%
  \BibitemOpen
  \bibfield  {author} {\bibinfo {author} {\bibnamefont {{Y. Akamatsu and A. Rothkopf}}},\ }\bibfield  {title} {\bibinfo {title} {Stochastic potential and quantum decoherence of heavy quarkonium in the quark-gluon plasma},\ }\href@noop {} {\bibfield  {journal} {\bibinfo  {journal} {Physical Review D}\ }\textbf {\bibinfo {volume} {85}},\ \bibinfo {pages} {105011} (\bibinfo {year} {2012})}\BibitemShut {NoStop}%
\bibitem [{\citenamefont {Akamatsu}(2013)}]{akamatsu20131}%
  \BibitemOpen
  \bibfield  {author} {\bibinfo {author} {\bibfnamefont {Y.}~\bibnamefont {Akamatsu}},\ }\bibfield  {title} {\bibinfo {title} {Real-time quantum dynamics of heavy-quark systems at high temperature},\ }\href@noop {} {\bibfield  {journal} {\bibinfo  {journal} {Physical Review D}\ }\textbf {\bibinfo {volume} {87}},\ \bibinfo {pages} {045016} (\bibinfo {year} {2013})}\BibitemShut {NoStop}%
\bibitem [{\citenamefont {{H. P. Breuer and F. Petruccione}}(2002)}]{breuer20021}%
  \BibitemOpen
  \bibfield  {author} {\bibinfo {author} {\bibnamefont {{H. P. Breuer and F. Petruccione}}},\ }\href@noop {} {\emph {\bibinfo {title} {{The theory of open quantum systems}}}}\ (\bibinfo {year} {2002})\BibitemShut {NoStop}%
\bibitem [{\citenamefont {Akamatsu}(2022)}]{Akamatsu:2020ypb}%
  \BibitemOpen
  \bibfield  {author} {\bibinfo {author} {\bibfnamefont {Y.}~\bibnamefont {Akamatsu}},\ }\bibfield  {title} {\bibinfo {title} {{Quarkonium in quark\textendash{}gluon plasma: Open quantum system approaches re-examined}},\ }\href {https://doi.org/10.1016/j.ppnp.2021.103932} {\bibfield  {journal} {\bibinfo  {journal} {Prog. Part. Nucl. Phys.}\ }\textbf {\bibinfo {volume} {123}},\ \bibinfo {pages} {103932} (\bibinfo {year} {2022})},\ \Eprint {https://arxiv.org/abs/2009.10559} {arXiv:2009.10559 [nucl-th]} \BibitemShut {NoStop}%
\bibitem [{\citenamefont {Yao}(2021)}]{Yao:2021lus}%
  \BibitemOpen
  \bibfield  {author} {\bibinfo {author} {\bibfnamefont {X.}~\bibnamefont {Yao}},\ }\bibfield  {title} {\bibinfo {title} {{Open quantum systems for quarkonia}},\ }\href {https://doi.org/10.1142/S0217751X21300106} {\bibfield  {journal} {\bibinfo  {journal} {Int. J. Mod. Phys. A}\ }\textbf {\bibinfo {volume} {36}},\ \bibinfo {pages} {2130010} (\bibinfo {year} {2021})},\ \Eprint {https://arxiv.org/abs/2102.01736} {arXiv:2102.01736 [hep-ph]} \BibitemShut {NoStop}%
\bibitem [{\citenamefont {Sharma}(2021)}]{Sharma:2021vvu}%
  \BibitemOpen
  \bibfield  {author} {\bibinfo {author} {\bibfnamefont {R.}~\bibnamefont {Sharma}},\ }\bibfield  {title} {\bibinfo {title} {{Quarkonium propagation in the quark\textendash{}gluon plasma}},\ }\href {https://doi.org/10.1140/epjs/s11734-021-00025-z} {\bibfield  {journal} {\bibinfo  {journal} {Eur. Phys. J. ST}\ }\textbf {\bibinfo {volume} {230}},\ \bibinfo {pages} {697} (\bibinfo {year} {2021})},\ \Eprint {https://arxiv.org/abs/2101.04268} {arXiv:2101.04268 [hep-ph]} \BibitemShut {NoStop}%
\bibitem [{\citenamefont {{N. Borghini and C. Gombeaud }}(2011)}]{borghini20111}%
  \BibitemOpen
  \bibfield  {author} {\bibinfo {author} {\bibnamefont {{N. Borghini and C. Gombeaud }}},\ }\bibfield  {title} {\bibinfo {title} {{Dynamical Evolution of Heavy Quarkonia in a Deconfined Medium}},\ }\href@noop {} {\  (\bibinfo {year} {2011})},\ \Eprint {https://arxiv.org/abs/1103.2945} {arXiv:1103.2945 [hep-ph]} \BibitemShut {NoStop}%
\bibitem [{\citenamefont {{N. Borghini and C. Gombeaud}}(2012)}]{borghini20121}%
  \BibitemOpen
  \bibfield  {author} {\bibinfo {author} {\bibnamefont {{N. Borghini and C. Gombeaud}}},\ }\bibfield  {title} {\bibinfo {title} {Heavy quarkonia in a medium as a quantum dissipative system: master-equation approach},\ }\href@noop {} {\bibfield  {journal} {\bibinfo  {journal} {The European Physical Journal C}\ }\textbf {\bibinfo {volume} {72}},\ \bibinfo {pages} {2000} (\bibinfo {year} {2012})}\BibitemShut {NoStop}%
\bibitem [{\citenamefont {Lindblad}(1976)}]{lindblad19761}%
  \BibitemOpen
  \bibfield  {author} {\bibinfo {author} {\bibfnamefont {G.}~\bibnamefont {Lindblad}},\ }\bibfield  {title} {\bibinfo {title} {On the generators of quantum dynamical semigroups},\ }\href {https://doi.org/10.1007/BF01608499} {\bibfield  {journal} {\bibinfo  {journal} {Communications in Mathematical Physics}\ }\textbf {\bibinfo {volume} {48}},\ \bibinfo {pages} {119} (\bibinfo {year} {1976})}\BibitemShut {NoStop}%
\bibitem [{\citenamefont {Gorini}\ \emph {et~al.}(1976)\citenamefont {Gorini}, \citenamefont {Kossakowski},\ and\ \citenamefont {Sudarshan}}]{Gorini:1975nb}%
  \BibitemOpen
  \bibfield  {author} {\bibinfo {author} {\bibfnamefont {V.}~\bibnamefont {Gorini}}, \bibinfo {author} {\bibfnamefont {A.}~\bibnamefont {Kossakowski}},\ and\ \bibinfo {author} {\bibfnamefont {E.~C.~G.}\ \bibnamefont {Sudarshan}},\ }\bibfield  {title} {\bibinfo {title} {{Completely Positive Dynamical Semigroups of N Level Systems}},\ }\href {https://doi.org/10.1063/1.522979} {\bibfield  {journal} {\bibinfo  {journal} {J. Math. Phys.}\ }\textbf {\bibinfo {volume} {17}},\ \bibinfo {pages} {821} (\bibinfo {year} {1976})}\BibitemShut {NoStop}%
\bibitem [{\citenamefont {{X. Yao and B. M\"{u}ller}}(2018)}]{yao20171}%
  \BibitemOpen
  \bibfield  {author} {\bibinfo {author} {\bibnamefont {{X. Yao and B. M\"{u}ller}}},\ }\bibfield  {title} {\bibinfo {title} {{Approach to equilibrium of quarkonium in quark-gluon plasma}},\ }\href {https://doi.org/10.1103/PhysRevC.97.049903, 10.1103/PhysRevC.97.014908} {\bibfield  {journal} {\bibinfo  {journal} {Phys. Rev.}\ }\textbf {\bibinfo {volume} {C97}},\ \bibinfo {pages} {014908} (\bibinfo {year} {2018})},\ \bibinfo {note} {[Erratum: Phys. Rev.C97,no.4,049903(2018)]},\ \Eprint {https://arxiv.org/abs/1709.03529} {arXiv:1709.03529 [hep-ph]} \BibitemShut {NoStop}%
\bibitem [{\citenamefont {{X. Yao and B. M\"{u}ller}}(2019)}]{yao20184}%
  \BibitemOpen
  \bibfield  {author} {\bibinfo {author} {\bibnamefont {{X. Yao and B. M\"{u}ller}}},\ }\bibfield  {title} {\bibinfo {title} {{Quarkonium inside the quark-gluon plasma: Diffusion, dissociation, recombination, and energy loss}},\ }\href {https://doi.org/10.1103/PhysRevD.100.014008} {\bibfield  {journal} {\bibinfo  {journal} {Phys. Rev.}\ }\textbf {\bibinfo {volume} {D100}},\ \bibinfo {pages} {014008} (\bibinfo {year} {2019})},\ \Eprint {https://arxiv.org/abs/1811.09644} {arXiv:1811.09644 [hep-ph]} \BibitemShut {NoStop}%
\bibitem [{\citenamefont {{X. Yao and T. Mehen}}(2019)}]{yao20191}%
  \BibitemOpen
  \bibfield  {author} {\bibinfo {author} {\bibnamefont {{X. Yao and T. Mehen}}},\ }\bibfield  {title} {\bibinfo {title} {Quarkonium in-medium transport equation derived from first principles},\ }\href {https://doi.org/10.1103/PhysRevD.99.096028} {\bibfield  {journal} {\bibinfo  {journal} {Phys. Rev. D}\ }\textbf {\bibinfo {volume} {99}},\ \bibinfo {pages} {096028} (\bibinfo {year} {2019})}\BibitemShut {NoStop}%
\bibitem [{\citenamefont {Akamatsu}(2015)}]{akamatsu20151}%
  \BibitemOpen
  \bibfield  {author} {\bibinfo {author} {\bibfnamefont {Y.}~\bibnamefont {Akamatsu}},\ }\bibfield  {title} {\bibinfo {title} {Heavy quark master equations in the lindblad form at high temperatures},\ }\href@noop {} {\bibfield  {journal} {\bibinfo  {journal} {Physical Review D}\ }\textbf {\bibinfo {volume} {91}},\ \bibinfo {pages} {056002} (\bibinfo {year} {2015})}\BibitemShut {NoStop}%
\bibitem [{\citenamefont {Kajimoto}\ \emph {et~al.}(2018)\citenamefont {Kajimoto}, \citenamefont {Akamatsu}, \citenamefont {Asakawa},\ and\ \citenamefont {Rothkopf}}]{akamatsu20181}%
  \BibitemOpen
  \bibfield  {author} {\bibinfo {author} {\bibfnamefont {S.}~\bibnamefont {Kajimoto}}, \bibinfo {author} {\bibfnamefont {Y.}~\bibnamefont {Akamatsu}}, \bibinfo {author} {\bibfnamefont {M.}~\bibnamefont {Asakawa}},\ and\ \bibinfo {author} {\bibfnamefont {A.}~\bibnamefont {Rothkopf}},\ }\bibfield  {title} {\bibinfo {title} {Dynamical dissociation of quarkonia by wave function decoherence},\ }\href@noop {} {\bibfield  {journal} {\bibinfo  {journal} {Physical Review D}\ }\textbf {\bibinfo {volume} {97}},\ \bibinfo {pages} {014003} (\bibinfo {year} {2018})}\BibitemShut {NoStop}%
\bibitem [{\citenamefont {Miura}\ \emph {et~al.}(2019)\citenamefont {Miura}, \citenamefont {Akamatsu}, \citenamefont {Asakawa},\ and\ \citenamefont {Rothkopf}}]{akamatsu20191}%
  \BibitemOpen
  \bibfield  {author} {\bibinfo {author} {\bibfnamefont {T.}~\bibnamefont {Miura}}, \bibinfo {author} {\bibfnamefont {Y.}~\bibnamefont {Akamatsu}}, \bibinfo {author} {\bibfnamefont {M.}~\bibnamefont {Asakawa}},\ and\ \bibinfo {author} {\bibfnamefont {A.}~\bibnamefont {Rothkopf}},\ }\bibfield  {title} {\bibinfo {title} {{Quantum Brownian motion of a heavy quark pair in the quark-gluon plasma}},\ }\href@noop {} {\  (\bibinfo {year} {2019})},\ \Eprint {https://arxiv.org/abs/1908.06293} {arXiv:1908.06293 [nucl-th]} \BibitemShut {NoStop}%
\bibitem [{\citenamefont {Sharma}\ and\ \citenamefont {Tiwari}(2020)}]{Sharma:2019xum}%
  \BibitemOpen
  \bibfield  {author} {\bibinfo {author} {\bibfnamefont {R.}~\bibnamefont {Sharma}}\ and\ \bibinfo {author} {\bibfnamefont {A.}~\bibnamefont {Tiwari}},\ }\bibfield  {title} {\bibinfo {title} {{Quantum evolution of quarkonia with correlated and uncorrelated noise}},\ }\href {https://doi.org/10.1103/PhysRevD.101.074004} {\bibfield  {journal} {\bibinfo  {journal} {Phys. Rev. D}\ }\textbf {\bibinfo {volume} {101}},\ \bibinfo {pages} {074004} (\bibinfo {year} {2020})},\ \Eprint {https://arxiv.org/abs/1912.07036} {arXiv:1912.07036 [hep-ph]} \BibitemShut {NoStop}%
\bibitem [{\citenamefont {Blaizot}\ \emph {et~al.}(2016)\citenamefont {Blaizot}, \citenamefont {De~Boni}, \citenamefont {Faccioli},\ and\ \citenamefont {Garberoglio}}]{blaizot20151}%
  \BibitemOpen
  \bibfield  {author} {\bibinfo {author} {\bibfnamefont {J.-P.}\ \bibnamefont {Blaizot}}, \bibinfo {author} {\bibfnamefont {D.}~\bibnamefont {De~Boni}}, \bibinfo {author} {\bibfnamefont {P.}~\bibnamefont {Faccioli}},\ and\ \bibinfo {author} {\bibfnamefont {G.}~\bibnamefont {Garberoglio}},\ }\bibfield  {title} {\bibinfo {title} {{Heavy quark bound states in a quark–gluon plasma: Dissociation and recombination}},\ }\href {https://doi.org/10.1016/j.nuclphysa.2015.10.011} {\bibfield  {journal} {\bibinfo  {journal} {Nucl. Phys.}\ }\textbf {\bibinfo {volume} {A946}},\ \bibinfo {pages} {49} (\bibinfo {year} {2016})},\ \Eprint {https://arxiv.org/abs/1503.03857} {arXiv:1503.03857 [nucl-th]} \BibitemShut {NoStop}%
\bibitem [{\citenamefont {{J. P. Blaizot and M. A. Escobedo}}(2018{\natexlab{a}})}]{blaizot20171}%
  \BibitemOpen
  \bibfield  {author} {\bibinfo {author} {\bibnamefont {{J. P. Blaizot and M. A. Escobedo}}},\ }\bibfield  {title} {\bibinfo {title} {{Quantum and classical dynamics of heavy quarks in a quark-gluon plasma}},\ }\href {https://doi.org/10.1007/JHEP06(2018)034} {\bibfield  {journal} {\bibinfo  {journal} {JHEP}\ }\textbf {\bibinfo {volume} {06}},\ \bibinfo {pages} {034}},\ \Eprint {https://arxiv.org/abs/1711.10812} {arXiv:1711.10812 [hep-ph]} \BibitemShut {NoStop}%
\bibitem [{\citenamefont {De~Boni}(2017)}]{DeBoni:2017ocl}%
  \BibitemOpen
  \bibfield  {author} {\bibinfo {author} {\bibfnamefont {D.}~\bibnamefont {De~Boni}},\ }\bibfield  {title} {\bibinfo {title} {{Fate of in-medium heavy quarks via a Lindblad equation}},\ }\href {https://doi.org/10.1007/JHEP08(2017)064} {\bibfield  {journal} {\bibinfo  {journal} {JHEP}\ }\textbf {\bibinfo {volume} {08}},\ \bibinfo {pages} {064}},\ \Eprint {https://arxiv.org/abs/1705.03567} {arXiv:1705.03567 [hep-ph]} \BibitemShut {NoStop}%
\bibitem [{\citenamefont {{J. P. Blaizot and M. A. Escobedo}}(2018{\natexlab{b}})}]{blaizot20181}%
  \BibitemOpen
  \bibfield  {author} {\bibinfo {author} {\bibnamefont {{J. P. Blaizot and M. A. Escobedo}}},\ }\bibfield  {title} {\bibinfo {title} {{Approach to equilibrium of a quarkonium in a quark-gluon plasma}},\ }\href {https://doi.org/10.1103/PhysRevD.98.074007} {\bibfield  {journal} {\bibinfo  {journal} {Phys. Rev.}\ }\textbf {\bibinfo {volume} {D98}},\ \bibinfo {pages} {074007} (\bibinfo {year} {2018}{\natexlab{b}})},\ \Eprint {https://arxiv.org/abs/1803.07996} {arXiv:1803.07996 [hep-ph]} \BibitemShut {NoStop}%
\bibitem [{\citenamefont {Katz}\ and\ \citenamefont {Gossiaux}(2014)}]{Katz:2013rpa}%
  \BibitemOpen
  \bibfield  {author} {\bibinfo {author} {\bibfnamefont {R.}~\bibnamefont {Katz}}\ and\ \bibinfo {author} {\bibfnamefont {P.~B.}\ \bibnamefont {Gossiaux}},\ }\bibfield  {title} {\bibinfo {title} {{Semi-classical approach to $J/\psi$ suppression in high energy heavy-ion collisions}},\ }\href {https://doi.org/10.1088/1742-6596/509/1/012095} {\bibfield  {journal} {\bibinfo  {journal} {J. Phys. Conf. Ser.}\ }\textbf {\bibinfo {volume} {509}},\ \bibinfo {pages} {012095} (\bibinfo {year} {2014})},\ \Eprint {https://arxiv.org/abs/1312.0881} {arXiv:1312.0881 [hep-ph]} \BibitemShut {NoStop}%
\bibitem [{\citenamefont {{P. B. Gossiaux and R. Katz}}(2016)}]{Gossiaux:2016htk}%
  \BibitemOpen
  \bibfield  {author} {\bibinfo {author} {\bibnamefont {{P. B. Gossiaux and R. Katz}}},\ }\bibfield  {title} {\bibinfo {title} {{Upsilon suppression in the Schrödinger–Langevin approach}},\ }\bibfield  {booktitle} {\emph {\bibinfo {booktitle} {{Proceedings, 25th International Conference on Ultra-Relativistic Nucleus-Nucleus Collisions (Quark Matter 2015): Kobe, Japan, September 27-October 3, 2015}}},\ }\href {https://doi.org/10.1016/j.nuclphysa.2016.04.017} {\bibfield  {journal} {\bibinfo  {journal} {Nucl. Phys.}\ }\textbf {\bibinfo {volume} {A956}},\ \bibinfo {pages} {737} (\bibinfo {year} {2016})},\ \Eprint {https://arxiv.org/abs/1601.01443} {arXiv:1601.01443 [hep-ph]} \BibitemShut {NoStop}%
\bibitem [{\citenamefont {Delorme}\ \emph {et~al.}(2024)\citenamefont {Delorme}, \citenamefont {Katz}, \citenamefont {Gousset}, \citenamefont {Gossiaux},\ and\ \citenamefont {Blaizot}}]{Delorme:2024rdo}%
  \BibitemOpen
  \bibfield  {author} {\bibinfo {author} {\bibfnamefont {S.}~\bibnamefont {Delorme}}, \bibinfo {author} {\bibfnamefont {R.}~\bibnamefont {Katz}}, \bibinfo {author} {\bibfnamefont {T.}~\bibnamefont {Gousset}}, \bibinfo {author} {\bibfnamefont {P.~B.}\ \bibnamefont {Gossiaux}},\ and\ \bibinfo {author} {\bibfnamefont {J.-P.}\ \bibnamefont {Blaizot}},\ }\bibfield  {title} {\bibinfo {title} {{Quarkonium dynamics in the quantum Brownian regime with non-abelian quantum master equations}},\ }\href {https://doi.org/10.1007/JHEP06(2024)060} {\bibfield  {journal} {\bibinfo  {journal} {JHEP}\ }\textbf {\bibinfo {volume} {06}},\ \bibinfo {pages} {060}},\ \Eprint {https://arxiv.org/abs/2402.04488} {arXiv:2402.04488 [hep-ph]} \BibitemShut {NoStop}%
\bibitem [{\citenamefont {Brambilla}\ \emph {et~al.}(2017{\natexlab{a}})\citenamefont {Brambilla}, \citenamefont {Escobedo}, \citenamefont {Soto},\ and\ \citenamefont {Vairo}}]{Brambilla:2016wgg}%
  \BibitemOpen
  \bibfield  {author} {\bibinfo {author} {\bibfnamefont {N.}~\bibnamefont {Brambilla}}, \bibinfo {author} {\bibfnamefont {M.~A.}\ \bibnamefont {Escobedo}}, \bibinfo {author} {\bibfnamefont {J.}~\bibnamefont {Soto}},\ and\ \bibinfo {author} {\bibfnamefont {A.}~\bibnamefont {Vairo}},\ }\bibfield  {title} {\bibinfo {title} {{Quarkonium suppression in heavy-ion collisions: an open quantum system approach}},\ }\href {https://doi.org/10.1103/PhysRevD.96.034021} {\bibfield  {journal} {\bibinfo  {journal} {Phys. Rev. D}\ }\textbf {\bibinfo {volume} {96}},\ \bibinfo {pages} {034021} (\bibinfo {year} {2017}{\natexlab{a}})},\ \Eprint {https://arxiv.org/abs/1612.07248} {arXiv:1612.07248 [hep-ph]} \BibitemShut {NoStop}%
\bibitem [{\citenamefont {Brambilla}\ \emph {et~al.}(2017{\natexlab{b}})\citenamefont {Brambilla}, \citenamefont {Escobedo}, \citenamefont {Soto},\ and\ \citenamefont {Vairo}}]{brambilla20171}%
  \BibitemOpen
  \bibfield  {author} {\bibinfo {author} {\bibfnamefont {N.}~\bibnamefont {Brambilla}}, \bibinfo {author} {\bibfnamefont {M.~A.}\ \bibnamefont {Escobedo}}, \bibinfo {author} {\bibfnamefont {J.}~\bibnamefont {Soto}},\ and\ \bibinfo {author} {\bibfnamefont {A.}~\bibnamefont {Vairo}},\ }\bibfield  {title} {\bibinfo {title} {Quarkonium suppression in heavy-ion collisions: An open quantum system approach},\ }\href {https://doi.org/10.1103/PhysRevD.96.034021} {\bibfield  {journal} {\bibinfo  {journal} {Phys. Rev. D}\ }\textbf {\bibinfo {volume} {96}},\ \bibinfo {pages} {034021} (\bibinfo {year} {2017}{\natexlab{b}})}\BibitemShut {NoStop}%
\bibitem [{\citenamefont {Brambilla}\ \emph {et~al.}(2021{\natexlab{a}})\citenamefont {Brambilla}, \citenamefont {Escobedo}, \citenamefont {Strickland}, \citenamefont {Vairo}, \citenamefont {Vander~Griend},\ and\ \citenamefont {Weber}}]{Brambilla:2020qwo}%
  \BibitemOpen
  \bibfield  {author} {\bibinfo {author} {\bibfnamefont {N.}~\bibnamefont {Brambilla}}, \bibinfo {author} {\bibfnamefont {M.~A.}\ \bibnamefont {Escobedo}}, \bibinfo {author} {\bibfnamefont {M.}~\bibnamefont {Strickland}}, \bibinfo {author} {\bibfnamefont {A.}~\bibnamefont {Vairo}}, \bibinfo {author} {\bibfnamefont {P.}~\bibnamefont {Vander~Griend}},\ and\ \bibinfo {author} {\bibfnamefont {J.~H.}\ \bibnamefont {Weber}},\ }\bibfield  {title} {\bibinfo {title} {{Bottomonium suppression in an open quantum system using the quantum trajectories method}},\ }\href {https://doi.org/10.1007/JHEP05(2021)136} {\bibfield  {journal} {\bibinfo  {journal} {JHEP}\ }\textbf {\bibinfo {volume} {05}},\ \bibinfo {pages} {136}},\ \Eprint {https://arxiv.org/abs/2012.01240} {arXiv:2012.01240 [hep-ph]} \BibitemShut {NoStop}%
\bibitem [{\citenamefont {Brambilla}\ \emph {et~al.}(2021{\natexlab{b}})\citenamefont {Brambilla}, \citenamefont {Escobedo}, \citenamefont {Strickland}, \citenamefont {Vairo}, \citenamefont {Vander~Griend},\ and\ \citenamefont {Weber}}]{Brambilla:2021wkt}%
  \BibitemOpen
  \bibfield  {author} {\bibinfo {author} {\bibfnamefont {N.}~\bibnamefont {Brambilla}}, \bibinfo {author} {\bibfnamefont {M.~A.}\ \bibnamefont {Escobedo}}, \bibinfo {author} {\bibfnamefont {M.}~\bibnamefont {Strickland}}, \bibinfo {author} {\bibfnamefont {A.}~\bibnamefont {Vairo}}, \bibinfo {author} {\bibfnamefont {P.}~\bibnamefont {Vander~Griend}},\ and\ \bibinfo {author} {\bibfnamefont {J.~H.}\ \bibnamefont {Weber}},\ }\bibfield  {title} {\bibinfo {title} {{Bottomonium production in heavy-ion collisions using quantum trajectories: Differential observables and momentum anisotropy}},\ }\href {https://doi.org/10.1103/PhysRevD.104.094049} {\bibfield  {journal} {\bibinfo  {journal} {Phys. Rev. D}\ }\textbf {\bibinfo {volume} {104}},\ \bibinfo {pages} {094049} (\bibinfo {year} {2021}{\natexlab{b}})},\ \Eprint {https://arxiv.org/abs/2107.06222} {arXiv:2107.06222 [hep-ph]} \BibitemShut {NoStop}%
\bibitem [{\citenamefont {Brambilla}\ \emph {et~al.}(2022{\natexlab{a}})\citenamefont {Brambilla}, \citenamefont {Escobedo}, \citenamefont {Islam}, \citenamefont {Strickland}, \citenamefont {Tiwari}, \citenamefont {Vairo},\ and\ \citenamefont {Vander~Griend}}]{Brambilla:2022ynh}%
  \BibitemOpen
  \bibfield  {author} {\bibinfo {author} {\bibfnamefont {N.}~\bibnamefont {Brambilla}}, \bibinfo {author} {\bibfnamefont {M.~A.}\ \bibnamefont {Escobedo}}, \bibinfo {author} {\bibfnamefont {A.}~\bibnamefont {Islam}}, \bibinfo {author} {\bibfnamefont {M.}~\bibnamefont {Strickland}}, \bibinfo {author} {\bibfnamefont {A.}~\bibnamefont {Tiwari}}, \bibinfo {author} {\bibfnamefont {A.}~\bibnamefont {Vairo}},\ and\ \bibinfo {author} {\bibfnamefont {P.}~\bibnamefont {Vander~Griend}},\ }\bibfield  {title} {\bibinfo {title} {{Heavy quarkonium dynamics at next-to-leading order in the binding energy over temperature}},\ }\href {https://doi.org/10.1007/JHEP08(2022)303} {\bibfield  {journal} {\bibinfo  {journal} {JHEP}\ }\textbf {\bibinfo {volume} {08}},\ \bibinfo {pages} {303}},\ \Eprint {https://arxiv.org/abs/2205.10289} {arXiv:2205.10289 [hep-ph]} \BibitemShut {NoStop}%
\bibitem [{\citenamefont {Redfield}(1957)}]{Redfield:1957}%
  \BibitemOpen
  \bibfield  {author} {\bibinfo {author} {\bibfnamefont {A.~G.}\ \bibnamefont {Redfield}},\ }\bibfield  {title} {\bibinfo {title} {On the theory of relaxation processes},\ }\href {https://doi.org/10.1147/rd.11.0019} {\bibfield  {journal} {\bibinfo  {journal} {IBM Journal of Research and Development}\ }\textbf {\bibinfo {volume} {1}},\ \bibinfo {pages} {19} (\bibinfo {year} {1957})}\BibitemShut {NoStop}%
\bibitem [{\citenamefont {Blum}(1981)}]{Blum1981DensityMT}%
  \BibitemOpen
  \bibfield  {author} {\bibinfo {author} {\bibfnamefont {K.}~\bibnamefont {Blum}},\ }\bibfield  {title} {\bibinfo {title} {Density matrix theory and applications}\ }(\bibinfo {year} {1981})\BibitemShut {NoStop}%
\bibitem [{\citenamefont {Breuer}\ \emph {et~al.}(1999)\citenamefont {Breuer}, \citenamefont {Kappler},\ and\ \citenamefont {Petruccione}}]{PhysRevA.59.1633}%
  \BibitemOpen
  \bibfield  {author} {\bibinfo {author} {\bibfnamefont {H.-P.}\ \bibnamefont {Breuer}}, \bibinfo {author} {\bibfnamefont {B.}~\bibnamefont {Kappler}},\ and\ \bibinfo {author} {\bibfnamefont {F.}~\bibnamefont {Petruccione}},\ }\bibfield  {title} {\bibinfo {title} {Stochastic wave-function method for non-markovian quantum master equations},\ }\href {https://doi.org/10.1103/PhysRevA.59.1633} {\bibfield  {journal} {\bibinfo  {journal} {Phys. Rev. A}\ }\textbf {\bibinfo {volume} {59}},\ \bibinfo {pages} {1633} (\bibinfo {year} {1999})}\BibitemShut {NoStop}%
\bibitem [{\citenamefont {Blaizot}\ and\ \citenamefont {Escobedo}(2021)}]{Blaizot:2021xqa}%
  \BibitemOpen
  \bibfield  {author} {\bibinfo {author} {\bibfnamefont {J.-P.}\ \bibnamefont {Blaizot}}\ and\ \bibinfo {author} {\bibfnamefont {M.~A.}\ \bibnamefont {Escobedo}},\ }\bibfield  {title} {\bibinfo {title} {{Phenomenological study of quarkonium suppression and the impact of the energy gap between singlets and octets}},\ }\href {https://doi.org/10.1103/PhysRevD.104.054034} {\bibfield  {journal} {\bibinfo  {journal} {Phys. Rev. D}\ }\textbf {\bibinfo {volume} {104}},\ \bibinfo {pages} {054034} (\bibinfo {year} {2021})},\ \Eprint {https://arxiv.org/abs/2106.15371} {arXiv:2106.15371 [hep-ph]} \BibitemShut {NoStop}%
\bibitem [{\citenamefont {et~al.}(2011)}]{chatrchyan20111}%
  \BibitemOpen
  \bibfield  {author} {\bibinfo {author} {\bibfnamefont {C.}~\bibnamefont {et~al.}} (\bibinfo {collaboration} {CMS Collaboration}),\ }\bibfield  {title} {\bibinfo {title} {{Indications of Suppression of Excited $\ensuremath{\Upsilon}$ States in Pb-Pb Collisions at $\sqrt{{s}_{\mathrm{NN}}}=2.76\text{ }\text{ }\mathrm{TeV}$}},\ }\href {https://doi.org/10.1103/PhysRevLett.107.052302} {\bibfield  {journal} {\bibinfo  {journal} {Phys. Rev. Lett.}\ }\textbf {\bibinfo {volume} {107}},\ \bibinfo {pages} {052302} (\bibinfo {year} {2011})}\BibitemShut {NoStop}%
\bibitem [{\citenamefont {Chatrchyan}\ \emph {et~al.}(2012)\citenamefont {Chatrchyan} \emph {et~al.}}]{CMS:2012gvv}%
  \BibitemOpen
  \bibfield  {author} {\bibinfo {author} {\bibfnamefont {S.}~\bibnamefont {Chatrchyan}} \emph {et~al.} (\bibinfo {collaboration} {CMS}),\ }\bibfield  {title} {\bibinfo {title} {{Observation of Sequential Upsilon Suppression in PbPb Collisions}},\ }\href {https://doi.org/10.1103/PhysRevLett.109.222301} {\bibfield  {journal} {\bibinfo  {journal} {Phys. Rev. Lett.}\ }\textbf {\bibinfo {volume} {109}},\ \bibinfo {pages} {222301} (\bibinfo {year} {2012})},\ \bibinfo {note} {[Erratum: Phys.Rev.Lett. 120, 199903 (2018)]},\ \Eprint {https://arxiv.org/abs/1208.2826} {arXiv:1208.2826 [nucl-ex]} \BibitemShut {NoStop}%
\bibitem [{\citenamefont {Khachatryan}\ \emph {et~al.}(2017)\citenamefont {Khachatryan} \emph {et~al.}}]{CMS:2016rpc}%
  \BibitemOpen
  \bibfield  {author} {\bibinfo {author} {\bibfnamefont {V.}~\bibnamefont {Khachatryan}} \emph {et~al.} (\bibinfo {collaboration} {CMS}),\ }\bibfield  {title} {\bibinfo {title} {{Suppression of $\Upsilon(1S), \Upsilon(2S)$ and $\Upsilon(3S)$ production in PbPb collisions at $\sqrt{s_{\rm NN}}$ = 2.76 TeV}},\ }\href {https://doi.org/10.1016/j.physletb.2017.04.031} {\bibfield  {journal} {\bibinfo  {journal} {Phys. Lett. B}\ }\textbf {\bibinfo {volume} {770}},\ \bibinfo {pages} {357} (\bibinfo {year} {2017})},\ \Eprint {https://arxiv.org/abs/1611.01510} {arXiv:1611.01510 [nucl-ex]} \BibitemShut {NoStop}%
\bibitem [{ATL(2019)}]{ATLAS:2019can}%
  \BibitemOpen
  \bibfield  {title} {\bibinfo {title} {{Production of $\varUpsilon(\mathrm{nS})$ mesons in Pb+Pb and $\it{pp}$ collisions at 5.02 TeV with ATLAS}},\ }\href@noop {} {\  (\bibinfo {year} {2019})}\BibitemShut {NoStop}%
\bibitem [{\citenamefont {Aad}\ \emph {et~al.}(2023)\citenamefont {Aad} \emph {et~al.}}]{ATLAS:2022exb}%
  \BibitemOpen
  \bibfield  {author} {\bibinfo {author} {\bibfnamefont {G.}~\bibnamefont {Aad}} \emph {et~al.} (\bibinfo {collaboration} {ATLAS}),\ }\bibfield  {title} {\bibinfo {title} {{Production of \ensuremath{\Upsilon}(nS) mesons in Pb+Pb and pp collisions at 5.02 TeV}},\ }\href {https://doi.org/10.1103/PhysRevC.107.054912} {\bibfield  {journal} {\bibinfo  {journal} {Phys. Rev. C}\ }\textbf {\bibinfo {volume} {107}},\ \bibinfo {pages} {054912} (\bibinfo {year} {2023})},\ \Eprint {https://arxiv.org/abs/2205.03042} {arXiv:2205.03042 [nucl-ex]} \BibitemShut {NoStop}%
\bibitem [{CMS(2022)}]{CMS:2022rna}%
  \BibitemOpen
  \bibfield  {title} {\bibinfo {title} {{Observation of the $\Upsilon\textrm{(3S)}$ meson and sequential suppression of $\Upsilon$ states in PbPb collisions at $\sqrt{\mathrm{s_{NN}}}=5.02~\mathrm{TeV}$}},\ }\href@noop {} {\  (\bibinfo {year} {2022})}\BibitemShut {NoStop}%
\bibitem [{\citenamefont {Tumasyan}\ \emph {et~al.}(2024)\citenamefont {Tumasyan} \emph {et~al.}}]{CMS:2023lfu}%
  \BibitemOpen
  \bibfield  {author} {\bibinfo {author} {\bibfnamefont {A.}~\bibnamefont {Tumasyan}} \emph {et~al.} (\bibinfo {collaboration} {CMS}),\ }\bibfield  {title} {\bibinfo {title} {{Observation of the $\Upsilon{}(3S)$ Meson and Suppression of $\Upsilon{}$ States in Pb-Pb Collisions at $\sqrt{\mathrm{s_{NN}}}=5.02$~TeV}},\ }\href {https://doi.org/10.1103/PhysRevLett.133.022302} {\bibfield  {journal} {\bibinfo  {journal} {Phys. Rev. Lett.}\ }\textbf {\bibinfo {volume} {133}},\ \bibinfo {pages} {022302} (\bibinfo {year} {2024})},\ \Eprint {https://arxiv.org/abs/2303.17026} {arXiv:2303.17026 [hep-ex]} \BibitemShut {NoStop}%
\bibitem [{\citenamefont {Strickland}\ and\ \citenamefont {Thapa}(2023)}]{Strickland:2023nfm}%
  \BibitemOpen
  \bibfield  {author} {\bibinfo {author} {\bibfnamefont {M.}~\bibnamefont {Strickland}}\ and\ \bibinfo {author} {\bibfnamefont {S.}~\bibnamefont {Thapa}},\ }\bibfield  {title} {\bibinfo {title} {{Bottomonium suppression at RHIC and LHC in an open quantum system approach}},\ }\href {https://doi.org/10.1103/PhysRevD.108.014031} {\bibfield  {journal} {\bibinfo  {journal} {Phys. Rev. D}\ }\textbf {\bibinfo {volume} {108}},\ \bibinfo {pages} {014031} (\bibinfo {year} {2023})},\ \Eprint {https://arxiv.org/abs/2305.17841} {arXiv:2305.17841 [hep-ph]} \BibitemShut {NoStop}%
\bibitem [{\citenamefont {Aboona}\ \emph {et~al.}(2023)\citenamefont {Aboona} \emph {et~al.}}]{STAR:2022rpk}%
  \BibitemOpen
  \bibfield  {author} {\bibinfo {author} {\bibfnamefont {B.}~\bibnamefont {Aboona}} \emph {et~al.} (\bibinfo {collaboration} {STAR}),\ }\bibfield  {title} {\bibinfo {title} {{Observation of sequential $\Upsilon$ suppression in Au+Au collisions at $\sqrt{s_{_\mathrm{NN}}}$ = 200 GeV with the STAR experiment}},\ }\href {https://doi.org/10.1103/PhysRevLett.130.112301} {\bibfield  {journal} {\bibinfo  {journal} {Phys. Rev. Lett.}\ }\textbf {\bibinfo {volume} {130}},\ \bibinfo {pages} {112301} (\bibinfo {year} {2023})},\ \Eprint {https://arxiv.org/abs/2207.06568} {arXiv:2207.06568 [nucl-ex]} \BibitemShut {NoStop}%
\bibitem [{\citenamefont {Aaij}\ \emph {et~al.}(2018)\citenamefont {Aaij} \emph {et~al.}}]{LHCb:2018psc}%
  \BibitemOpen
  \bibfield  {author} {\bibinfo {author} {\bibfnamefont {R.}~\bibnamefont {Aaij}} \emph {et~al.} (\bibinfo {collaboration} {LHCb}),\ }\bibfield  {title} {\bibinfo {title} {{Study of $\Upsilon$ production in $p$Pb collisions at $\sqrt{s_{NN}}=8.16$ TeV}},\ }\href {https://doi.org/10.1007/JHEP11(2018)194} {\bibfield  {journal} {\bibinfo  {journal} {JHEP}\ }\textbf {\bibinfo {volume} {11}},\ \bibinfo {pages} {194}},\ \bibinfo {note} {[Erratum: JHEP 02, 093 (2020)]},\ \Eprint {https://arxiv.org/abs/1810.07655} {arXiv:1810.07655 [hep-ex]} \BibitemShut {NoStop}%
\bibitem [{\citenamefont {Aaboud}\ \emph {et~al.}(2018)\citenamefont {Aaboud} \emph {et~al.}}]{ATLAS:2017prf}%
  \BibitemOpen
  \bibfield  {author} {\bibinfo {author} {\bibfnamefont {M.}~\bibnamefont {Aaboud}} \emph {et~al.} (\bibinfo {collaboration} {ATLAS}),\ }\bibfield  {title} {\bibinfo {title} {{Measurement of quarkonium production in proton\textendash{}lead and proton\textendash{}proton collisions at $5.02~\mathrm {TeV}$ with the ATLAS detector}},\ }\href {https://doi.org/10.1140/epjc/s10052-018-5624-4} {\bibfield  {journal} {\bibinfo  {journal} {Eur. Phys. J. C}\ }\textbf {\bibinfo {volume} {78}},\ \bibinfo {pages} {171} (\bibinfo {year} {2018})},\ \Eprint {https://arxiv.org/abs/1709.03089} {arXiv:1709.03089 [nucl-ex]} \BibitemShut {NoStop}%
\bibitem [{\citenamefont {Acharya}\ \emph {et~al.}(2020)\citenamefont {Acharya} \emph {et~al.}}]{ALICE:2019qie}%
  \BibitemOpen
  \bibfield  {author} {\bibinfo {author} {\bibfnamefont {S.}~\bibnamefont {Acharya}} \emph {et~al.} (\bibinfo {collaboration} {ALICE}),\ }\bibfield  {title} {\bibinfo {title} {{$\Upsilon$ production in p\textendash{}Pb collisions at $\sqrt{s_{NN}}$=8.16 TeV}},\ }\href {https://doi.org/10.1016/j.physletb.2020.135486} {\bibfield  {journal} {\bibinfo  {journal} {Phys. Lett. B}\ }\textbf {\bibinfo {volume} {806}},\ \bibinfo {pages} {135486} (\bibinfo {year} {2020})},\ \Eprint {https://arxiv.org/abs/1910.14405} {arXiv:1910.14405 [nucl-ex]} \BibitemShut {NoStop}%
\bibitem [{\citenamefont {Tumasyan}\ \emph {et~al.}(2022)\citenamefont {Tumasyan} \emph {et~al.}}]{CMS:2022wfi}%
  \BibitemOpen
  \bibfield  {author} {\bibinfo {author} {\bibfnamefont {A.}~\bibnamefont {Tumasyan}} \emph {et~al.} (\bibinfo {collaboration} {CMS}),\ }\bibfield  {title} {\bibinfo {title} {{Nuclear modification of $\Upsilon$ states in pPb collisions at $\sqrt{s_\mathrm{NN}}$ = 5.02 TeV}},\ }\href {https://doi.org/10.1016/j.physletb.2022.137397} {\bibfield  {journal} {\bibinfo  {journal} {Phys. Lett. B}\ }\textbf {\bibinfo {volume} {835}},\ \bibinfo {pages} {137397} (\bibinfo {year} {2022})},\ \Eprint {https://arxiv.org/abs/2202.11807} {arXiv:2202.11807 [hep-ex]} \BibitemShut {NoStop}%
\bibitem [{\citenamefont {Strickland}\ \emph {et~al.}(2024)\citenamefont {Strickland}, \citenamefont {Thapa},\ and\ \citenamefont {Vogt}}]{Strickland:2024oat}%
  \BibitemOpen
  \bibfield  {author} {\bibinfo {author} {\bibfnamefont {M.}~\bibnamefont {Strickland}}, \bibinfo {author} {\bibfnamefont {S.}~\bibnamefont {Thapa}},\ and\ \bibinfo {author} {\bibfnamefont {R.}~\bibnamefont {Vogt}},\ }\bibfield  {title} {\bibinfo {title} {{Bottomonium suppression in 5.02 and 8.16~TeV p-Pb collisions}},\ }\href {https://doi.org/10.1103/PhysRevD.109.096016} {\bibfield  {journal} {\bibinfo  {journal} {Phys. Rev. D}\ }\textbf {\bibinfo {volume} {109}},\ \bibinfo {pages} {096016} (\bibinfo {year} {2024})},\ \Eprint {https://arxiv.org/abs/2401.16704} {arXiv:2401.16704 [nucl-th]} \BibitemShut {NoStop}%
\bibitem [{Note2()}]{Note2}%
  \BibitemOpen
  \bibinfo {note} {The master equations are derived for a medium in global equilibrium. For phenomenology in an expanding medium, local thermodynamic variables are used to describe ${\protect \rm {E}}$ in the neighbourhood of ${\protect \rm {S}}$.}\BibitemShut {Stop}%
\bibitem [{\citenamefont {Yao}\ and\ \citenamefont {Mehen}(2021)}]{Yao:2020eqy}%
  \BibitemOpen
  \bibfield  {author} {\bibinfo {author} {\bibfnamefont {X.}~\bibnamefont {Yao}}\ and\ \bibinfo {author} {\bibfnamefont {T.}~\bibnamefont {Mehen}},\ }\bibfield  {title} {\bibinfo {title} {{Quarkonium Semiclassical Transport in Quark-Gluon Plasma: Factorization and Quantum Correction}},\ }\href {https://doi.org/10.1007/JHEP02(2021)062} {\bibfield  {journal} {\bibinfo  {journal} {JHEP}\ }\textbf {\bibinfo {volume} {02}},\ \bibinfo {pages} {062}},\ \Eprint {https://arxiv.org/abs/2009.02408} {arXiv:2009.02408 [hep-ph]} \BibitemShut {NoStop}%
\bibitem [{\citenamefont {Banerjee}\ \emph {et~al.}(2022{\natexlab{a}})\citenamefont {Banerjee}, \citenamefont {Gavai}, \citenamefont {Datta},\ and\ \citenamefont {Majumdar}}]{Banerjee:2022gen}%
  \BibitemOpen
  \bibfield  {author} {\bibinfo {author} {\bibfnamefont {D.}~\bibnamefont {Banerjee}}, \bibinfo {author} {\bibfnamefont {R.}~\bibnamefont {Gavai}}, \bibinfo {author} {\bibfnamefont {S.}~\bibnamefont {Datta}},\ and\ \bibinfo {author} {\bibfnamefont {P.}~\bibnamefont {Majumdar}},\ }\bibfield  {title} {\bibinfo {title} {{Temperature dependence of the static quark diffusion coefficient}},\ }\href@noop {} {\  (\bibinfo {year} {2022}{\natexlab{a}})},\ \Eprint {https://arxiv.org/abs/2206.15471} {arXiv:2206.15471 [hep-ph]} \BibitemShut {NoStop}%
\bibitem [{\citenamefont {Messiah}(1979)}]{Messiah:1979eg}%
  \BibitemOpen
  \bibfield  {author} {\bibinfo {author} {\bibfnamefont {A.}~\bibnamefont {Messiah}},\ }\href@noop {} {\emph {\bibinfo {title} {{QUANTUM MECHANICS. VOL. 2 (GERMAN TRANSLATION)}}}}\ (\bibinfo {year} {1979})\BibitemShut {NoStop}%
\bibitem [{\citenamefont {Alberico}\ \emph {et~al.}(2013)\citenamefont {Alberico}, \citenamefont {Beraudo}, \citenamefont {De~Pace}, \citenamefont {Molinari}, \citenamefont {Monteno}, \citenamefont {Nardi}, \citenamefont {Prino},\ and\ \citenamefont {Sitta}}]{alberico20131}%
  \BibitemOpen
  \bibfield  {author} {\bibinfo {author} {\bibfnamefont {W.~M.}\ \bibnamefont {Alberico}}, \bibinfo {author} {\bibfnamefont {A.}~\bibnamefont {Beraudo}}, \bibinfo {author} {\bibfnamefont {A.}~\bibnamefont {De~Pace}}, \bibinfo {author} {\bibfnamefont {A.}~\bibnamefont {Molinari}}, \bibinfo {author} {\bibfnamefont {M.}~\bibnamefont {Monteno}}, \bibinfo {author} {\bibfnamefont {M.}~\bibnamefont {Nardi}}, \bibinfo {author} {\bibfnamefont {F.}~\bibnamefont {Prino}},\ and\ \bibinfo {author} {\bibfnamefont {M.}~\bibnamefont {Sitta}},\ }\bibfield  {title} {\bibinfo {title} {Heavy flavors in aa collisions: production, transport and final spectra},\ }\href {https://doi.org/10.1140/epjc/s10052-013-2481-z} {\bibfield  {journal} {\bibinfo  {journal} {The European Physical Journal C}\ }\textbf {\bibinfo {volume} {73}},\ \bibinfo {pages} {2481} (\bibinfo {year} {2013})}\BibitemShut {NoStop}%
\bibitem [{\citenamefont {Sirunyan}\ \emph {et~al.}(2019)\citenamefont {Sirunyan}, \citenamefont {Tumasyan} \emph {et~al.}}]{CMS:2019270}%
  \BibitemOpen
  \bibfield  {author} {\bibinfo {author} {\bibfnamefont {A.}~\bibnamefont {Sirunyan}}, \bibinfo {author} {\bibfnamefont {A.}~\bibnamefont {Tumasyan}}, \emph {et~al.} (\bibinfo {collaboration} {CMS}),\ }\bibfield  {title} {\bibinfo {title} {{Measurement of nuclear modification factors of $\Upsilon(1S)$, $\Upsilon(2S)$ and $\Upsilon(3S)$ mesons in PbPb collisions at $\sqrt{{s}_{\mathrm{NN}}}=5.02$~TeV}},\ }\href {https://doi.org/https://doi.org/10.1016/j.physletb.2019.01.006} {\bibfield  {journal} {\bibinfo  {journal} {Physics Letters B}\ }\textbf {\bibinfo {volume} {790}},\ \bibinfo {pages} {270} (\bibinfo {year} {2019})}\BibitemShut {NoStop}%
\bibitem [{\citenamefont {Miller}\ \emph {et~al.}(2007)\citenamefont {Miller}, \citenamefont {Reygers}, \citenamefont {Sanders},\ and\ \citenamefont {Steinberg}}]{Miller:2007ri}%
  \BibitemOpen
  \bibfield  {author} {\bibinfo {author} {\bibfnamefont {M.~L.}\ \bibnamefont {Miller}}, \bibinfo {author} {\bibfnamefont {K.}~\bibnamefont {Reygers}}, \bibinfo {author} {\bibfnamefont {S.~J.}\ \bibnamefont {Sanders}},\ and\ \bibinfo {author} {\bibfnamefont {P.}~\bibnamefont {Steinberg}},\ }\bibfield  {title} {\bibinfo {title} {{Glauber modeling in high energy nuclear collisions}},\ }\href {https://doi.org/10.1146/annurev.nucl.57.090506.123020} {\bibfield  {journal} {\bibinfo  {journal} {Ann. Rev. Nucl. Part. Sci.}\ }\textbf {\bibinfo {volume} {57}},\ \bibinfo {pages} {205} (\bibinfo {year} {2007})},\ \Eprint {https://arxiv.org/abs/nucl-ex/0701025} {arXiv:nucl-ex/0701025} \BibitemShut {NoStop}%
\bibitem [{\citenamefont {Lin}\ \emph {et~al.}(2005)\citenamefont {Lin}, \citenamefont {Ko}, \citenamefont {Li}, \citenamefont {Zhang},\ and\ \citenamefont {Pal}}]{PhysRevC.72.064901}%
  \BibitemOpen
  \bibfield  {author} {\bibinfo {author} {\bibfnamefont {Z.-W.}\ \bibnamefont {Lin}}, \bibinfo {author} {\bibfnamefont {C.~M.}\ \bibnamefont {Ko}}, \bibinfo {author} {\bibfnamefont {B.-A.}\ \bibnamefont {Li}}, \bibinfo {author} {\bibfnamefont {B.}~\bibnamefont {Zhang}},\ and\ \bibinfo {author} {\bibfnamefont {S.}~\bibnamefont {Pal}},\ }\bibfield  {title} {\bibinfo {title} {Multiphase transport model for relativistic heavy ion collisions},\ }\href {https://doi.org/10.1103/PhysRevC.72.064901} {\bibfield  {journal} {\bibinfo  {journal} {Phys. Rev. C}\ }\textbf {\bibinfo {volume} {72}},\ \bibinfo {pages} {064901} (\bibinfo {year} {2005})}\BibitemShut {NoStop}%
\bibitem [{\citenamefont {Kim}\ \emph {et~al.}(2025)\citenamefont {Kim}, \citenamefont {Park}, \citenamefont {Hong}, \citenamefont {Hong}, \citenamefont {Kim}, \citenamefont {Kim}, \citenamefont {Kweon}, \citenamefont {Lee}, \citenamefont {Lim},\ and\ \citenamefont {Seo}}]{PhysRevC.111.014902}%
  \BibitemOpen
  \bibfield  {author} {\bibinfo {author} {\bibfnamefont {J.}~\bibnamefont {Kim}}, \bibinfo {author} {\bibfnamefont {J.}~\bibnamefont {Park}}, \bibinfo {author} {\bibfnamefont {B.}~\bibnamefont {Hong}}, \bibinfo {author} {\bibfnamefont {J.}~\bibnamefont {Hong}}, \bibinfo {author} {\bibfnamefont {E.-J.}\ \bibnamefont {Kim}}, \bibinfo {author} {\bibfnamefont {Y.}~\bibnamefont {Kim}}, \bibinfo {author} {\bibfnamefont {M.}~\bibnamefont {Kweon}}, \bibinfo {author} {\bibfnamefont {S.~H.}\ \bibnamefont {Lee}}, \bibinfo {author} {\bibfnamefont {S.}~\bibnamefont {Lim}},\ and\ \bibinfo {author} {\bibfnamefont {J.}~\bibnamefont {Seo}},\ }\bibfield  {title} {\bibinfo {title} {{Investigation of suppression of $\Upsilon(nS)$ in relativistic heavy-ion collisions at $\sqrt{{s}_{NN}}=200$ GeV and 5.02 TeV}},\ }\href {https://doi.org/10.1103/PhysRevC.111.014902} {\bibfield  {journal} {\bibinfo  {journal} {Phys. Rev. C}\ }\textbf {\bibinfo {volume} {111}},\ \bibinfo {pages} {014902} (\bibinfo {year} {2025})}\BibitemShut
  {NoStop}%
\bibitem [{\citenamefont {{ S. Gupta and R. Sharma }}(2014)}]{gupta20141}%
  \BibitemOpen
  \bibfield  {author} {\bibinfo {author} {\bibnamefont {{ S. Gupta and R. Sharma }}},\ }\bibfield  {title} {\bibinfo {title} {{Thermalization of quarkonia at energies available at the CERN Large Hadron Collider}},\ }\href {https://doi.org/10.1103/PhysRevC.89.057901} {\bibfield  {journal} {\bibinfo  {journal} {Phys. Rev.}\ }\textbf {\bibinfo {volume} {C89}},\ \bibinfo {pages} {057901} (\bibinfo {year} {2014})},\ \Eprint {https://arxiv.org/abs/1401.2930} {arXiv:1401.2930 [nucl-th]} \BibitemShut {NoStop}%
\bibitem [{\citenamefont {Kumar}\ \emph {et~al.}(2023)\citenamefont {Kumar}, \citenamefont {Sarkar}, \citenamefont {Bhaduri},\ and\ \citenamefont {Jaiswal}}]{Kumar:2023acr}%
  \BibitemOpen
  \bibfield  {author} {\bibinfo {author} {\bibfnamefont {D.}~\bibnamefont {Kumar}}, \bibinfo {author} {\bibfnamefont {N.}~\bibnamefont {Sarkar}}, \bibinfo {author} {\bibfnamefont {P.~P.}\ \bibnamefont {Bhaduri}},\ and\ \bibinfo {author} {\bibfnamefont {A.}~\bibnamefont {Jaiswal}},\ }\bibfield  {title} {\bibinfo {title} {{Examination of thermalization of quarkonia at energies available at the CERN Large Hadron Collider}},\ }\href {https://doi.org/10.1103/PhysRevC.107.064906} {\bibfield  {journal} {\bibinfo  {journal} {Phys. Rev. C}\ }\textbf {\bibinfo {volume} {107}},\ \bibinfo {pages} {064906} (\bibinfo {year} {2023})},\ \Eprint {https://arxiv.org/abs/2303.02900} {arXiv:2303.02900 [hep-ph]} \BibitemShut {NoStop}%
\bibitem [{\citenamefont {Brambilla}\ \emph {et~al.}(2018)\citenamefont {Brambilla}, \citenamefont {Escobedo}, \citenamefont {Soto},\ and\ \citenamefont {Vairo}}]{brambilla20181}%
  \BibitemOpen
  \bibfield  {author} {\bibinfo {author} {\bibfnamefont {N.}~\bibnamefont {Brambilla}}, \bibinfo {author} {\bibfnamefont {M.~A.}\ \bibnamefont {Escobedo}}, \bibinfo {author} {\bibfnamefont {J.}~\bibnamefont {Soto}},\ and\ \bibinfo {author} {\bibfnamefont {A.}~\bibnamefont {Vairo}},\ }\bibfield  {title} {\bibinfo {title} {Heavy quarkonium suppression in a fireball},\ }\href {https://doi.org/10.1103/PhysRevD.97.074009} {\bibfield  {journal} {\bibinfo  {journal} {Phys. Rev. D}\ }\textbf {\bibinfo {volume} {97}},\ \bibinfo {pages} {074009} (\bibinfo {year} {2018})}\BibitemShut {NoStop}%
\bibitem [{\citenamefont {Brambilla}\ \emph {et~al.}(2019)\citenamefont {Brambilla}, \citenamefont {Escobedo}, \citenamefont {Vairo},\ and\ \citenamefont {Vander~Griend}}]{Brambilla:2019tpt}%
  \BibitemOpen
  \bibfield  {author} {\bibinfo {author} {\bibfnamefont {N.}~\bibnamefont {Brambilla}}, \bibinfo {author} {\bibfnamefont {M.~A.}\ \bibnamefont {Escobedo}}, \bibinfo {author} {\bibfnamefont {A.}~\bibnamefont {Vairo}},\ and\ \bibinfo {author} {\bibfnamefont {P.}~\bibnamefont {Vander~Griend}},\ }\bibfield  {title} {\bibinfo {title} {{Transport coefficients from in medium quarkonium dynamics}},\ }\href {https://doi.org/10.1103/PhysRevD.100.054025} {\bibfield  {journal} {\bibinfo  {journal} {Phys. Rev.}\ }\textbf {\bibinfo {volume} {D100}},\ \bibinfo {pages} {054025} (\bibinfo {year} {2019})},\ \Eprint {https://arxiv.org/abs/1903.08063} {arXiv:1903.08063 [hep-ph]} \BibitemShut {NoStop}%
\bibitem [{\citenamefont {Kim}\ \emph {et~al.}(2018)\citenamefont {Kim}, \citenamefont {Petreczky},\ and\ \citenamefont {Rothkopf}}]{Kim:2018yhk}%
  \BibitemOpen
  \bibfield  {author} {\bibinfo {author} {\bibfnamefont {S.}~\bibnamefont {Kim}}, \bibinfo {author} {\bibfnamefont {P.}~\bibnamefont {Petreczky}},\ and\ \bibinfo {author} {\bibfnamefont {A.}~\bibnamefont {Rothkopf}},\ }\bibfield  {title} {\bibinfo {title} {{Quarkonium in-medium properties from realistic lattice NRQCD}},\ }\href {https://doi.org/10.1007/JHEP11(2018)088} {\bibfield  {journal} {\bibinfo  {journal} {JHEP}\ }\textbf {\bibinfo {volume} {11}},\ \bibinfo {pages} {088}},\ \Eprint {https://arxiv.org/abs/1808.08781} {arXiv:1808.08781 [hep-lat]} \BibitemShut {NoStop}%
\bibitem [{\citenamefont {Larsen}\ \emph {et~al.}(2019)\citenamefont {Larsen}, \citenamefont {Meinel}, \citenamefont {Mukherjee},\ and\ \citenamefont {Petreczky}}]{Larsen:2019bwy}%
  \BibitemOpen
  \bibfield  {author} {\bibinfo {author} {\bibfnamefont {R.}~\bibnamefont {Larsen}}, \bibinfo {author} {\bibfnamefont {S.}~\bibnamefont {Meinel}}, \bibinfo {author} {\bibfnamefont {S.}~\bibnamefont {Mukherjee}},\ and\ \bibinfo {author} {\bibfnamefont {P.}~\bibnamefont {Petreczky}},\ }\bibfield  {title} {\bibinfo {title} {{Thermal broadening of bottomonia: Lattice nonrelativistic QCD with extended operators}},\ }\href {https://doi.org/10.1103/PhysRevD.100.074506} {\bibfield  {journal} {\bibinfo  {journal} {Phys. Rev. D}\ }\textbf {\bibinfo {volume} {100}},\ \bibinfo {pages} {074506} (\bibinfo {year} {2019})},\ \Eprint {https://arxiv.org/abs/1908.08437} {arXiv:1908.08437 [hep-lat]} \BibitemShut {NoStop}%
\bibitem [{\citenamefont {Banerjee}\ \emph {et~al.}(2012)\citenamefont {Banerjee}, \citenamefont {Datta}, \citenamefont {Gavai},\ and\ \citenamefont {Majumdar}}]{datta20121}%
  \BibitemOpen
  \bibfield  {author} {\bibinfo {author} {\bibfnamefont {D.}~\bibnamefont {Banerjee}}, \bibinfo {author} {\bibfnamefont {S.}~\bibnamefont {Datta}}, \bibinfo {author} {\bibfnamefont {R.}~\bibnamefont {Gavai}},\ and\ \bibinfo {author} {\bibfnamefont {P.}~\bibnamefont {Majumdar}},\ }\bibfield  {title} {\bibinfo {title} {Heavy quark momentum diffusion coefficient from lattice qcd},\ }\href {https://doi.org/10.1103/PhysRevD.85.014510} {\bibfield  {journal} {\bibinfo  {journal} {Phys. Rev. D}\ }\textbf {\bibinfo {volume} {85}},\ \bibinfo {pages} {014510} (\bibinfo {year} {2012})}\BibitemShut {NoStop}%
\bibitem [{\citenamefont {Ding}\ \emph {et~al.}(2011)\citenamefont {Ding}, \citenamefont {Francis}, \citenamefont {Kaczmarek}, \citenamefont {Karsch}, \citenamefont {Satz},\ and\ \citenamefont {Soldner}}]{Ding:2011hr}%
  \BibitemOpen
  \bibfield  {author} {\bibinfo {author} {\bibfnamefont {H.~T.}\ \bibnamefont {Ding}}, \bibinfo {author} {\bibfnamefont {A.}~\bibnamefont {Francis}}, \bibinfo {author} {\bibfnamefont {O.}~\bibnamefont {Kaczmarek}}, \bibinfo {author} {\bibfnamefont {F.}~\bibnamefont {Karsch}}, \bibinfo {author} {\bibfnamefont {H.}~\bibnamefont {Satz}},\ and\ \bibinfo {author} {\bibfnamefont {W.}~\bibnamefont {Soldner}},\ }\bibfield  {title} {\bibinfo {title} {{Heavy Quark diffusion from lattice QCD spectral functions}},\ }\href {https://doi.org/10.1088/0954-3899/38/12/124070} {\bibfield  {journal} {\bibinfo  {journal} {J. Phys. G}\ }\textbf {\bibinfo {volume} {38}},\ \bibinfo {pages} {124070} (\bibinfo {year} {2011})},\ \Eprint {https://arxiv.org/abs/1107.0311} {arXiv:1107.0311 [nucl-th]} \BibitemShut {NoStop}%
\bibitem [{\citenamefont {Banerjee}\ \emph {et~al.}(2022{\natexlab{b}})\citenamefont {Banerjee}, \citenamefont {Datta},\ and\ \citenamefont {Laine}}]{Banerjee:2022uge}%
  \BibitemOpen
  \bibfield  {author} {\bibinfo {author} {\bibfnamefont {D.}~\bibnamefont {Banerjee}}, \bibinfo {author} {\bibfnamefont {S.}~\bibnamefont {Datta}},\ and\ \bibinfo {author} {\bibfnamefont {M.}~\bibnamefont {Laine}},\ }\bibfield  {title} {\bibinfo {title} {{Lattice study of a magnetic contribution to heavy quark momentum diffusion}},\ }\href {https://doi.org/10.1007/JHEP08(2022)128} {\bibfield  {journal} {\bibinfo  {journal} {JHEP}\ }\textbf {\bibinfo {volume} {08}},\ \bibinfo {pages} {128}},\ \Eprint {https://arxiv.org/abs/2204.14075} {arXiv:2204.14075 [hep-lat]} \BibitemShut {NoStop}%
\bibitem [{\citenamefont {Brambilla}\ \emph {et~al.}(2022{\natexlab{b}})\citenamefont {Brambilla}, \citenamefont {Leino}, \citenamefont {Philipsen}, \citenamefont {Reisinger}, \citenamefont {Vairo},\ and\ \citenamefont {Wagner}}]{Brambilla:2021wqs}%
  \BibitemOpen
  \bibfield  {author} {\bibinfo {author} {\bibfnamefont {N.}~\bibnamefont {Brambilla}}, \bibinfo {author} {\bibfnamefont {V.}~\bibnamefont {Leino}}, \bibinfo {author} {\bibfnamefont {O.}~\bibnamefont {Philipsen}}, \bibinfo {author} {\bibfnamefont {C.}~\bibnamefont {Reisinger}}, \bibinfo {author} {\bibfnamefont {A.}~\bibnamefont {Vairo}},\ and\ \bibinfo {author} {\bibfnamefont {M.}~\bibnamefont {Wagner}},\ }\bibfield  {title} {\bibinfo {title} {{Lattice gauge theory computation of the static force}},\ }\href {https://doi.org/10.1103/PhysRevD.105.054514} {\bibfield  {journal} {\bibinfo  {journal} {Phys. Rev. D}\ }\textbf {\bibinfo {volume} {105}},\ \bibinfo {pages} {054514} (\bibinfo {year} {2022}{\natexlab{b}})},\ \Eprint {https://arxiv.org/abs/2106.01794} {arXiv:2106.01794 [hep-lat]} \BibitemShut {NoStop}%
\bibitem [{\citenamefont {Brambilla}\ \emph {et~al.}(2022{\natexlab{c}})\citenamefont {Brambilla}, \citenamefont {Chung}, \citenamefont {Vairo},\ and\ \citenamefont {Wang}}]{Brambilla:2021egm}%
  \BibitemOpen
  \bibfield  {author} {\bibinfo {author} {\bibfnamefont {N.}~\bibnamefont {Brambilla}}, \bibinfo {author} {\bibfnamefont {H.~S.}\ \bibnamefont {Chung}}, \bibinfo {author} {\bibfnamefont {A.}~\bibnamefont {Vairo}},\ and\ \bibinfo {author} {\bibfnamefont {X.-P.}\ \bibnamefont {Wang}},\ }\bibfield  {title} {\bibinfo {title} {{QCD static force in gradient flow}},\ }\href {https://doi.org/10.1007/JHEP01(2022)184} {\bibfield  {journal} {\bibinfo  {journal} {JHEP}\ }\textbf {\bibinfo {volume} {01}},\ \bibinfo {pages} {184}},\ \Eprint {https://arxiv.org/abs/2111.07811} {arXiv:2111.07811 [hep-ph]} \BibitemShut {NoStop}%
\bibitem [{\citenamefont {Moore}\ and\ \citenamefont {Teaney}(2005)}]{moore20051}%
  \BibitemOpen
  \bibfield  {author} {\bibinfo {author} {\bibfnamefont {G.~D.}\ \bibnamefont {Moore}}\ and\ \bibinfo {author} {\bibfnamefont {D.}~\bibnamefont {Teaney}},\ }\bibfield  {title} {\bibinfo {title} {How much do heavy quarks thermalize in a heavy ion collision?},\ }\href {https://doi.org/10.1103/PhysRevC.71.064904} {\bibfield  {journal} {\bibinfo  {journal} {Phys. Rev. C}\ }\textbf {\bibinfo {volume} {71}},\ \bibinfo {pages} {064904} (\bibinfo {year} {2005})}\BibitemShut {NoStop}%
\bibitem [{\citenamefont {Mustafa}(2005)}]{mustafa20051}%
  \BibitemOpen
  \bibfield  {author} {\bibinfo {author} {\bibfnamefont {M.~G.}\ \bibnamefont {Mustafa}},\ }\bibfield  {title} {\bibinfo {title} {Energy loss of charm quarks in the quark-gluon plasma: Collisional vs radiative losses},\ }\href {https://doi.org/10.1103/PhysRevC.72.014905} {\bibfield  {journal} {\bibinfo  {journal} {Phys. Rev. C}\ }\textbf {\bibinfo {volume} {72}},\ \bibinfo {pages} {014905} (\bibinfo {year} {2005})}\BibitemShut {NoStop}%
\bibitem [{\citenamefont {Caron-Huot}\ and\ \citenamefont {Moore}(2008)}]{Caron-Huot:2007rwy}%
  \BibitemOpen
  \bibfield  {author} {\bibinfo {author} {\bibfnamefont {S.}~\bibnamefont {Caron-Huot}}\ and\ \bibinfo {author} {\bibfnamefont {G.~D.}\ \bibnamefont {Moore}},\ }\bibfield  {title} {\bibinfo {title} {{Heavy quark diffusion in perturbative QCD at next-to-leading order}},\ }\href {https://doi.org/10.1103/PhysRevLett.100.052301} {\bibfield  {journal} {\bibinfo  {journal} {Phys. Rev. Lett.}\ }\textbf {\bibinfo {volume} {100}},\ \bibinfo {pages} {052301} (\bibinfo {year} {2008})},\ \Eprint {https://arxiv.org/abs/0708.4232} {arXiv:0708.4232 [hep-ph]} \BibitemShut {NoStop}%
\bibitem [{\citenamefont {{C. H. Simon and G. D. Moore}}(2008)}]{huot20081}%
  \BibitemOpen
  \bibfield  {author} {\bibinfo {author} {\bibnamefont {{C. H. Simon and G. D. Moore}}},\ }\bibfield  {title} {\bibinfo {title} {Heavy quark diffusion in {QCD} and $\cal{N}$ = 4 {SYM} at next-to-leading order},\ }\href {https://doi.org/10.1088/1126-6708/2008/02/081} {\bibfield  {journal} {\bibinfo  {journal} {Journal of High Energy Physics}\ }\textbf {\bibinfo {volume} {2008}},\ \bibinfo {pages} {081} (\bibinfo {year} {2008})}\BibitemShut {NoStop}%
\bibitem [{\citenamefont {Kapusta}\ and\ \citenamefont {Gale}(2006)}]{kapusta_gale_2006}%
  \BibitemOpen
  \bibfield  {author} {\bibinfo {author} {\bibfnamefont {J.~I.}\ \bibnamefont {Kapusta}}\ and\ \bibinfo {author} {\bibfnamefont {C.}~\bibnamefont {Gale}},\ }\href {https://doi.org/10.1017/CBO9780511535130} {\emph {\bibinfo {title} {Finite-Temperature Field Theory: Principles and Applications}}},\ \bibinfo {edition} {2nd}\ ed.,\ Cambridge Monographs on Mathematical Physics\ (\bibinfo  {publisher} {Cambridge University Press},\ \bibinfo {year} {2006})\BibitemShut {NoStop}%
\bibitem [{Note3()}]{Note3}%
  \BibitemOpen
  \bibinfo {note} {The analysis is a little formal since the integral over $d^3q$ is UV-divergent and needs to be regularized by an ultraviolet completion of the theory~\cite {brambilla20131}. This detail does not change the arguments.}\BibitemShut {Stop}%
\bibitem [{\citenamefont {Bellac}(2011)}]{Bellac:2011kqa}%
  \BibitemOpen
  \bibfield  {author} {\bibinfo {author} {\bibfnamefont {M.~L.}\ \bibnamefont {Bellac}},\ }\href {https://doi.org/10.1017/CBO9780511721700} {\emph {\bibinfo {title} {{Thermal Field Theory}}}},\ Cambridge Monographs on Mathematical Physics\ (\bibinfo  {publisher} {Cambridge University Press},\ \bibinfo {year} {2011})\BibitemShut {NoStop}%
\bibitem [{\citenamefont {{Grandchamp L. and Rapp R.}}(2002)}]{grandchamp20021}%
  \BibitemOpen
  \bibfield  {author} {\bibinfo {author} {\bibnamefont {{Grandchamp L. and Rapp R.}}},\ }\bibfield  {title} {\bibinfo {title} {{Charmonium suppression and regeneration from SPS to RHIC}},\ }\href {https://doi.org/10.1016/S0375-9474(02)01027-8} {\bibfield  {journal} {\bibinfo  {journal} {Nucl. Phys.}\ }\textbf {\bibinfo {volume} {A709}},\ \bibinfo {pages} {415} (\bibinfo {year} {2002})},\ \Eprint {https://arxiv.org/abs/hep-ph/0205305} {arXiv:hep-ph/0205305 [hep-ph]} \BibitemShut {NoStop}%
\bibitem [{\citenamefont {{ L. Grandchamp, R. Rapp and G. E. Brown}}(2004)}]{grandchamp20041}%
  \BibitemOpen
  \bibfield  {author} {\bibinfo {author} {\bibnamefont {{ L. Grandchamp, R. Rapp and G. E. Brown}}},\ }\bibfield  {title} {\bibinfo {title} {{In medium effects on charmonium production in heavy ion collisions}},\ }\href {https://doi.org/10.1103/PhysRevLett.92.212301} {\bibfield  {journal} {\bibinfo  {journal} {Phys. Rev. Lett.}\ }\textbf {\bibinfo {volume} {92}},\ \bibinfo {pages} {212301} (\bibinfo {year} {2004})},\ \Eprint {https://arxiv.org/abs/hep-ph/0306077} {arXiv:hep-ph/0306077 [hep-ph]} \BibitemShut {NoStop}%
\bibitem [{\citenamefont {Greco}\ \emph {et~al.}(2004)\citenamefont {Greco}, \citenamefont {Ko},\ and\ \citenamefont {Rapp}}]{greco20031}%
  \BibitemOpen
  \bibfield  {author} {\bibinfo {author} {\bibfnamefont {V.}~\bibnamefont {Greco}}, \bibinfo {author} {\bibfnamefont {C.~M.}\ \bibnamefont {Ko}},\ and\ \bibinfo {author} {\bibfnamefont {R.}~\bibnamefont {Rapp}},\ }\bibfield  {title} {\bibinfo {title} {{Quark coalescence for charmed mesons in ultrarelativistic heavy ion collisions}},\ }\href {https://doi.org/10.1016/j.physletb.2004.06.064} {\bibfield  {journal} {\bibinfo  {journal} {Phys. Lett.}\ }\textbf {\bibinfo {volume} {B595}},\ \bibinfo {pages} {202} (\bibinfo {year} {2004})},\ \Eprint {https://arxiv.org/abs/nucl-th/0312100} {arXiv:nucl-th/0312100 [nucl-th]} \BibitemShut {NoStop}%
\bibitem [{\citenamefont {Du}\ \emph {et~al.}(2017)\citenamefont {Du}, \citenamefont {He},\ and\ \citenamefont {Rapp}}]{du20171}%
  \BibitemOpen
  \bibfield  {author} {\bibinfo {author} {\bibfnamefont {X.}~\bibnamefont {Du}}, \bibinfo {author} {\bibfnamefont {M.}~\bibnamefont {He}},\ and\ \bibinfo {author} {\bibfnamefont {R.}~\bibnamefont {Rapp}},\ }\bibfield  {title} {\bibinfo {title} {{Color Screening and Regeneration of Bottomonia in High-Energy Heavy-Ion Collisions}},\ }\href {https://doi.org/10.1103/PhysRevC.96.054901} {\bibfield  {journal} {\bibinfo  {journal} {Phys. Rev.}\ }\textbf {\bibinfo {volume} {C96}},\ \bibinfo {pages} {054901} (\bibinfo {year} {2017})},\ \Eprint {https://arxiv.org/abs/1706.08670} {arXiv:1706.08670 [hep-ph]} \BibitemShut {NoStop}%
\end{thebibliography}%
\end{document}